\documentclass[aps,prd,nofootinbib,showpacs,amsmath,amssymb,10pt]{revtex4}
\usepackage{graphicx}
\usepackage{amsfonts}
\usepackage{amssymb}
\usepackage{mathtools}
\usepackage{dsfont}
\usepackage{wrapfig}
\usepackage{longtable}
\usepackage{dcolumn}
\usepackage{multirow}

\begin{document}

\newcommand{\GeV}{\ensuremath{\, \mathrm{GeV}}}
\newcommand{\cm}{\ensuremath{\, \mathrm{cm}}}
\newcommand{\s}{\ensuremath{\, \mathrm{s}}}
\newcommand{\mub}{\ensuremath{{\bar{\mu}}}}
\newcommand{\nub}{\ensuremath{{\bar{\nu}}}}
\newcommand{\rhob}{\ensuremath{{\bar{\rho}}}}
\newcommand{\sigmab}{\ensuremath{{\bar{\sigma}}}}
\newcommand{\lambdab}{\ensuremath{{\bar{\lambda}}}}
\newcommand{\vev}{\ensuremath{{\langle v \rangle}}}


\title{Dark matter coupling to electroweak gauge and Higgs bosons:\\ an effective field theory approach}

\author{Jing-Yuan Chen} \email{chjy@uchicago.edu}

\author{Edward W.\ Kolb} \email{Rocky.Kolb@uchicago.edu} 

\author{Lian-Tao Wang} \email{liantaow@uchicago.edu}

\affiliation{Enrico Fermi Institute and Kavli Institute for Cosmological Physics, the University of Chicago, Chicago, Illinois \ \ 60637-1433 }

\begin{abstract}
If dark matter is a new species of particle produced in the early universe as a cold thermal relic (a weakly-interacting massive particle---WIMP), its present abundance, its scattering with matter in direct-detection experiments, its present-day annihilation signature in indirect-detection experiments, and its production and detection at colliders, depend crucially on the WIMP coupling to standard-model (SM) particles.  It is usually assumed that the WIMP couples to the SM sector through its interactions with quarks and leptons.  In this paper we explore the possibility that the WIMP coupling to the SM sector is via electroweak gauge and Higgs bosons.  In the absence of an ultraviolet-complete particle-physics model, we employ effective field theory to describe the WIMP--SM coupling. We consider both scalars and Dirac fermions as possible dark-matter candidates. Starting with an exhaustive list of operators up to dimension 8, we present detailed calculation of dark-matter annihilations to all possible final states, including $\gamma \gamma $, $\gamma Z$, $\gamma h$, $ZZ$, $Zh$, $W^+ W^-$, $hh$, and $f\bar{f}$, and demonstrate the correlations among them. We compute the mass scale of the effective field theory necessary to obtain the correct dark-matter mass density, and well as the resulting photon line signals. 
\end{abstract}

\pacs{98.70.Cq, 95.35.+d, 95.30.Cq, 95.55.Ka, 95.85.Ry}

\date{\today}

\maketitle

\section{Introduction \label{introduction}}
The identity of the dark-matter (DM) particle is one of the main outstanding mysteries in cosmology and particle physics. Many candidates have been proposed. Among the numerous possibilities, we choose in this paper to focus on the hypothesis that DM is a cold thermal relic of the early universe, a weakly interacting massive particle (WIMP).  It is well known that the thermal relic abundance of such a dark-matter particle, with mass in the GeV to TeV range, can provide the requisite DM mass density. At the same time, the weak but non-vanishing couplings between such a DM particle and  Standard Model (SM) particles open up the possibility of detecting dark matter through astronomical observations (indirect detection), terrestrial experiments (direct detection), and collider searches at the LHC. 
 
Both the thermal relic abundance and the signal in indirect-detection observations are controlled by the rate at which dark-matter particles annihilate into SM final states. In this paper, we focus on the scenario in which DM  particles annihilate only (or dominantly) to two-body final states containing SM electroweak gauge bosons and Higgs.  While the ultimate goal would be to understand the complete theory in which the DM is embedded and that would describe the DM--SM interactions, until the ultimate goal is realized we can study aspects of DM by employing an effective field theory (EFT) to describe the (non-renormalizable, low-energy) DM--SM interactions.  

In this paper, we consider dark-matter annihilation dominated by all possible SM di-boson final states, $\gamma \gamma$, $\gamma Z$, $\gamma h$, $ZZ$, $Zh$, $W^+ W^-$, and $hh$. As we will discuss in detail later, sometimes gauge invariance requires us to include Standard Model fermions, $f\bar{f}$, as final states. Dark-matter annihilation into fermions gives rise to a large class of indirect-detection signals, such as high-energy positron, antiproton, or neutrino fluxes. However, estimation of such signals (possibly with the exception of the neutrino flux) typically suffers from large astrophysical uncertainties. At the same time, DM annihilation to two-body final states involving one or two mono-chromatic photons (photon ``lines'') likely provides the cleanest indirect signal of dark matter. Therefore, the scenario we consider has the best chance of producing an indisputable discovery of a WIMP through indirect detection. SM gauge invariance usually implies correlation between the photon channels, $\gamma \gamma$, $\gamma h$, and $\gamma Z$, and other di-boson channels $ZZ$, $W^+ W^-$, and $Zh$. In addition, annihilation into the $hh$ final state can also give interesting  signals.  In this case, the annihilation could proceed through the so-called Higgs portal operator $J_{\rm DM} H^\dagger H$, in which $J_{\rm DM}$ is a SM singlet. It has no direct correlation with the other di-boson channels. 

It is not hard to imagine how our scenario can be (approximately) realized. We first note that due to gauge and Lorentz invariance, couplings with only the dark matter particle and the Standard Model particles must be non-renormalizable. The only exception is scalar dark matter with Higgs portal couplings. Therefore, in a complete theory, additional new states are almost always required to mediate dark matter annihilation. The choice of the properties of these new states, such as their spin and gauge quantum numbers,  can lead to very different dominant annihilation channels. Hence, the dominance of di-boson final states is at least as plausible as that of the fermionic final states. There are other potential arguments for di-boson dominance. For example, di-boson final states can come from the SM gauge field strength tensors, which have mass dimension two.  At the same time, SM fermonic final states come from operators with lowest mass dimension three.  The annihilation into di-bosons can also dominate if the annihilation to fermions is suppressed  by the velocity of DM particles. A well known example of this case is the Majorana fermion dark matter, such as the  bino LSP in the MSSM. We also note a possibility in which the DM particle has enhanced coupling to the top quark, but it is too light to annihilate into $t \bar{t}$ \cite{Jackson:2009kg} and annihilation proceeds through a loop process into SM di-boson final states. 

We will parameterize our ignorance of the detailed physics of dark-matter annihilation by using an effective field theory approach, manifestly preserving SM gauge symmetries. We assume that the only light states at the weak scale are the DM and the SM particles. In this case, we can integrate out the unknown new physics which mediates the DM interaction with SM states, and study the signal of  DM annihilation using  non-renormalizable operators of the form ${\cal O} \propto J_{\rm DM} \cdot J_{\rm SM}$, where the SM current $J_{\rm SM}$ consists of $W^{\pm}$, $Z$, $\gamma$, and $h$. We could, in principle, also include gluons as possible final states.  However, in practice, the signal of this final state is similar to that of the SM quarks. For simplicity, we will not consider it further in our paper.  We will only consider the cases in which both $J_{\rm DM}$  and $J_{\rm SM}$ are SM gauge singlets. Moreover, we assume $J_{\rm DM}$ only consists of the neutral DM particle. In principle, the DM particle could also be part of a multiplet. Therefore, we are implicitly assuming that the additional states in the DM multiplet become heavy after electroweak symmetry breaking and can be integrated out.  

Our study is motivated by the recent claims of a $\gamma$-ray line of energy around 130 GeV in the Fermi data \cite{Abdo:2010nc,Fermi_sypo,Bringmann:2012vr,Weniger:2012tx,Su:2012ft,Rajaraman:2012db}, which could be interpreted as a line from dark-matter annihilation at the galactic center. But we consider the full range of possibilities and will not be restricted only to the parameter space that can give such a signal.  

In the following, we first categorize all effective operators (up to dimension 8), suppressed by the new physics scale $\Lambda$. We then proceed to compute the annihilation rates into all possible two-body SM final states. We will focus on $2 \to 2$ annihilation processes: Processes with a larger number of particles in the final state, such as $2 \to 3$ processes, while possibly important in certain cases,  are typically subdominant in comparison to $2\rightarrow 2$ processes due to phase space suppression. We then derive the value of $\Lambda$ required to have the correct thermal relic abundance. We also compute the strength of the indirect detection signals, with particular focus on a possible photon line from the galactic center. 

Some of the topics studied in this paper have been considered earlier \cite{Cotta:2012nj,Rajaraman:2012fu}. However, there are some differences in emphasis between these works and our current work. In particular, Ref.\ \cite{Cotta:2012nj} considered dark matter candidates that are charged under $SU(2)_L$, as well as effective operators that are not invariant under $SU(2)_L$; these constructions allow the possibility that the UV complete theory may have already undergone electroweak symmetry breaking (EWSB). In contrast, we assume that the light Higgs field is solely responsible for electroweak symmetry breaking, and include it in effective flied theory.  Ref.\ \cite{Rajaraman:2012fu} mainly focuses on the gamma-ray line. While we are also motivated by the gamma-ray line, we will also consider other signals.  We also include a detailed study of the thermal relic abundance.

This paper is organized as follows:  In the next section we describe the notation used for the particles and couplings in the theories we study.  Section \ref{factors} discusses the types of operators, as well as the contributions of the initial and final state operators to the square of the matrix element.  Section \ref{bunchotables} contains the results for the annihilation cross sections for all the two-body final states for all of the various operators.  Section \ref{lines} calculates the present-day annihilation cross section into photons for sample masses of the DM, assuming the mass parameter of the EFT is the value necessary to result in the observed dark-matter density.  We conclude in Sec.\ \ref{conclusions}.

\section{Particles and Couplings \label{particles} }

We will consider the possibility that the WIMP is a complex scalar particle, $\phi$, or a Dirac or Majorana fermion, $\chi$.  We denote the WIMP mass as $M$.  The momenta of the annihilating WIMPs are denoted by $p$ and $p'$.  The spins of the initial-state fermionic WIMPS, if present, will be denoted as $s$ and $s'$.  We will consider various bilinears of fermionic WIMPS.  We use bilinears formed with $\mathds{1}$, $\gamma^5\equiv -(i/4!)\epsilon^{\mu\nu\rho\sigma}\gamma_\mu \gamma_\nu \gamma_\rho \gamma_\sigma$, $\gamma^\mu$, $\gamma^{\mu 5}\equiv \gamma^\mu \gamma^5$, and $\gamma^{\mu\nu}\equiv (i/2)[\gamma^\mu, \gamma^\nu]$.

For Majorana fermion WIMPs, using the basis $\bar{v}^{s'}(p')=i\left[u^{s'}(p')\right]^T \gamma^2\gamma^0$, the only difference from Dirac fermions is that a particle is identified with its antiparticle. It is conventional and convenient to put a factor of $1/2$ for Majorana fermions into the bilinears $\bar{\chi}...\chi$, due to the identification of particle and antiparticle. Using the fact that $\gamma^{\mu \, T} \gamma^0 \gamma^2=\gamma^2 \gamma^0 \gamma^\mu$, we have for Majorana WIMPs $\frac{1}{2}\bar{\chi}\gamma^\mu\chi = 0$ and $\frac{1}{2}\bar{\chi}\gamma^{\mu\nu}\chi = 0$, and the contributions of $\frac{1}{2}\bar{\chi}\chi$, $\frac{1}{2}\bar{\chi}\gamma^5\chi$, and $\frac{1}{2}\bar{\chi}\gamma^{\mu 5}\chi$ to $\mathcal{M}$ are the same as those of Dirac fermions, but without the factor $1/2$.

The SM Higgs doublet is denoted as $H$, with vacuum expectation value (vev) $\vev/\sqrt{2}$, where $\vev=246\ \GeV$.  The physical Higgs boson is denoted as $h$.  We will always work in the unitary gauge.

Electroweak gauge bosons are the $SU(2)_W$ gauge fields $W^a$ for $a=1, 2, 3$, and the the $U(1)_Y$ hypercharge gauge field $B$.  After the electroweak symmetry breaking change of basis, we denote $A_{\mu\nu} \equiv 2\partial_{[\mu} A_{\nu]}$ where $A$ may be a photon, $W$, or $Z$ boson, depending on the context.\footnote{Throughout, we use the notation for symmetrization and antisymmetrization of indices $2A_{[\mu}B_{\nu]}=A_\mu B_\nu - A_\nu B_\mu$ and $2A_{\{\mu}B_{\nu\}}=A_\mu B_\nu + A_\nu B_\mu$} Note that $A_{\mu\nu}$ is defined with derivatives only, \textit{i.e.,} without the structure constant term. The CP-violating electric dipole tensors are $\widetilde{W}^a_{\,\mu\nu}\equiv W^{a\;\rho\sigma}\epsilon_{\rho\sigma\mu\nu}/2$ and $\widetilde{B}_{\,\mu\nu}\equiv B^{\rho\sigma}\epsilon_{\rho\sigma\mu\nu}/2$. If the final state is two vector bosons $AA'$, then $m$ and $m'$ will denote their masses, $k$ and $k'$ will denote their momenta, and $r$ and $r'$ will denote their polarizations. If the final state is $Ah$, $k$ will denote the momentum of $A$ and $k'$ will denote the momentum of $h$. 

We denote by $C$ the coefficients in the EW mixing matrix:
\begin{eqnarray}
\left(
\begin{array}{c} 
W^1 \\ W^2 \\ W^3 \\ B 
\end{array}
\right) 
& = &
\left(
\begin{array}{cccc}
C_{1 W^+} & C_{1 W^-} &    0    &      0      \\ 
C_{2 W^+} & C_{2 W^-} &    0    &      0      \\
    0     &     0     &  C_{3Z} & C_{3\gamma} \\
    0     &     0     &  C_{YZ} & C_{Y\gamma}
\end{array}
\right) 
\left(
\begin{array}{c}
  W^+ \\
  W^- \\
  Z   \\
\gamma
\end{array}
\right) 
= 
\left(\begin{array}{cccc} 
1/\sqrt{2} &  1/\sqrt{2} &       0       &   0   \\
i/\sqrt{2} & -i/\sqrt{2} &       0       &   0   \\
    0      &      0      &  \cos\theta_W & \sin\theta_W \\
    0      &      0      & -\sin\theta_W & \cos\theta_W 
\end{array} 
\right)
\left(
\begin{array}{c}
  W^+ \\
  W^- \\
  Z   \\
\gamma
\end{array}
\right) \ .
\end{eqnarray}
\section{Initial and Final-State Matrix Element Factors \label{factors} }

Each SM gauge-invariant vertex operator can be written as a WIMP factor, which contains the initial state of WIMPs, multiplied by a SM factor, which contains SM particles. A  consequence is if there is only one vertex operator responsible for WIMP annihilation, then in computing the square of the matrix element, $\left|\mathcal{M}\right|^2$, we can compute the contributions from the WIMP factor and the SM factor separately. However, if a linear combination of vertex operators are responsible for WIMP annihilation, the different operators can connect the same set of initial and final states, and $\left|\mathcal{M}\right|^2$ will consist of interference terms in addition to one term from each operator.  Here, we avoid this complication by  ignoring possible linear combinations of multiple vertex operators. 

The initial-state WIMP factors are rather simple.  For scalar WIMPs we can form $J_{DM}$ operators with mass-dimension 2 or mass-dimension 3.  For fermion WIMPs, $J_{DM}$ operators have mass dimension 3.  The initial-state WIMP factors are discussed and listed in Section \ref{initial}.  By assumption, the final-state SM factors will only contain gauge bosons or Higgs bosons (or fermions in the case of tensor operators as discussed in Sec.\ \ref{4mixed}). They have mass-dimension 2, 4, or 5.  They are discussed and listed in Section \ref{final}.

Given a vertex operator, there might be 2-to-3 or 2-to-4 annihilation processes. However, we will only consider 2-to-2 processes as 3-particle or 4-particle final states are suppressed by phase-space factors. This also helps us avoid the complicated 3-particle or 4-particle final state phase space integrals.

We can classify the possible terms as products of scalar/pseudoscalar terms,
vector/axial vector terms, and tensor terms that can produce a di-boson final state.  Since the DM mass we consider is close to the electroweak symmetry breaking scale and the Higgs boson mass, we will preserve the manifest $SU(2)_L \times U(1)_Y$ gauge symmetry in the operators we consider to maintain a sensible power counting in the EFT. We group possible operators into Hermitian combinations. 

The scalar/pseudoscalar terms are
\begin{equation}
\left. 
\parbox{28pt}{\hspace*{0.32cm}$\phi^\dagger\phi$\\ 
              \hspace*{0.47cm}$\bar{\chi}\chi$ \\ 
                             $\bar{\chi}i\gamma^5\chi$} 
\right\} 
\times 
\left\{  
\parbox{8.7cm}{$H^\dagger H$\hspace{30pt}
with final state \ $hh$\\
             $B_{\mu\nu}\ B^{\mu\nu}$\hspace{14.0pt}
with final states $\gamma\gamma$, $\gamma Z$, $ZZ$\\ 
             $B_{\mu\nu}\ \widetilde{B}^{ \mu\nu}$\hspace{14.0pt}
with final states $\gamma\gamma$, $\gamma Z$, $ZZ$\\ 
             $W^a_{\,\mu\nu}\ W^{a\,\mu\nu}$\hspace{1.8pt} 
with final states $\gamma\gamma$, $\gamma Z$, $ZZ$, $W^+W^-$\\ 
             $W^a_{\,\mu\nu}\ \widetilde{W}^{a\,\mu\nu}$\hspace{1.9pt}
with final states $\gamma\gamma$, $\gamma Z$, $ZZ$, $W^+W^-$ \ .}
\right. 
\end{equation}
The $H^\dagger H$ final state can appear in a renormalizible mass-dimension 4 operator (with a $\phi^\dagger\phi$ WIMP operator) or a mass-dimension 5 operator (with $\bar{\chi}\chi$ or $\bar{\chi}i\gamma^5\chi$ WIMP operators).  This is the so called Higgs portal. Since we are interested in operators which can, in principle, give a photon line, we will not consider the operator $\phi^\dagger\phi H^\dagger H$ as part of our EFT.  There are a total of 12 possible terms that lead to a photon in the final state: four terms of mass-dimension 6 and eight terms of mass-dimension 7.  The initial-state contributions to $\left|\mathcal{M}\right|^2$ are given in Table \ref{WIMPmatrix}, and the final-state contributions to $\left|\mathcal{M}\right|^2$ are given in Table \ref{gaugematrix}.  

Now we turn to the vector/axial vector terms.  First consider the WIMP factor $\phi^\dagger \partial^\mu \phi + h.c.$  Nonvanishing terms are
\begin{equation}
\left(\phi^\dagger \partial^\mu \phi + h.c. \right)
\times 
\left\{
\parbox{11.5cm}{
$\phantom{i} \left(B_{\lambda\mu} Y_H \, H^\dagger D^\lambda H + h.c. \right)$
with final state $Zh$ \\
$\phantom{i} \left(W^a_{\ \lambda\mu} \, H^\dagger t^aD^\lambda H + h.c. \right)$
with final state $Zh$ \\ 
$i \left(B_{\lambda\mu} Y_H \, H^\dagger D^\lambda H - h.c. \right)$
with final states $\gamma Z$, $ZZ$ \\ 
$i \left(\widetilde{B}_{\lambda\mu} Y_H \, H^\dagger D^\lambda H - h.c. \right)$
with final states $\gamma Z$, $ZZ$ \\
$i \left(W^a_{\ \lambda\mu}  \, H^\dagger t^aD^\lambda H - h.c. \right)$
with final states $\gamma Z$, $ZZ$, $W^+W^-$ \\ 
$i \left(\widetilde{W}^a_{\ \lambda\mu}  \, H^\dagger t^aD^\lambda H - h.c. \right)$
with final states $\gamma Z$, $ZZ$, $W^+W^-$ \ ,
}
\right. 
\end{equation}
leading to four terms with a photon in the final state. We choose the hypercharge normalization so that $Y_H=1/2$. 
The operator $\left(\phi^\dagger \partial^\mu \phi + h.c. \right)\times \left(\widetilde{B}_{\lambda\mu} Y_H \, H^\dagger D^\lambda H + h.c. \right)$ vanishes. This can be seen by expressing it as $\partial^\mu(\phi^\dagger\phi) \partial^\lambda(H^\dagger H) \widetilde{B}_{\lambda\mu} Y_H$. Integrating by parts, one moves $\partial^\mu$ onto $\widetilde{B}_{\lambda\mu}$ and $\partial^\lambda(H^\dagger H)$. $\partial^\mu\widetilde{B}_{\lambda\mu}$ vanishes identically, and $\partial^\mu\partial^\lambda(H^\dagger H)\widetilde{B}_{\lambda\mu}$ is a contraction between a term symmetric in $\{\lambda\mu\}$ with a term antisymmetric in $[\lambda\mu]$ and therefore vanishes.
A similar argument applies to the term $\left(\phi^\dagger \partial^\mu \phi + h.c. \right)\times \left(\widetilde{W}^a_{\ \lambda\mu}  \, H^\dagger t^a D^\lambda H + h.c. \right)$. We also note that operator $\left(\phi^\dagger \partial^\mu \phi + h.c. \right)\times \left({B}_{\lambda\mu} Y_H \, H^\dagger D^\lambda H + h.c. \right)$ does not produce $\gamma h$ final state at tree level, because $\partial^\mu A_{\lambda \mu}$ vanishes for an on-shell photon.

Now consider the remaining three WIMP vector operators $i\left(\phi^\dagger \partial^\mu \phi - h.c. \right)$, $\bar{\chi}\gamma^\mu\chi$, and $\bar{\chi}\gamma^{\mu 5}\chi$.  All terms result in a photon in the final state.  The operators are
\begin{equation}
\left. 
\parbox{70pt}{$i\left(\phi^\dagger \partial^\mu \phi - h.c. \right)$\\ 
             \hspace*{43pt}$\bar{\chi}\gamma^\mu\chi$\\
             \hspace*{39pt}$\bar{\chi}\gamma^{\mu 5}\chi$ } 
\right\} 
\times 
\left\{  
\parbox{11.5cm}{
$\phantom{i}\left(B_{\lambda\mu} Y_H \, H^\dagger D^\lambda H + h.c. \right)$
with final states $\gamma h$, $Z h$\\ 
$\phantom{i}\left(\widetilde{B}_{\lambda\mu} Y_H \, H^\dagger D^\lambda H + h.c. \right)$
with final states $\gamma h$, $Z h$ \\ 
$i \left(B_{\lambda\mu} Y_H \, H^\dagger D^\lambda H - h.c. \right)$
with final states $\gamma Z$, $ZZ$ \\ 
$i \left(\widetilde{B}_{\lambda\mu} Y_H \, H^\dagger D^\lambda H - h.c. \right)$
with final states $\gamma Z$, $ZZ$ \\
$\phantom{i}\left(W^a_{\ \lambda\mu} \, H^\dagger t^aD^\lambda H + h.c. \right)$ 
with final states $\gamma h$, $Z h$, $W^+W^-$ \\ 
$\phantom{i}\left(\widetilde{W}^a_{\ \lambda\mu} \, H^\dagger t^aD^\lambda H + h.c. \right)$
with final states $\gamma h$, $Z h$, $W^+W^-$ \\ 
$i \left(W^a_{\ \lambda\mu} \, H^\dagger t^aD^\lambda H - h.c. \right)$
with final states $\gamma Z$, $ZZ$, $W^+W^-$ \\ 
$i \left(\widetilde{W}^a_{\ \lambda\mu}  \, H^\dagger t^aD^\lambda H - h.c. \right)$
with final states $\gamma Z$, $ZZ$, $W^+W^-$ \ ,
}
\right. 
\end{equation}
for a total of 24 terms.   

The initial-state contributions to $\left|\mathcal{M}\right|^2$ are given in Table \ref{WIMPmatrix}, and the final-state contributions to $\left|\mathcal{M}\right|^2$ are given in Table \ref{gaugehiggs5}. 

Finally, consider tensor-like couplings.  There are 4 possible mass-dimension 7 terms, and two possible mass-dimension 5 terms.\footnote{Terms $\bar{\chi}\gamma^{\mu\nu}\chi B_{\lambda\mu} B^\lambda_{\ \nu}$ and $\bar{\chi}\gamma^{\mu\nu} \chi W^a_{\ \lambda\mu} W^{a\,\lambda}_{\ \ \ \, \nu}$ vanish because they are a product of an antisymmetric tensor ($\gamma^{\mu\nu}$) and a symmetric tensor (\textit{e.g.,} $B_{\lambda\mu} B^\lambda_{\ \nu}$).  Though less transparent, terms of the form $ \widetilde{B}_{\lambda\mu} B^{\lambda}_{\ \nu}$ and $ \widetilde{W}^a_{\ \lambda\mu} W^{a\,\lambda}_{\ \ \ \nu}$ are only non-zero  when $\mu=\nu$.
To see this, notice that $2B^Y_{\lambda\mu} \widetilde{B}^{Y\, \lambda}_{\ \ \ \, \nu}=B^Y_{[\lambda\mu]} B^Y_{[\rho\sigma]} \epsilon^{\rho\sigma\lambda}_{\ \ \ \ \nu}$, which is antisymmetric in both $[\mu\rho]$ and $[\mu\sigma]$ (see this by exchanging $\lambda\leftrightarrow\rho$ and $\lambda\leftrightarrow\sigma$). Since $\mu$ is antisymmetric with $\lambda, \rho, \sigma$, the only non-vanishing terms are those with $\mu=\nu$.}  The non-zero tensor terms are of the form
\begin{equation}
\parbox{1cm}{$\bar{\chi}\gamma^{\mu\nu} \chi$ } 
\times 
\left\{  
\parbox{250pt}{
$B_{\mu\nu}$ 
\hspace*{34pt} with final states $Z h$, $W^+W^-$, $f\bar{f}$\\ 
$\widetilde{B}_{\mu\nu}$
\hspace*{34pt}  with final states $Z h$, $W^+W^-$, $f\bar{f}$\\
$B_{\mu\nu} Y_H \, H^\dagger H$ 
with final states $\gamma h$, $Z h$, $W^+W^-$, $f\bar{f}$\\ 
$\widetilde{B}_{\mu\nu} Y_H \, H^\dagger H$ 
with final states $\gamma h$, $Z h$, $W^+W^-$, $f\bar{f}$\\ 
$W^a_{\ \mu\nu} H^\dagger t^a H$
with final states $\gamma h$, $Z h$, $W^+W^-$, $f\bar{f}$\\ 
$\widetilde{W}^a_{\ \mu\nu} H^\dagger t^a H$
with final states $\gamma h$, $Z h$, $W^+W^-$, $f\bar{f}$ \ ,
}
\right. 
\end{equation}
for a total of four terms leading to a photon in the final state. Although we are focusing on operators which  lead to di-boson final states, we have to include the fermonic final states in this case, as dictated by the structure of this class of operators (see Sec.\ \ref{4mixed}). 

The initial-state contributions to $\left|\mathcal{M}\right|^2$ are again given in Table \ref{WIMPmatrix},  and the final-state contributions to $\left|\mathcal{M}\right|^2$ are given in Table \ref{gaugehiggs4}. 

The 12 scalar/pseudoscalar plus 28 vector/axial vector plus 4 tensor terms lead to a total of 44 possible terms with photons in the final state.  Some processes have two annihilation modes containing photons, including $\gamma \gamma, \ \gamma Z,$ or $\gamma h$. Therefore,  a photon line with a given energy can arise from different processes.  An example using 130 GeV as the energy of the line is given in Table \ref{energies}.  

All final states for all the possible terms are calculated and given in Sec.\ \ref{bunchotables}.  The scalar/pseudoscalar terms are given in Tables \ref{resultone} through \ref{resultthreefive}, sorted by initial operators.  The vector/axial vector terms are given in Tables \ref{resultfourpp} through \ref{resultsevenm}, again sorted by initial operators.  Finally the tensor results are in Tables \ref{resulteight}-\ref{resultten}.

\subsection{Initial-state WIMP factors \label{initial}}

The initial-state WIMP factors appearing in annihilation matrix elements are shown in Table \ref{WIMPmatrix}, along with their contribution to $\left|\mathcal{M}\right|^2$.  The mass-dimension (either 2 or 3) is indicated.

\renewcommand*\arraystretch{1.5}
\begin{table}
\caption{\label{WIMPmatrix} Contributions to the matrix element from initial states. For fermions, the spins $s, s'$ have been averaged.}
\begin{ruledtabular}
\begin{tabular}{ccccc}
Mass dimension & Operator & Contribution to $\left|\mathcal{M}\right|^2$ & Notes \\ \hline\hline
2
&
$\phi^\dagger \phi$
&
1
&
\\
\hline
\multirow{2}{*}{3}
&
$\left(\phi^\dagger \partial^\mu \phi + h.c. \right)$
&
$(p+p')^\mu (p+p')^\mub$
\\
&
$i\left(\phi^\dagger \partial^\mu \phi - h.c. \right)$
&
$(p-p')^\mu (p-p')^\mub$ 
&
\\
[1ex]
\hline
\multirow{5}{*}{3} 
&
$\bar{\chi}\chi$
&
$p\cdot p'-M^2$ 
\\
&
$\bar{\chi}i\gamma^5\chi$ 
&
$p\cdot p'+M^2$ 
\\
&
$\bar{\chi}\gamma^\mu\chi$
&
$2p^{\{\mu} p'^{\mub\}} -g^{\mu\mub}(p\cdot p' +M^2)$
&
$a$
\\
&
$\bar{\chi}\gamma^{\mu 5} \chi$
&
$2p^{\{\mu} p'^{\mub\}} -g^{\mu\mub}(p\cdot p' - M^2)$
&
\\
&
$\bar{\chi}\gamma^{\mu\nu} \chi$
&
$2(p\cdot p' -M^2)(g^{\mu[\mub}g^{\nub]\nu}) - 4 p^{[\mu}  g^{\nu][\nub} p'^{\mub]}  
                                             - 4 p'^{[\mu} g^{\nu][\nub} p^{\mub]} $
&
$a$
\end{tabular}
$^a$ Operator vanishes for Majorana fermions.
\end{ruledtabular}
\end{table}

Note that $p\cdot p'-M^2 \propto s-4M^2$, where $s=(p+p')^2=(k+k')^2$ is the center-of-mass energy.  In the nonrelativistic (NR) limit, $s\rightarrow 4M^2 + M^2v^2$, so the scalar operator $\bar{\chi}\chi$ will have an NR annihilation cross section proportional to $v^2$.

\subsection{Final-state SM factors \label{final}}

\subsubsection{Final-state SM operators coupling to scalar/pseudoscalar DM operators \label{gauge}}

\begin{figure*}[th]
\begin{center}
\includegraphics[width=.47\linewidth,keepaspectratio]{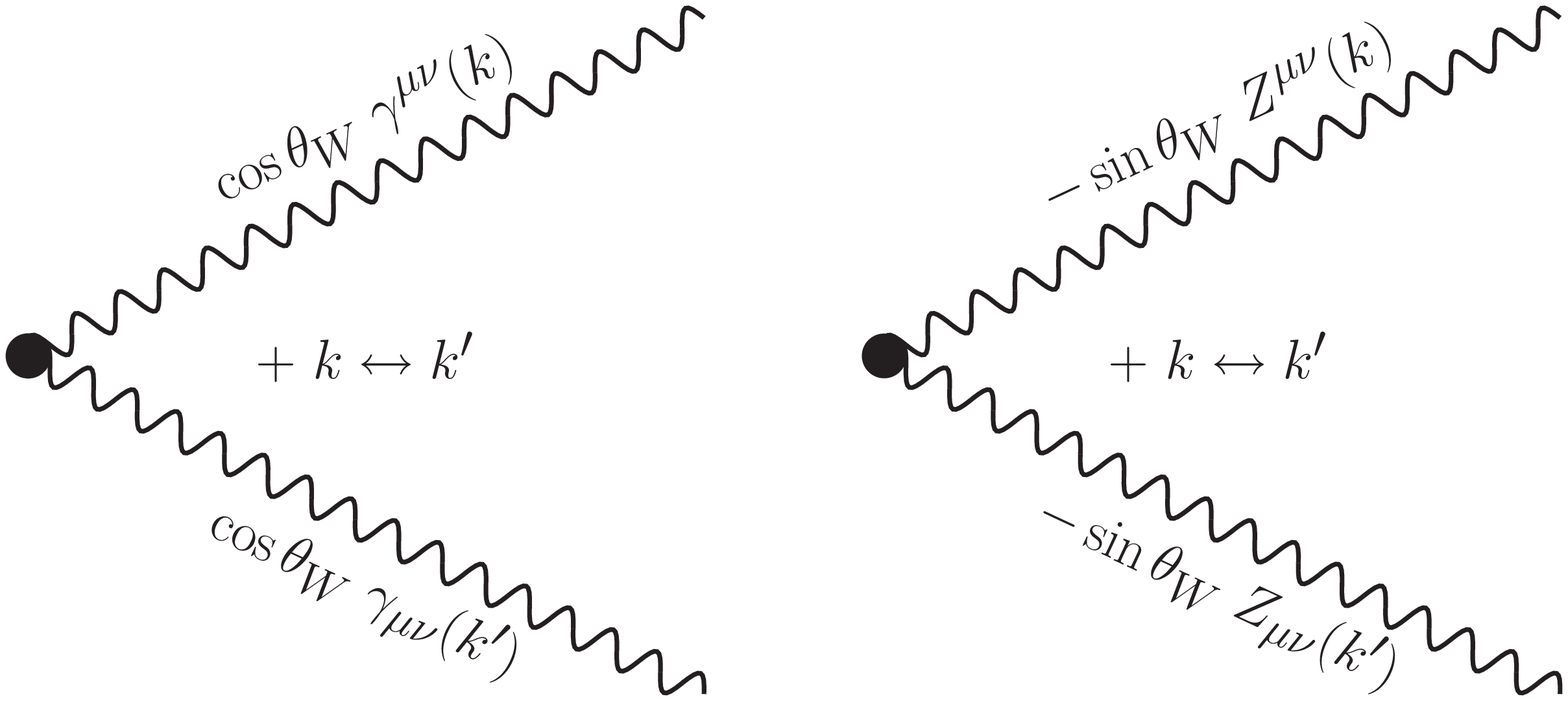}
\hspace*{24pt}
\includegraphics[width=.42\linewidth,keepaspectratio]{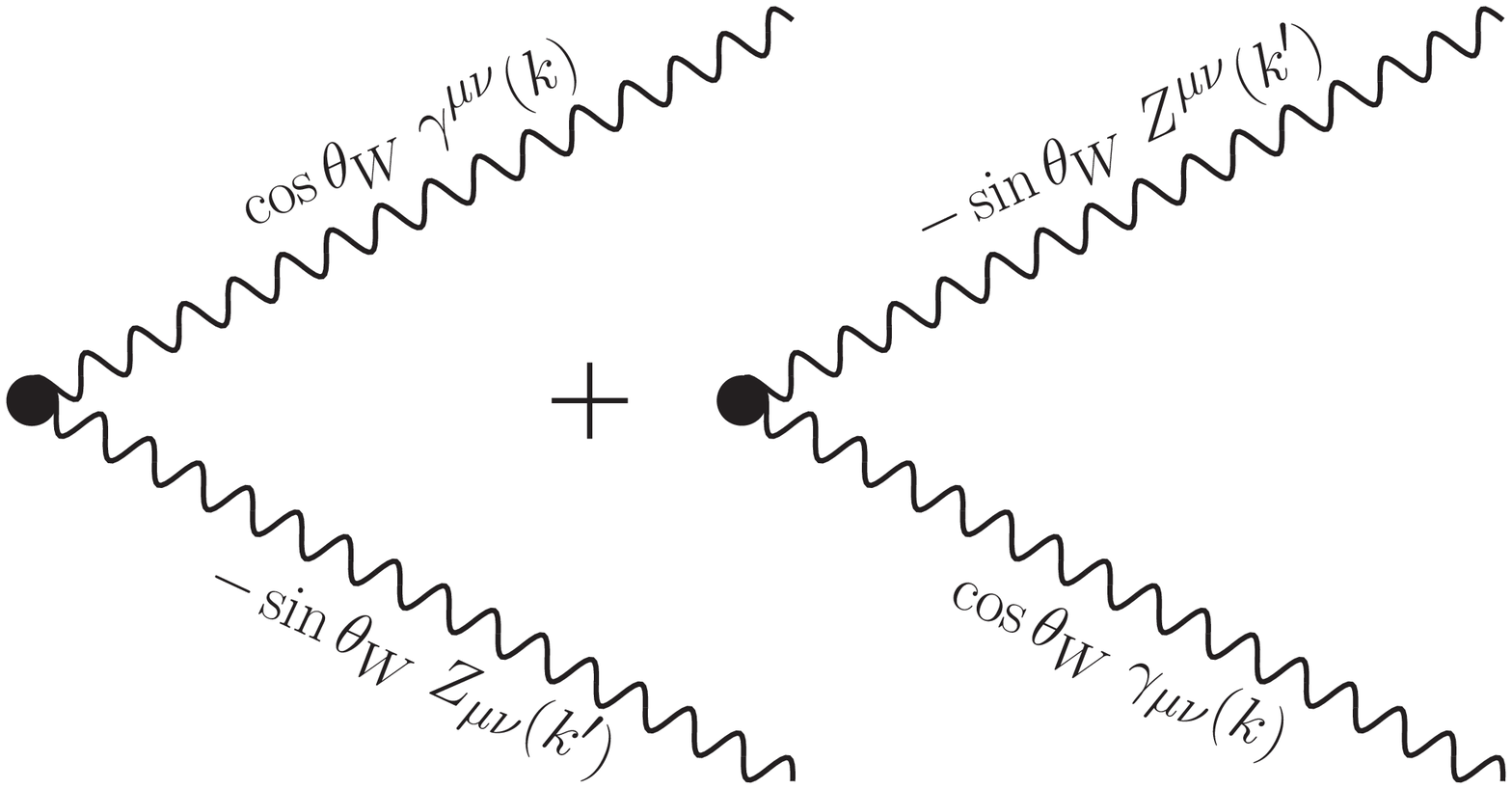}
\caption{Final-state dimension-4 diagrams of the form $B^{\mu\nu}B_{\mu\nu}$ that couple to scalar or pseudoscalar WIMP factors.\label{fig:BmunuBmunu}}
\end{center}
\end{figure*}

\begin{figure*}
\begin{center}
\includegraphics[width=.47\linewidth,keepaspectratio]{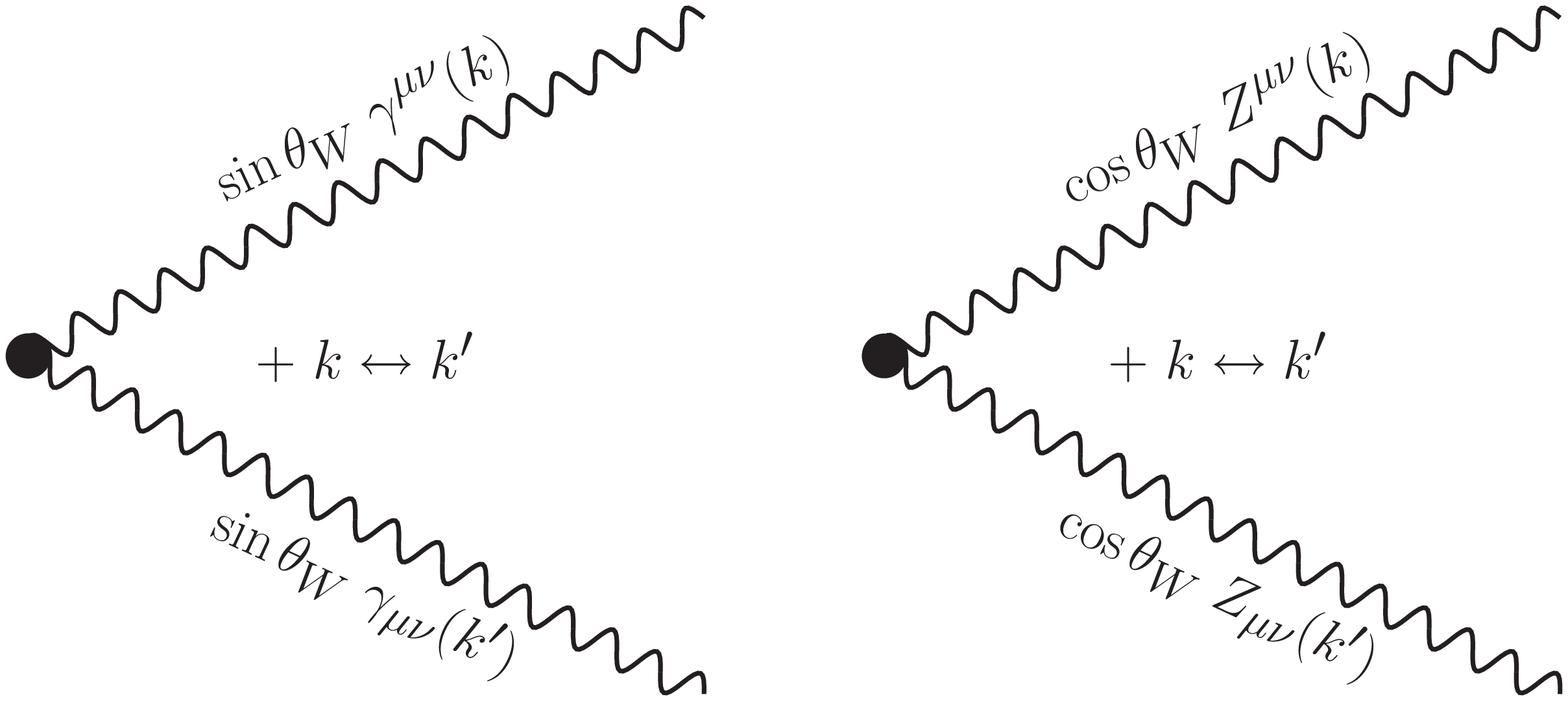} \hspace*{24pt}
\includegraphics[width=.42\linewidth,keepaspectratio]{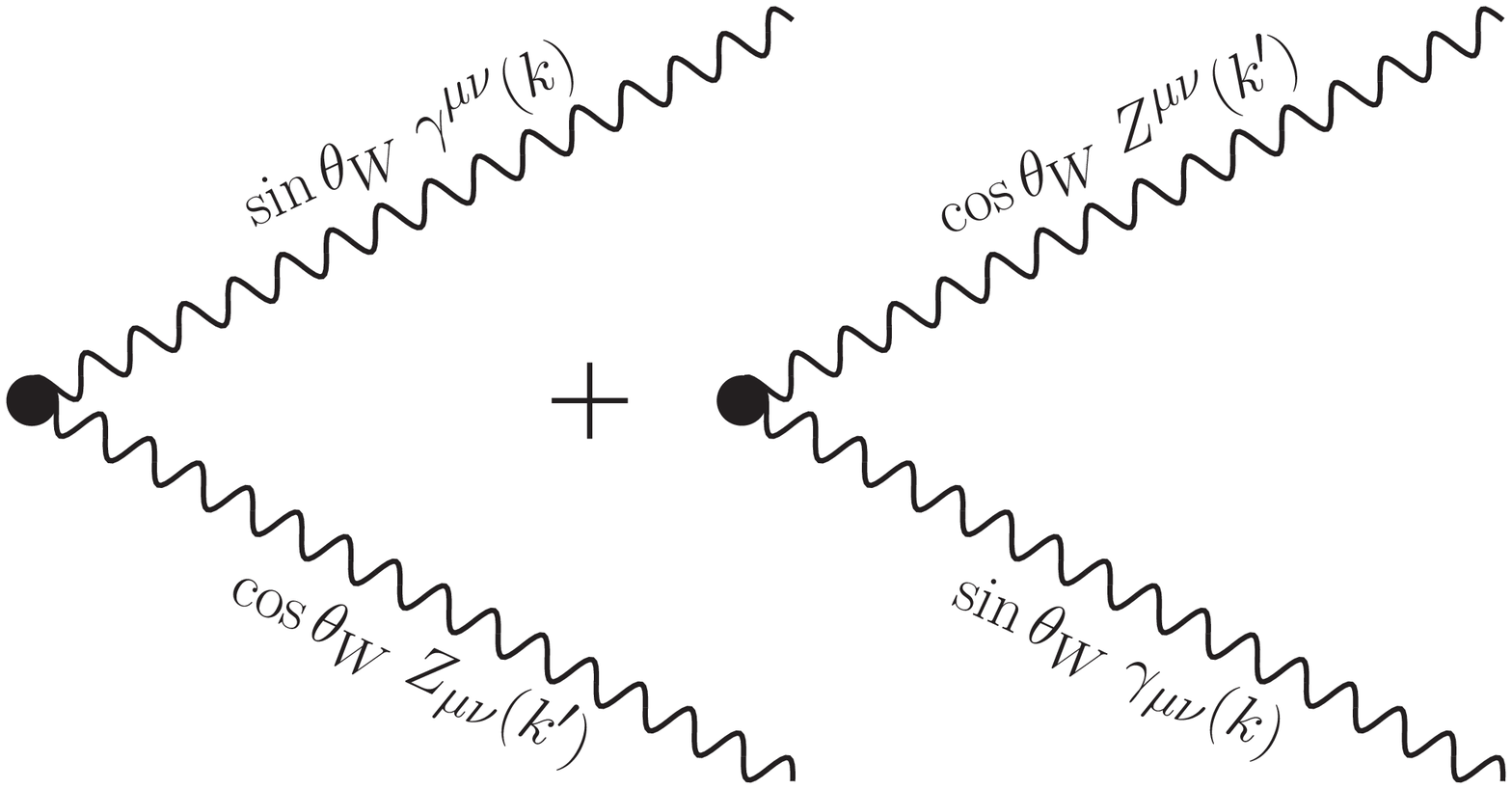} \\
\vspace*{24pt}
\includegraphics[width=.42\linewidth,keepaspectratio]{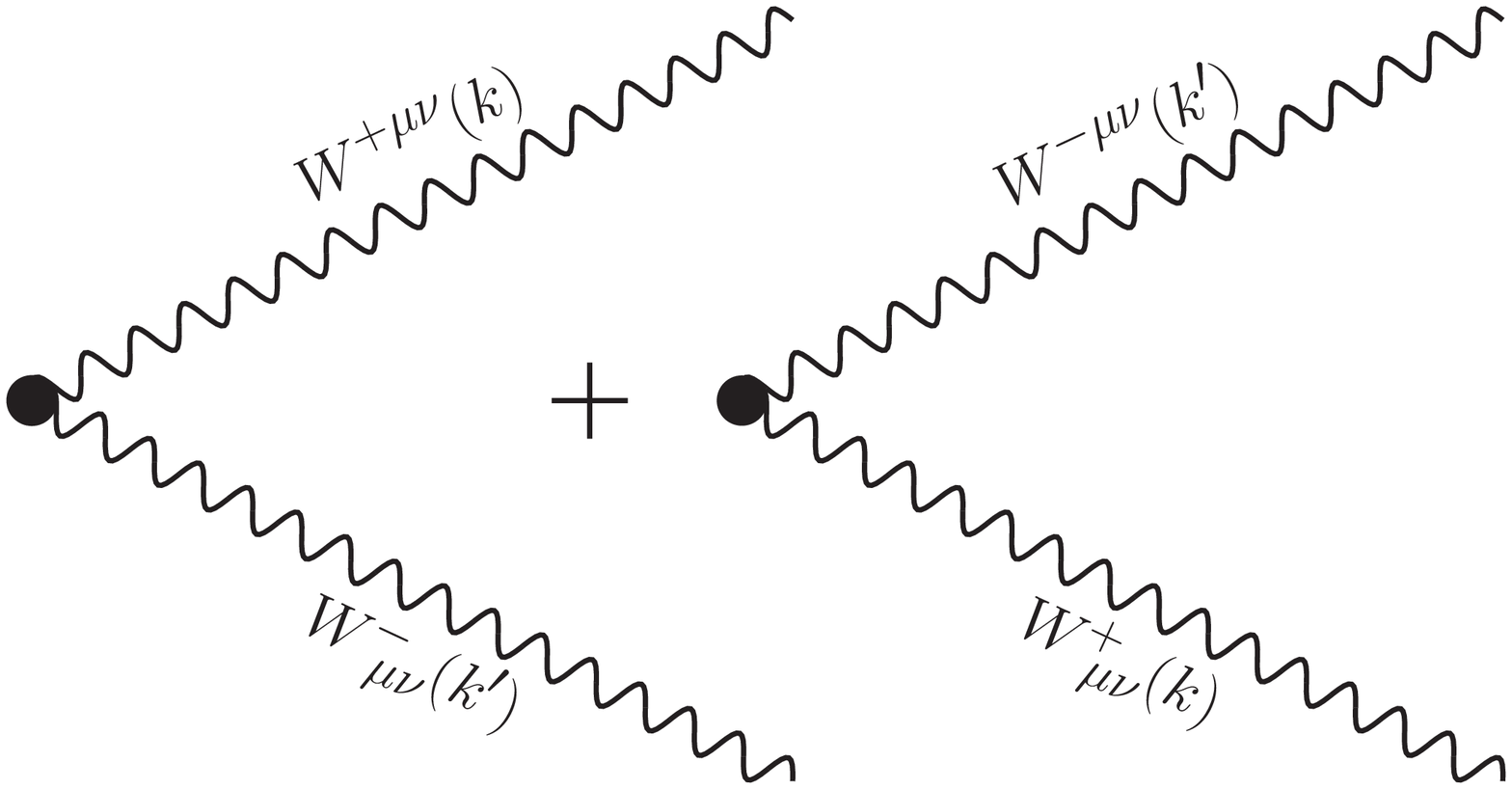}
\caption{Final-state dimension-4 diagrams of the form $W^{a\mu\nu}W^a_{\ \mu\nu}$ that couple to scalar or pseudoscalar WIMP factors.\label{fig:WmunuWmunu}}
\end{center}
\end{figure*}

Appearing in the matrix elements will be terms of the form $BB$, $B\widetilde{B}$, $W^a W^a$, $W^a\widetilde{W}^a$, or $H^\dagger H$.  The first four factors represent mass-dimension 4 operators, and the final contribution to $\left|\mathcal{M}\right|^2$ has mass-dimension 2.  After electroweak symmetry breaking we must change basis to express the matrix element in terms of $\gamma$'s, $Z$'s, or $W$'s.  Terms will appear as $C_{YA}C_{YA'}A A'$ or $C_{aA}C_{aA'}A A'$, where the $A$ and $A'$ can be either a photon or a $Z$ boson (or $W$ bosons in the case of $SU(2)_W$) and the $C$s are the coefficients of the EW mixing matrix.  The diagrams representing the possible final states are shown in Figs.\ \ref{fig:BmunuBmunu} and \ref{fig:WmunuWmunu}.  In these, and in all such diagrams, it should be understood that there are processes with gauge bosons replaced by their duals. The contributions to the matrix elements are give in Table \ref{gaugematrix}.

Another possible ``scalar'' final state is $H^\dagger H$, which has mass-dimension 2.  In the unitary gauge $H^\dagger H=\left(\vev+h\right)^2/2$.  This is also listed in Table \ref{gaugematrix}.  Of course, there will not be a photon in the two-body final state.


\renewcommand*\arraystretch{2.0}
\begin{table}
\caption{\label{gaugematrix} Contributions to the matrix element for SM final states that couple to scalar or pseudoscalar DM operators.  The mass dimension of the SM operator is indicated. The polarizations have been summed over.  The EW mixing-matrix factors are not explicitly included.  $A$ and $\widetilde{A}$ can be $\gamma$ or $Z$ for the $U(1)_Y$ case and additionally $W^+W^-$ for the $SU(2)_W$ case.}
\begin{ruledtabular}
\begin{tabular}{cccc}
Mass Dimension & Operator & Appears in $\mathcal{M}$ as & Contribution to $\left|\mathcal{M}\right|^2$  \\ 
\hline\hline
\multirow{2}{*}{$4$}
&
\multirow{2}{*}{$B_{\mu\nu} B^{\mu\nu};\ 
                 W^a_{\ \mu\nu} W^{a\,\mu\nu}$}
&
$A_{\mu \nu}A'^{\mu\nu}$ \ \ \ \ if $A=A'$
&
\multirow{2}{*}{$32 \left[(k\cdot k')^2 + m^2 m'^2/2 \right]$}
\\
&
&
$2 A_{\mu\{\nu} A'^{\mu \nu\}}$  if $A\neq A'$
&
\\
\hline
\multirow{2}{*}{$4$}
&
\multirow{2}{*}{$B_{\mu\nu} \widetilde{B}^{ \mu\nu};\ 
                 W^a_{\ \mu\nu} \widetilde{W}^{a\, \mu\nu}$}
&
$A_{\mu \nu}\widetilde{A}'^{\mu\nu}$ \ \ \ \ if $A=A'$
&
\multirow{2}{*}{$32 \left[(k\cdot k')^2 - m^2 m'^2/2 \right]$}
\\
&
&
$2 A_{\mu\{\nu} \widetilde{A}'^{\mu \nu\}}$ if $A\neq A'$
&
\\
\hline
$2$
&
$H^\dagger H$
&
$hh$
&
$\dfrac{1}{4}$
\\
[1ex]
\end{tabular}
\end{ruledtabular}
\end{table}

\subsubsection{Final-state SM operators coupling to vector/axial vector DM operators \label{5mixed}}

Mass-dimension 5 mixed Higgs/gauge-boson factors will be of the form of a gauge field times a product of the Higgs field times a derivative of the Higgs field.

\begin{figure*}[ht]
\begin{center}
\includegraphics[width=.47\linewidth,keepaspectratio]{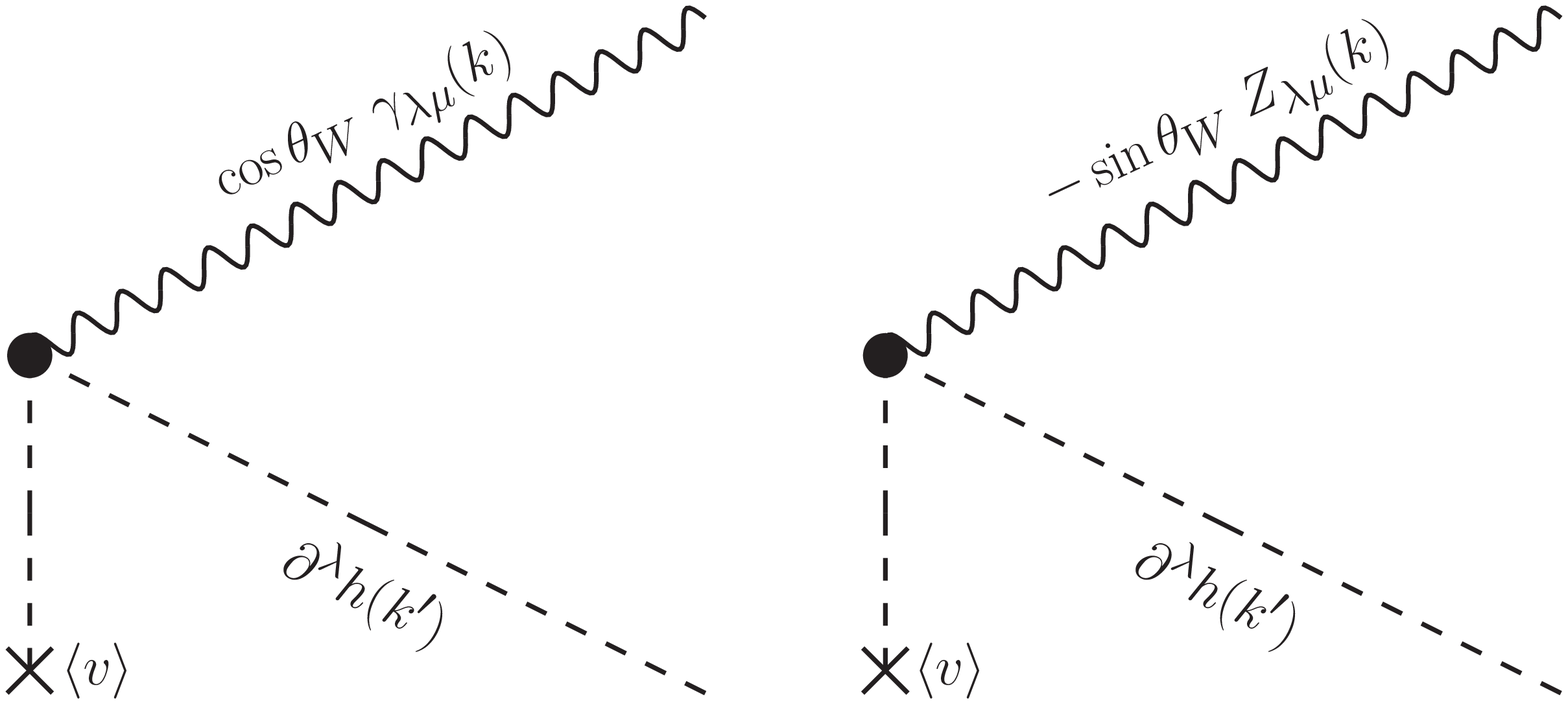}
\caption{Final-state dimension-5 diagrams of the form $(B_{\lambda\mu}Y_HH^\dagger D^\lambda H+h.c.)$ that couple to vector or axial-vector WIMP factors.\label{fig:BlambdamuHDlambdaHp}}
\end{center}
\end{figure*}

First consider the mass-dimension 5 terms coupling to hypercharge.  One of the possible terms is of the form $\left(B_{\lambda\mu} Y_H \, H^\dagger D^\lambda H + h.c. \right)$.  After symmetry breaking the terms responsible for 2-to-2 annihilation processes are $C_{YA} \, \vev A_{\lambda\mu} \partial^\lambda h /2$, where $A$ can be either a photon or a $Z$ boson.  For the operator with $B_{\lambda\mu}$ replaced by $\widetilde{B}_{\lambda\mu}$, the result is the same with $A_{\lambda\mu}$ replaced by $\widetilde{A}_{\lambda\mu}$.  These diagrams are illustrated in Fig.\ \ref{fig:BlambdamuHDlambdaHp}.

\begin{figure*}
\begin{center}
\includegraphics[width=.47\linewidth,keepaspectratio]{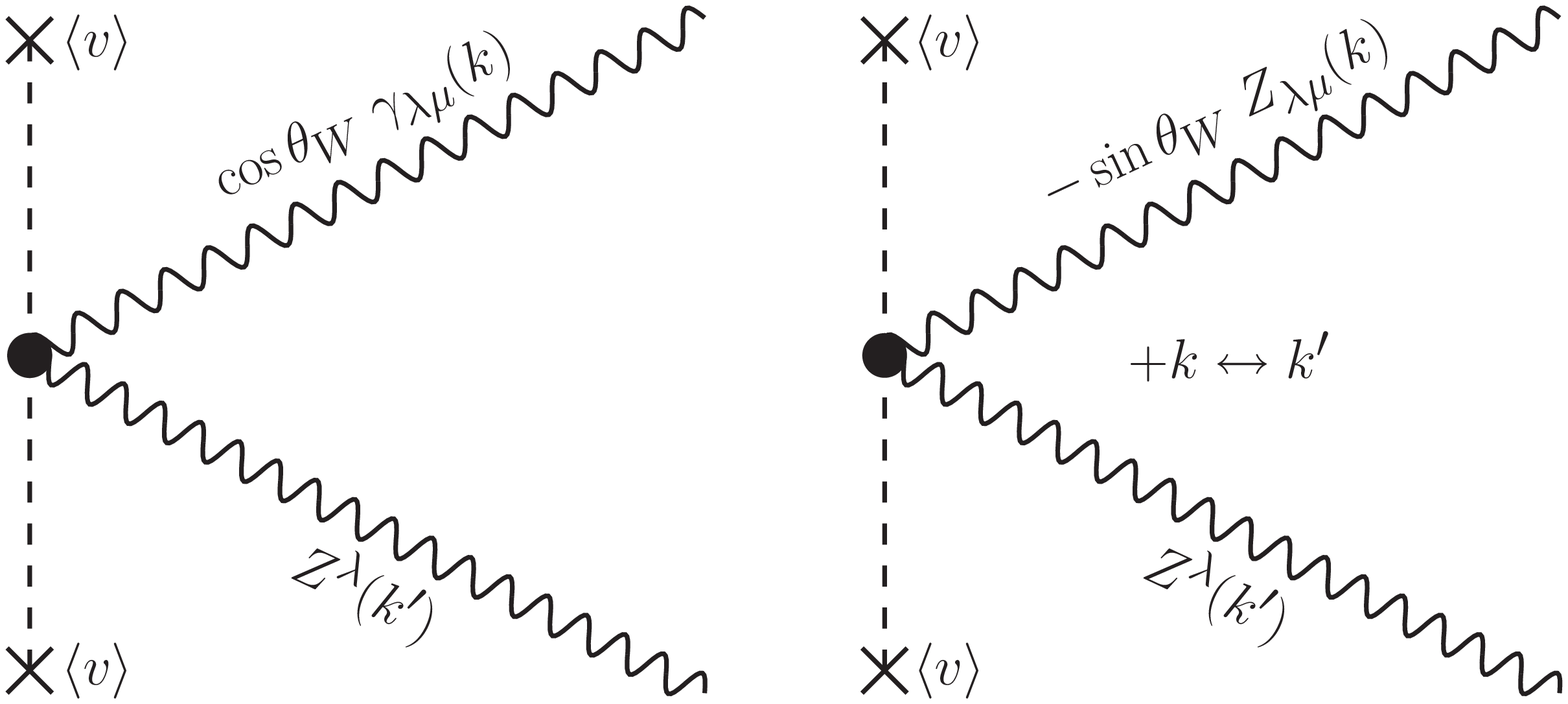}
\caption{Final-state dimension-5 diagrams of the form $(B_{\ \lambda\mu}Y_HH^\dagger D^\lambda H-h.c.)$ that couple to vector or axial-vector WIMP factors.\label{fig:BlambdamuHDlambdaHm}}
\end{center}
\end{figure*}

The other possible hypercharge term is $i \left(B_{\lambda\mu} Y_H \, H^\dagger D^\lambda H - h.c. \right)$.  This operator leads to terms of the form $-C_{YA} \, m_Z \vev A_{\lambda\mu} Z^\lambda /2$, where $A$ can be either photon or $Z$ boson.  Note that no Higgs is produced in 2-to-2 process for this operator.   Again, for the operator with $B_{\lambda\mu}$ replaced by $\widetilde{B}_{\lambda\mu}$, the result is the same with $A_{\lambda\mu}$ replaced by $\widetilde{A}_{\lambda\mu}$.  These final states are shown in Fig.\ \ref{fig:BlambdamuHDlambdaHm}.

\begin{figure*}
\begin{center}
\includegraphics[width=.47\linewidth,keepaspectratio]{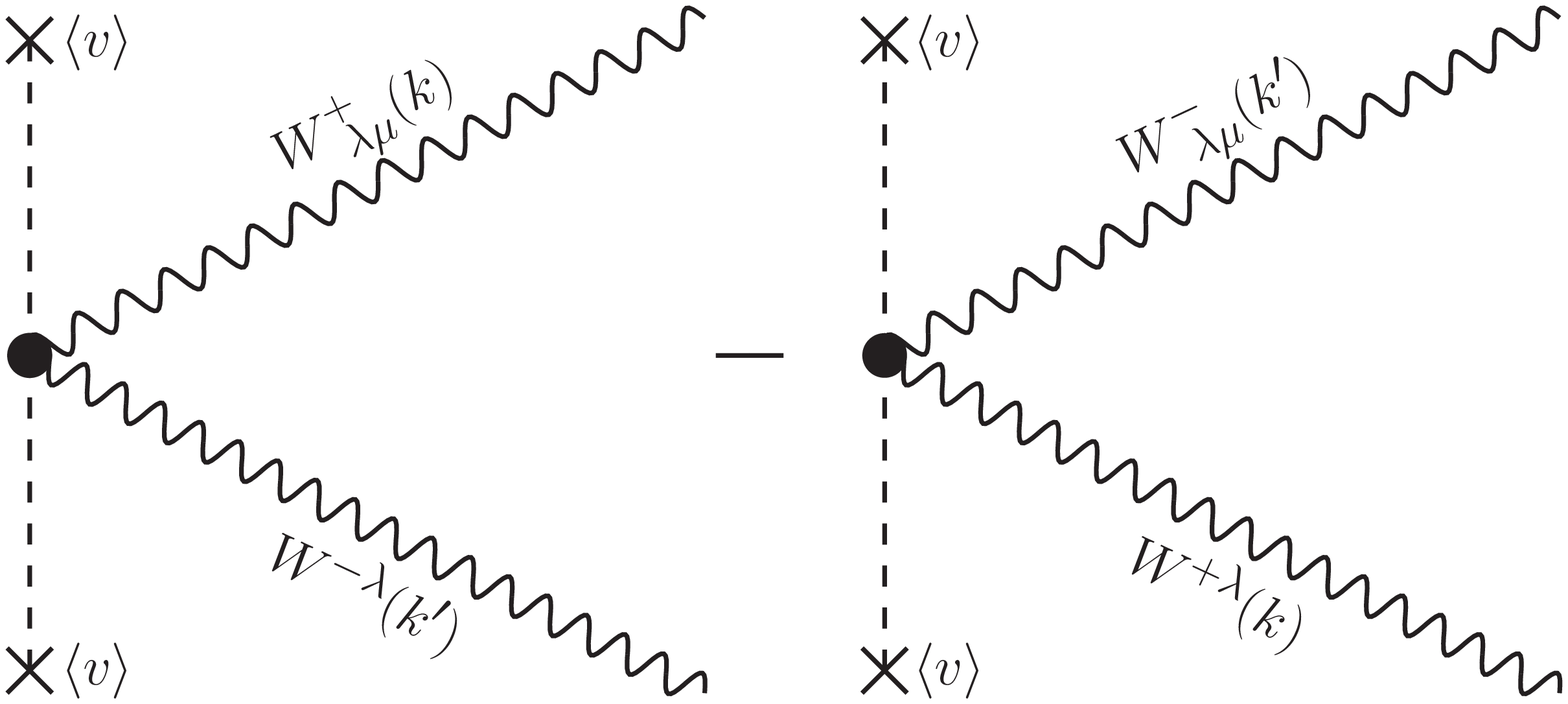} \hspace{24pt}
\includegraphics[width=.47\linewidth,keepaspectratio]{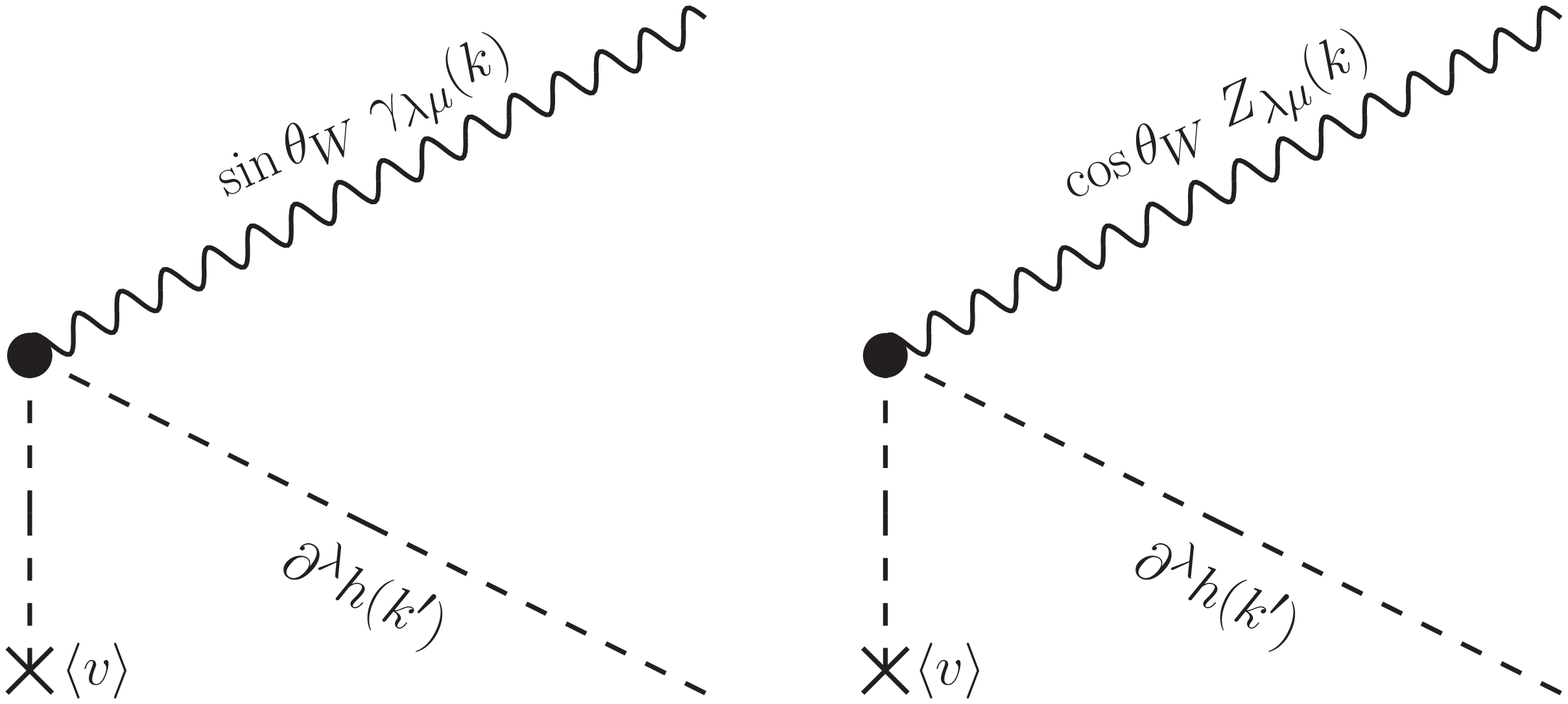}
\caption{Final-state dimension-5 diagrams of the form $(W^a_{\ \lambda\mu}H^\dagger t^a D^\lambda H+h.c.)$ that couple to vector or axial-vector WIMP factors.\label{fig:WlambdamuHDlambdaHp}}
\end{center}
\end{figure*}

\begin{figure*}
\begin{center}
\includegraphics[width=.47\linewidth,keepaspectratio]{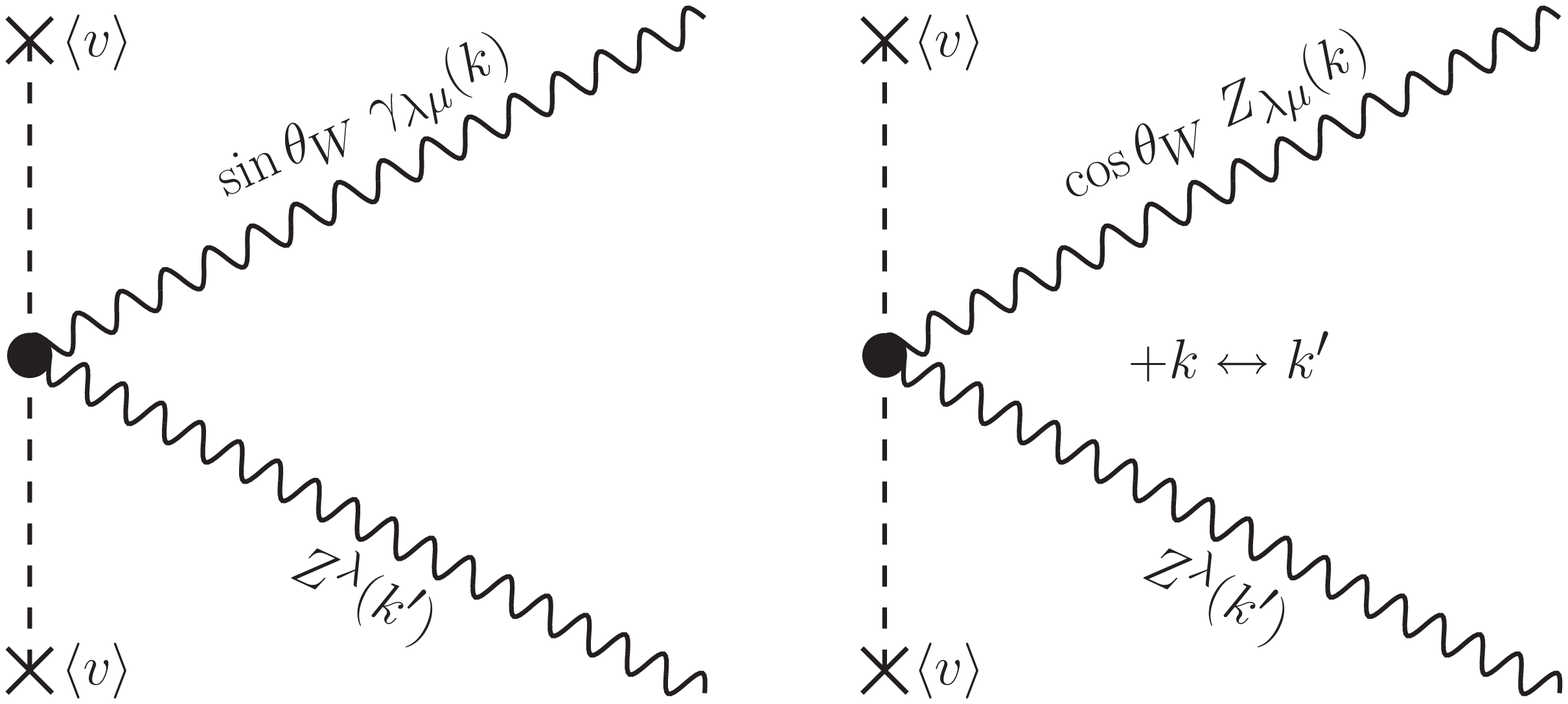} \hspace*{24pt}
\includegraphics[width=.47\linewidth,keepaspectratio]{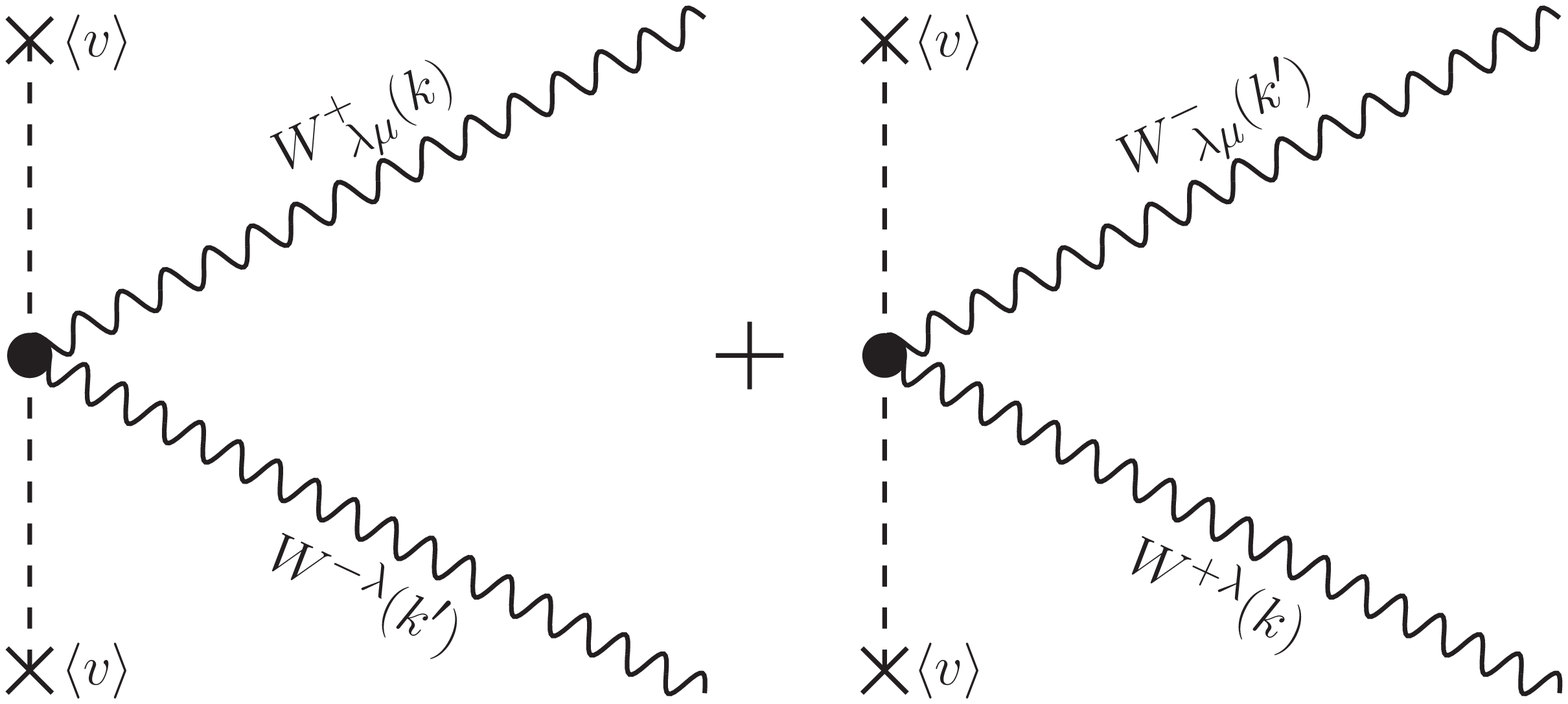}
\caption{Final-state dimension-5 diagrams of the form $(W^a_{\ \lambda\mu}H^\dagger t^aD^\lambda H-h.c.)$ that couple to vector or axial-vector WIMP factors.\label{fig:WlambdamuHDlambdaHm}}
\end{center}
\end{figure*}

Now consider the mass-dimension 5 terms coupling to $SU(2)_W$ fields.  One of the possible terms is of the form $\left(W^a_{\ \lambda\mu} H^\dagger t^a D^\lambda H + h.c. \right)$.  After symmetry breaking,  terms involving a Higgs are $-C_{3A}\vev A_{\lambda\mu}\partial^\lambda h/2$, where $A$ can be a photon or a $Z$.  These terms also have an annihilation channel into $W^+W^-$, which has a factor $i\vev m_W\left(W^+_{\ \lambda\mu}W^{-\lambda} -W^-_{\ \lambda\mu}W^{+\lambda}\right)/2$. (We ignore the structure constant term from the field strength tensor, since we are only interested in tree-level 2-to-2 processes.)  The associated final-state diagrams are shown in Figs.\ \ref{fig:WlambdamuHDlambdaHp} and \ref{fig:WlambdamuHDlambdaHm}.

The terms are given in Table \ref{gaugehiggs5}.  All the terms have the correct structure to couple to $\bar{\chi}\gamma^\mu \chi$, $\bar{\chi}\gamma^{\mu 5} \chi$, or $\phi^\dagger\partial^\mu\phi$ terms.  Also, all terms are proportional to the Higgs vacuum expectation value $\vev^2$ and some terms are proportional to $m_Z^2$ or $m_W^2$; we do not explicitly include these factors in the table, nor do we include factors of the electroweak mixing angles. All proper factors are included in the presentation of our final results in the next section. 

\begin{table}
\caption{\label{gaugehiggs5} Mass-dimension 5 contributions to the matrix element for mixed Higgs/gauge-boson final states.  The polarizations have been summed over.  The EW mixing-matrix factors are not explicitly included. These terms couple to vector/axial-vector WIMP operators.}
\begin{ruledtabular}
\begin{tabular}{ccc}
Operators & Appears in $\mathcal{M}$ as & Contribution to $\left|\mathcal{M}\right|^2$  \\ 
\hline\hline
$\left(B_{\lambda\mu} Y_H \, H^\dagger D^\lambda H + h.c. \right)$
&
$A_{\lambda\mu} \partial^\lambda h \ \ [A=\gamma\ \mathrm{or}\ Z]$
&
$-k_\mu k_\mub \, m_h^2 + 2(k\cdot k')k_{\{\mu}k'_{\mub\}} -(k\cdot k')^2 g_{\mu\mub}$
\\
\hline
\multirow{2}{*}{$\left(\widetilde{B}_{\lambda\mu} Y_H \, H^\dagger D^\lambda H + h.c. \right)$}
&
\multirow{2}{*}{$\widetilde{A}_{\lambda\mu} \partial^\lambda h \ \ [A=\gamma\ \mathrm{or}\ Z]$}
&
\multirow{2}{*}{\parbox{6cm}{$-k_\mu k_\mub \, m_h^2 - k'_\mu k'_\mub \, m^2 + 2(k\cdot k')k_{\{\mu}k'_{\mub\}}$\\
 $\ \ \ + \left(m^2 m_h^2 - (k\cdot k')^2 \right) g_{\mu\mub}$}}
\\
&
&
\\
\hline
\multirow{2}{*}{$i \left(B_{\lambda\mu} Y_H \, H^\dagger D^\lambda H - h.c. \right)$}
&
$A_{\lambda\mu} Z^\lambda \ \ [A=\gamma]$
&
$k_\mu k_\mub -\dfrac{(k\cdot k')^2}{m_Z^2} g_{\mu\mub} + 2\dfrac{k\cdot k'}{m_Z^2}k_{\{\mu}k'_{\mub\}}$
\\
&
$A_{\lambda\mu} Z^\lambda \ \ [A = Z]$
&
$k_\mu k_\mub + k'_\mu k'_\mub + 2\left(m_Z^2-\dfrac{(k\cdot k')^2}{m_Z^2}\right)g_{\mu\mub} + 2\left(2\dfrac{k\cdot k'}{m_Z^2}+3\right)\dfrac{k\cdot k'}{m_Z^2}k_{\{\mu}k'_{\mub\}}$
\\
[1ex]
\hline
\multirow{2}{*}{$i \left(\widetilde{B}_{\lambda\mu} Y_H \, H^\dagger D^\lambda H - h.c. \right)$}
&
$\widetilde{A}_{\lambda\mu} Z^\lambda \ \ [A=\gamma]$
&
$k_\mu k_\mub  - \dfrac{(k\cdot k')^2}{m_Z^2} g_{\mu\mub} + 2\dfrac{k\cdot k'}{m_Z^2}k_{\{\mu}k'_{\mub\}}$
\\
&
$\widetilde{A}_{\lambda\mu} Z^\lambda \ \ [A = Z]$
&
$- 2\left(m_Z - \dfrac{k\cdot k'}{m_Z}\right)^2 g_{\mu\mub} + 4 \left( \dfrac{k\cdot k'}{m_Z^2} - 1 \right) k_{\{\mu}k'_{\mub\}}$
\\
[1ex]
\hline
\multirow{2}{*}{$\left(W^a_{\ \lambda\mu}  \, H^\dagger t^aD^\lambda H + h.c. \right)$}
&
$A_{\lambda\mu} \partial^\lambda h \ \ [A=\gamma\ \mathrm{or}\ Z]$
&
$-k_\mu k_\mub \, m_h^2 + 2(k\cdot k')k_{\{\mu}k'_{\mub\}} -(k\cdot k')^2 g_{\mu\mub}$
\\
&
$W^+W^-$
&
$k_\mu k_\mub + k'_\mu k'_\mub +2\left(m_W^2-\dfrac{(k\cdot k')^2}{m_W^2}\right)g_{\mu\mub}+ 2\left(2\dfrac{k\cdot k'}{m_W^2}-3\right)\dfrac{k\cdot k'}{m_Z^2}k_{\{\mu}k'_{\mub\}}$
\\
[1ex]
\hline
\multirow{3}{*}{$\left(\widetilde{W}^a_{\ \lambda\mu}  \, H^\dagger t^aD^\lambda H + h.c. \right)$}
&
\multirow{2}{*}{$\widetilde{A}_{\lambda\mu} \partial^\lambda h \ \ [A=\gamma\ \mathrm{or}\ Z]$}
&
\multirow{2}{*}{\parbox{6cm}{$-k_\mu k_\mub \, m_h^2 - k'_\mu k'_\mub \, m^2 + 2(k\cdot k')k_{\{\mu}k'_{\mub\}}$\\
 $\ \ \ + \left(m^2 m_h^2 - (k\cdot k')^2 \right) g_{\mu\mub}$}}
\\
& & \\
&
$W^+W^-$
&
$ 4\left(\dfrac{k\cdot k'}{m_W^2}+1\right)k_{\{\mu}k'_{\mub\}} - 2\left(\dfrac{k\cdot k'}{m_W}+m_W\right)^2 $
\\
[1ex]
\hline
\multirow{3}{*}{$i \left(W^a_{\ \lambda\mu}  \, H^\dagger t^aD^\lambda H - h.c. \right)$}
&
$A_{\lambda\mu} Z^\lambda \ \ [A=\gamma]$
&
$k_\mu k_\mub  - \dfrac{(k\cdot k')^2}{m_Z^2} g_{\mu\mub} + 2\dfrac{k\cdot k'}{m_Z^2}k_{\{\mu}k'_{\mub\}}$
\\
&
$A_{\lambda\mu} Z^\lambda \ \ [A=Z]$
&
$k_\mu k_\mub + k'_\mu k'_\mub + 2\left(m_Z^2-\dfrac{(k\cdot k')^2}{m_Z^2}\right)g_{\mu\mub} + 2\left(2\dfrac{k\cdot k'}{m_Z^2}+3\right)\dfrac{k\cdot k'}{m_Z^2}k_{\{\mu}k'_{\mub\}}$
\\
&
$W^+W^-$
&
$k_\mu k_\mub + k'_\mu k'_\mub +2\left(m_W^2-\dfrac{(k\cdot k')^2}{m_W^2}\right)g_{\mu\mub}+ 2\left(2\dfrac{k\cdot k'}{m_W^2}+3\right)\dfrac{k\cdot k'}{m_Z^2}k_{\{\mu}k'_{\mub\}}$
\\
[1ex]
\hline
\multirow{3}{*}{$i \left(\widetilde{W}^a_{\ \lambda\mu}  \, H^\dagger t^aD^\lambda H - h.c. \right)$}
&
$\widetilde{A}_{\lambda\mu} Z^\lambda \ \ [A=\gamma]$
&
$k_\mu k_\mub  - \dfrac{(k\cdot k')^2}{m_Z^2} g_{\mu\mub} + 2\dfrac{k\cdot k'}{m_Z^2}k_{\{\mu}k'_{\mub\}}$
\\
&
$\widetilde{A}_{\lambda\mu} Z^\lambda \ \ [A=Z]$
&
$- 2\left(m_Z - \dfrac{k\cdot k'}{m_Z}\right)^2 g_{\mu\mub} + 4 \left( \dfrac{k\cdot k'}{m_Z^2} - 1 \right) k_{\{\mu}k'_{\mub\}}$
\\
&
$W^+W^-$
&
$ 4\left(\dfrac{k\cdot k'}{m_W^2}-1\right)k_{\{\mu}k'_{\mub\}} - 2\left(\dfrac{k\cdot k'}{m_W}-m_W\right)^2 $
\end{tabular}
\end{ruledtabular}
\end{table}

\subsubsection{Final-state SM operators coupling to tensor DM operators \label{4mixed}}

\begin{figure*}
\begin{center}
\includegraphics[width=.22\linewidth,keepaspectratio]{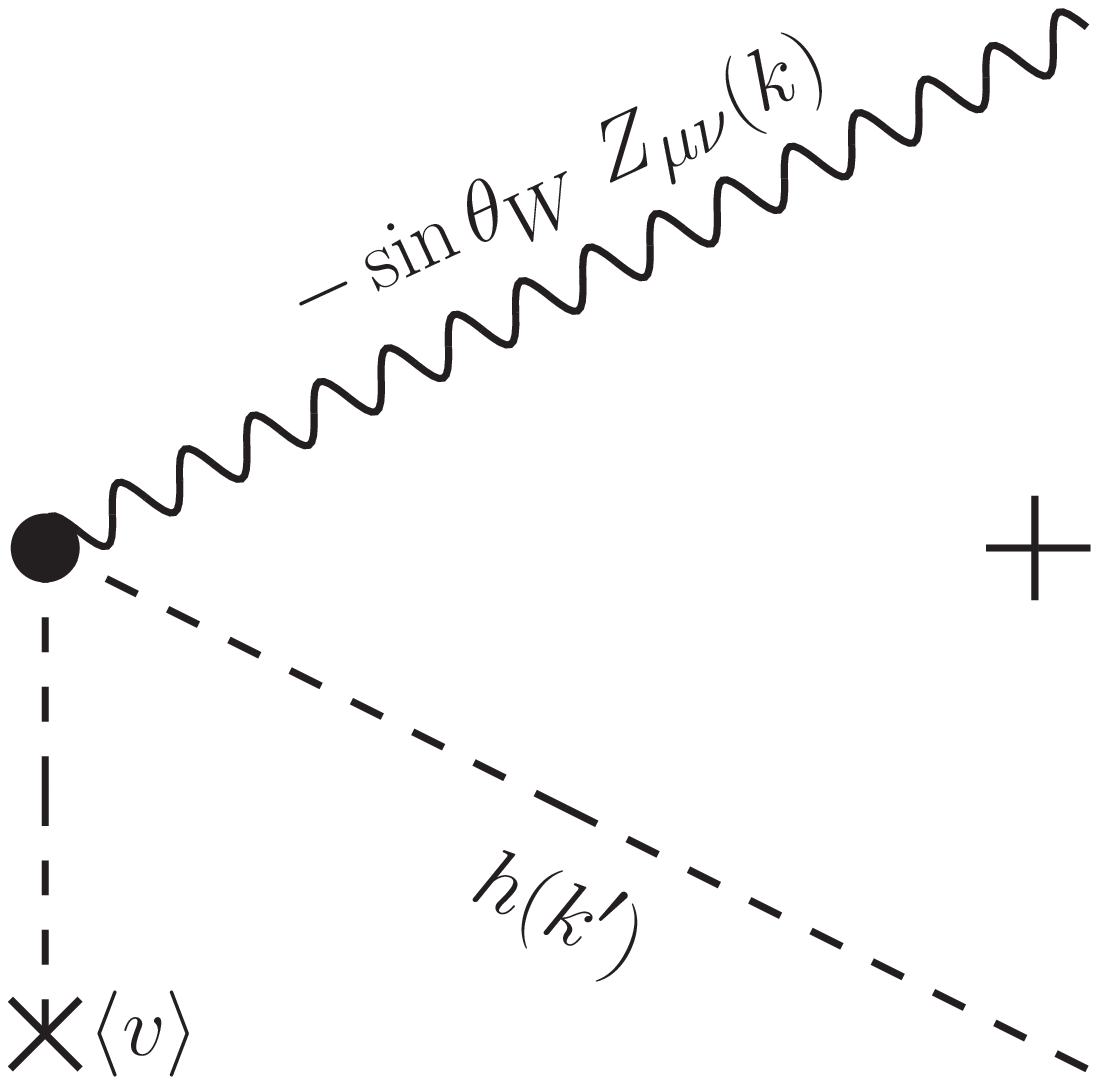} \
\includegraphics[width=.22\linewidth,keepaspectratio]{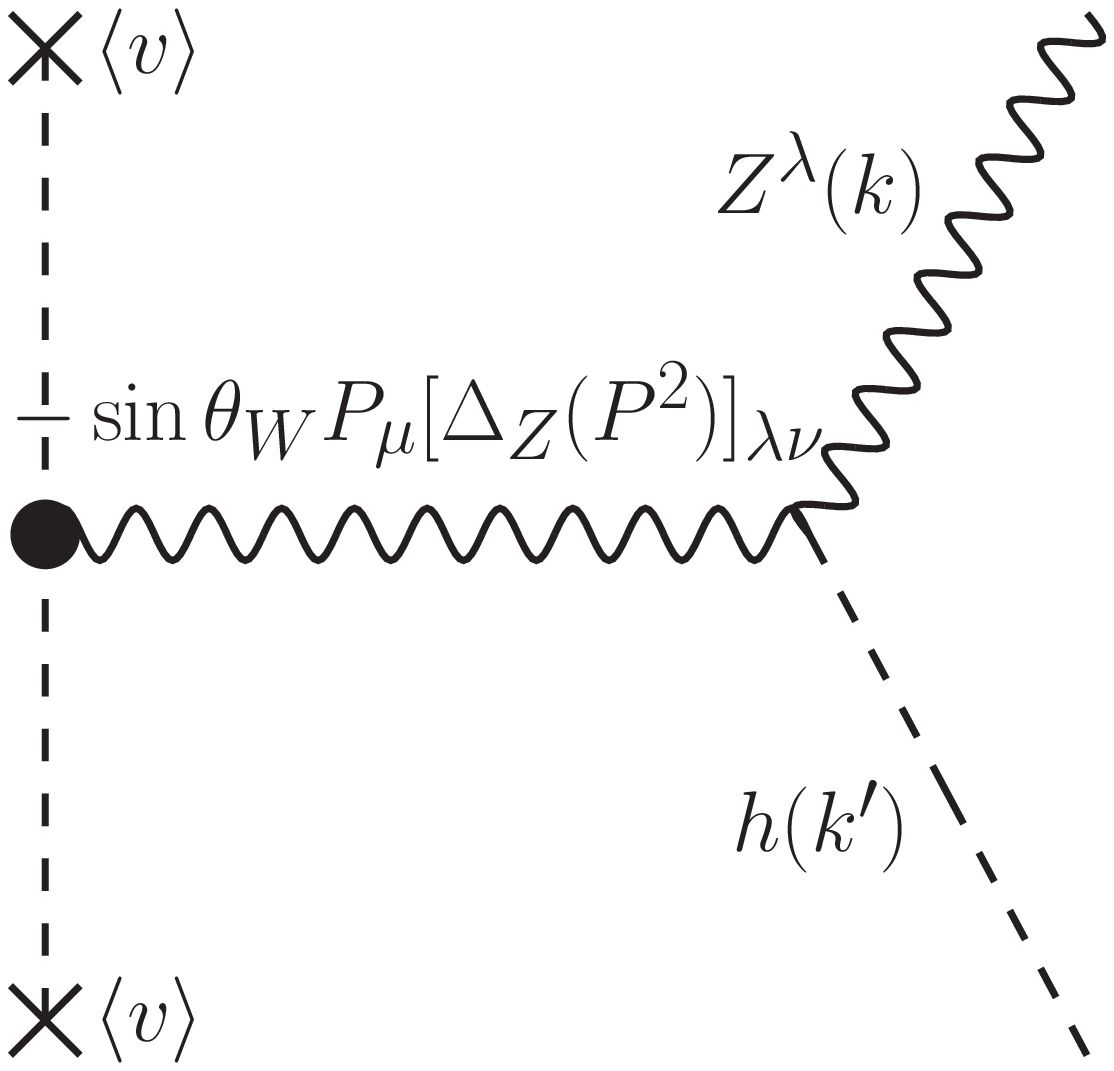} \hspace*{24pt}
\includegraphics[width=.47\linewidth,keepaspectratio]{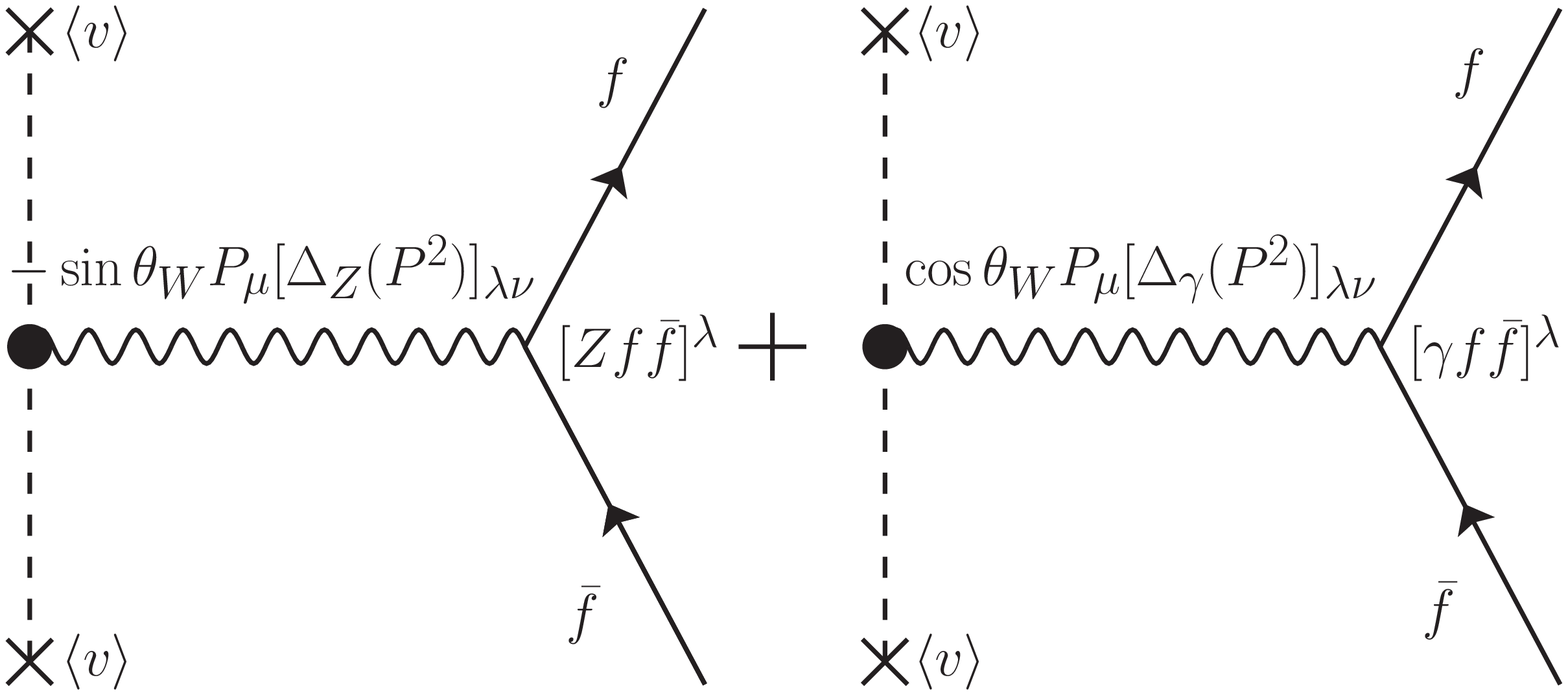} \\
\vspace*{16pt}
\includegraphics[width=.2\linewidth,keepaspectratio]{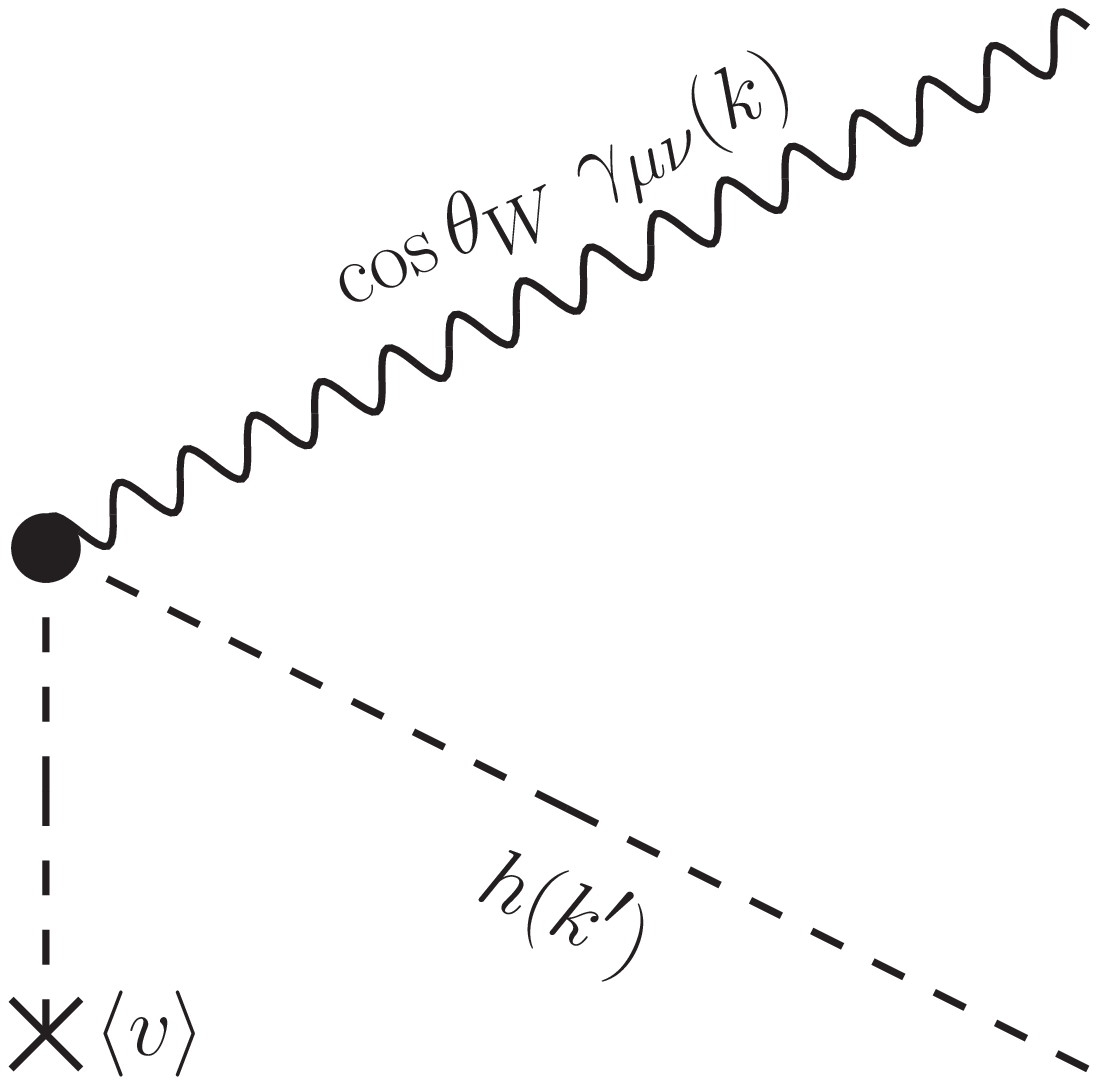} \hspace*{48pt}
\includegraphics[width=.47\linewidth,keepaspectratio]{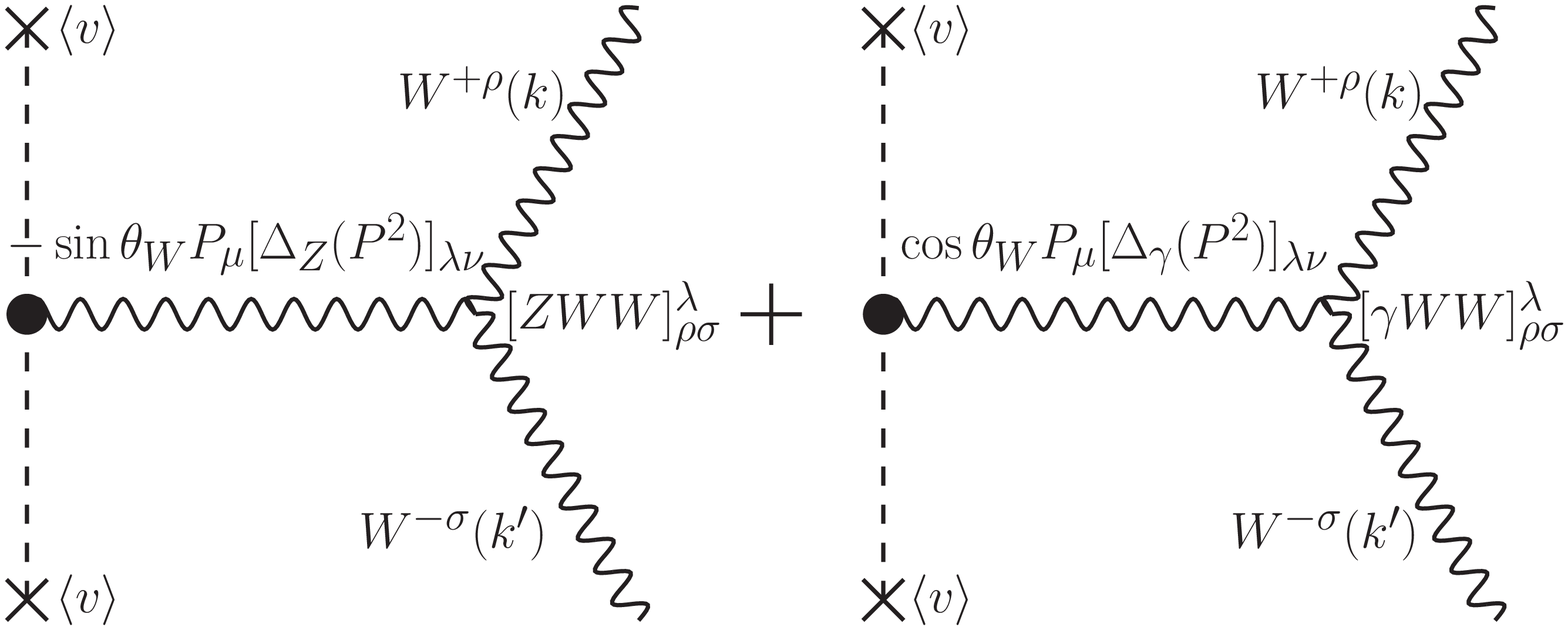}
\caption{Final-state dimension-4 diagrams of the form $B_{\mu\nu}Y_HH^\dagger H$ that couple to tensor WIMP factors. We define $P=p+p'$. The vertex factors $[Zf\bar{f}]$, $[\gamma\bar{f}f]$, $[ZWW]$, and $[\gamma WW]$ are defined in the text.\label{fig:BmunuHHp}}
\end{center}
\end{figure*}

\begin{figure*}
\begin{center}
\includegraphics[width=.22\linewidth,keepaspectratio]{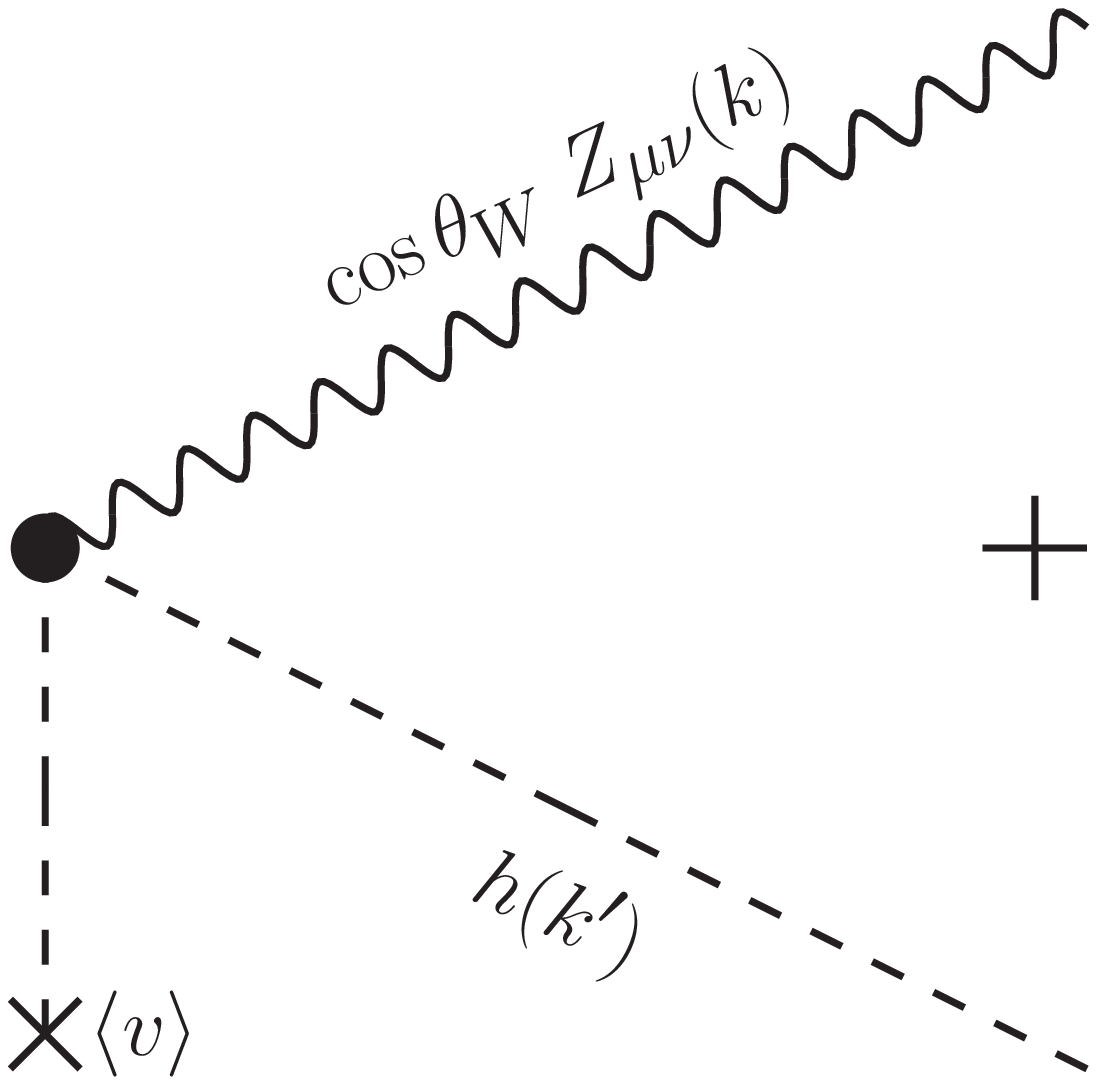}
\includegraphics[width=.235\linewidth,keepaspectratio]{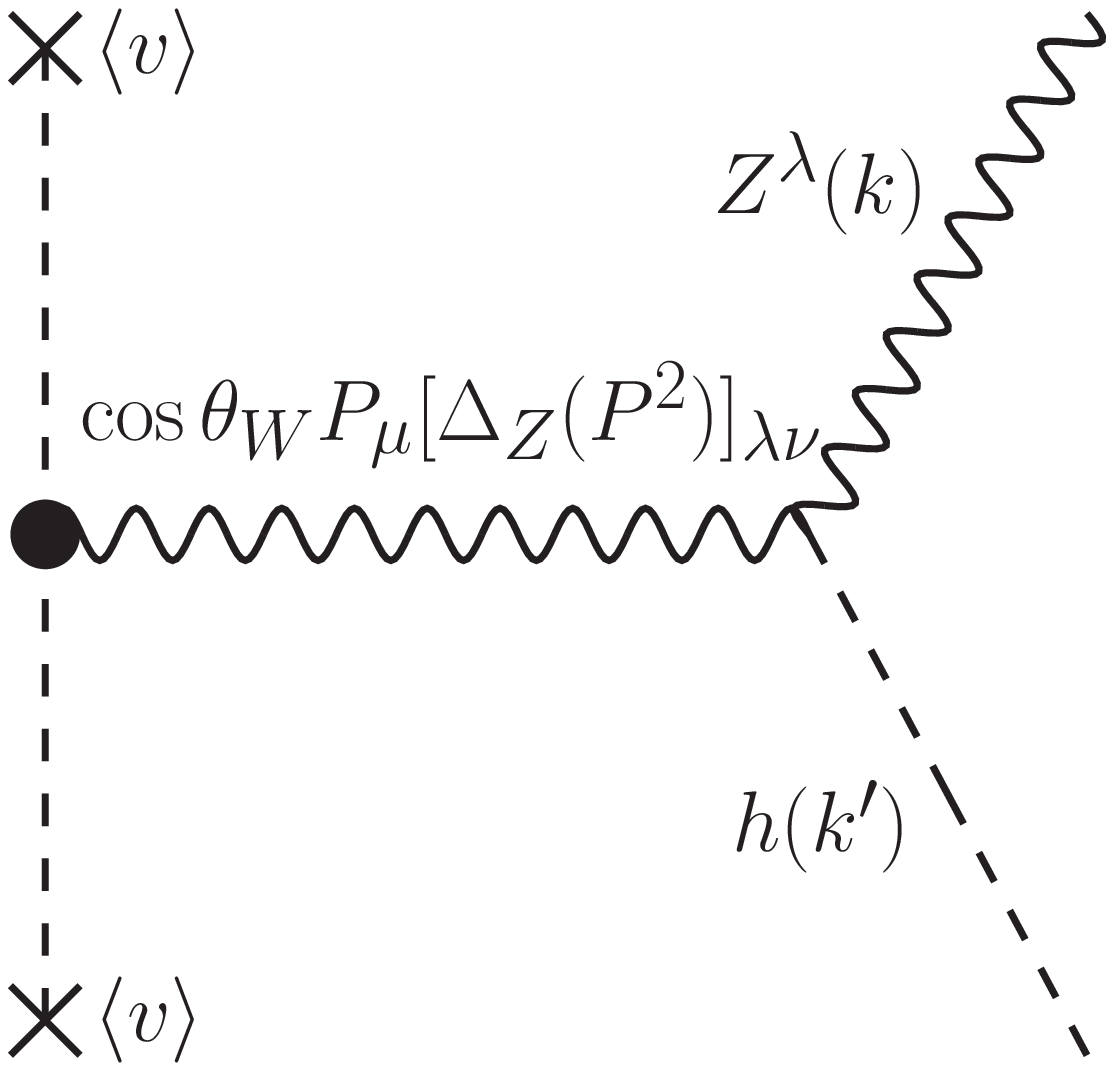} \hspace*{24pt}
\includegraphics[width=.47\linewidth,keepaspectratio]{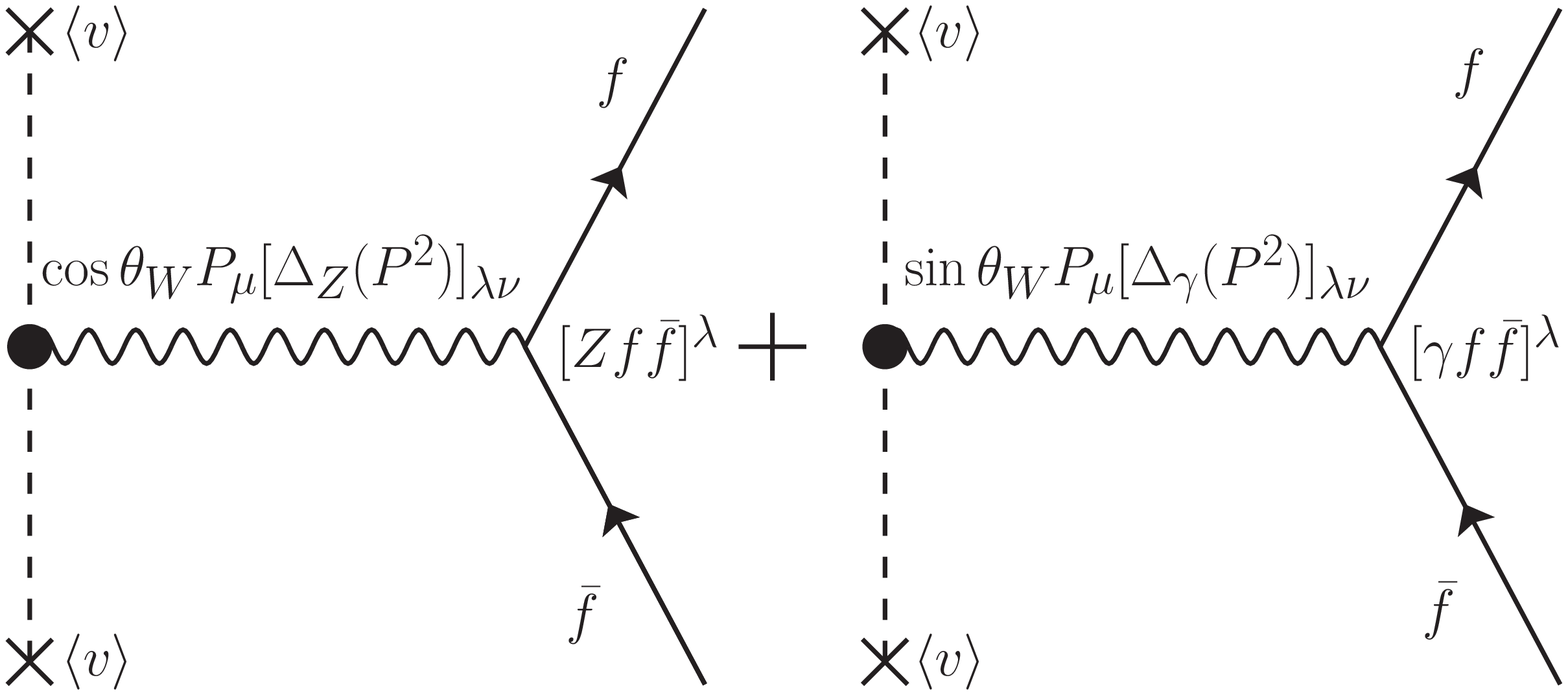}\\
\vspace*{16pt}
\includegraphics[width=.2\linewidth,keepaspectratio]{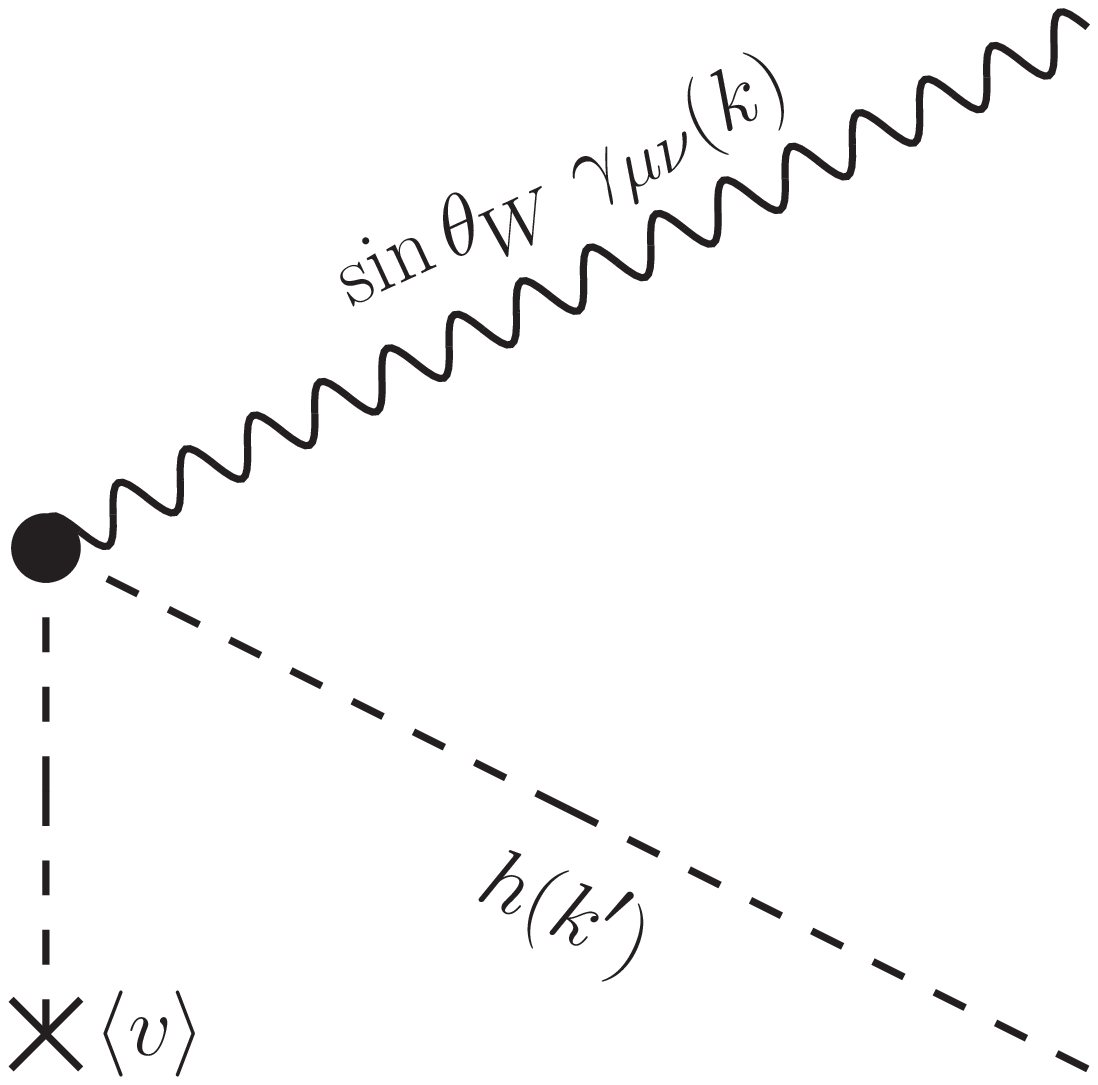} \hspace*{48pt} \includegraphics[width=.47\linewidth,keepaspectratio]{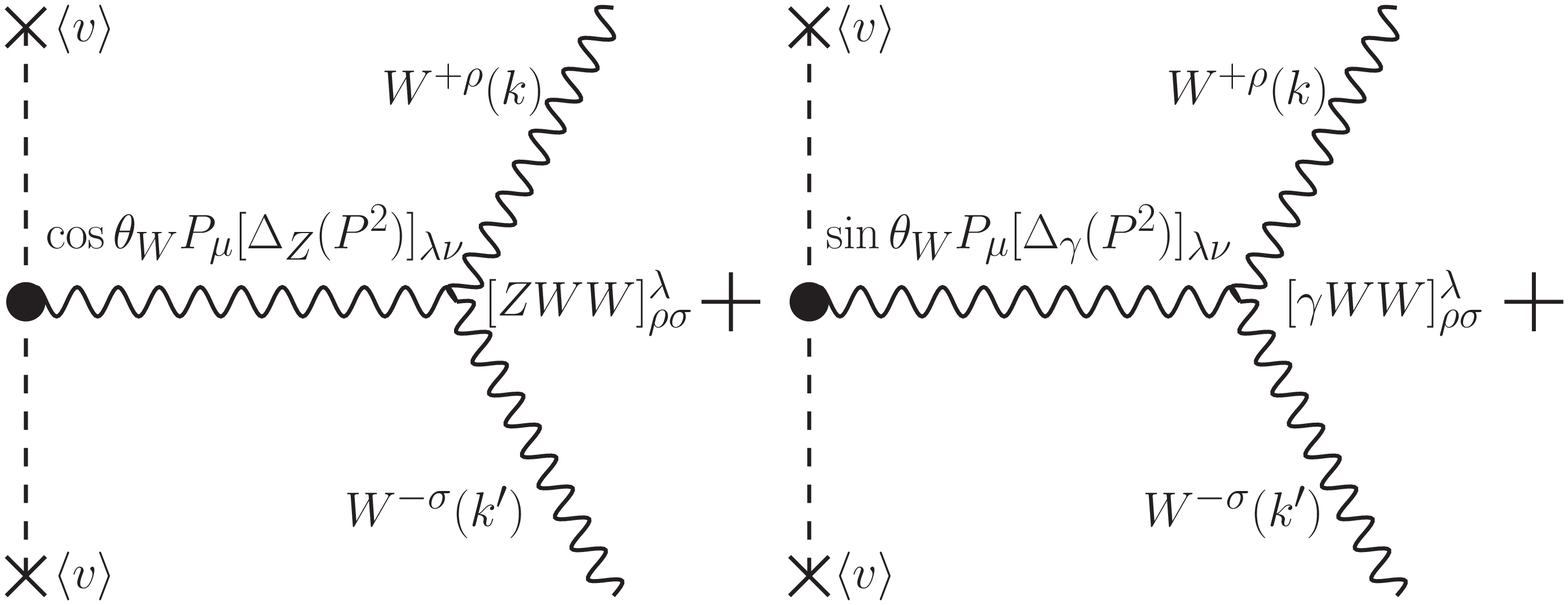}
\includegraphics[width=.2\linewidth,keepaspectratio]{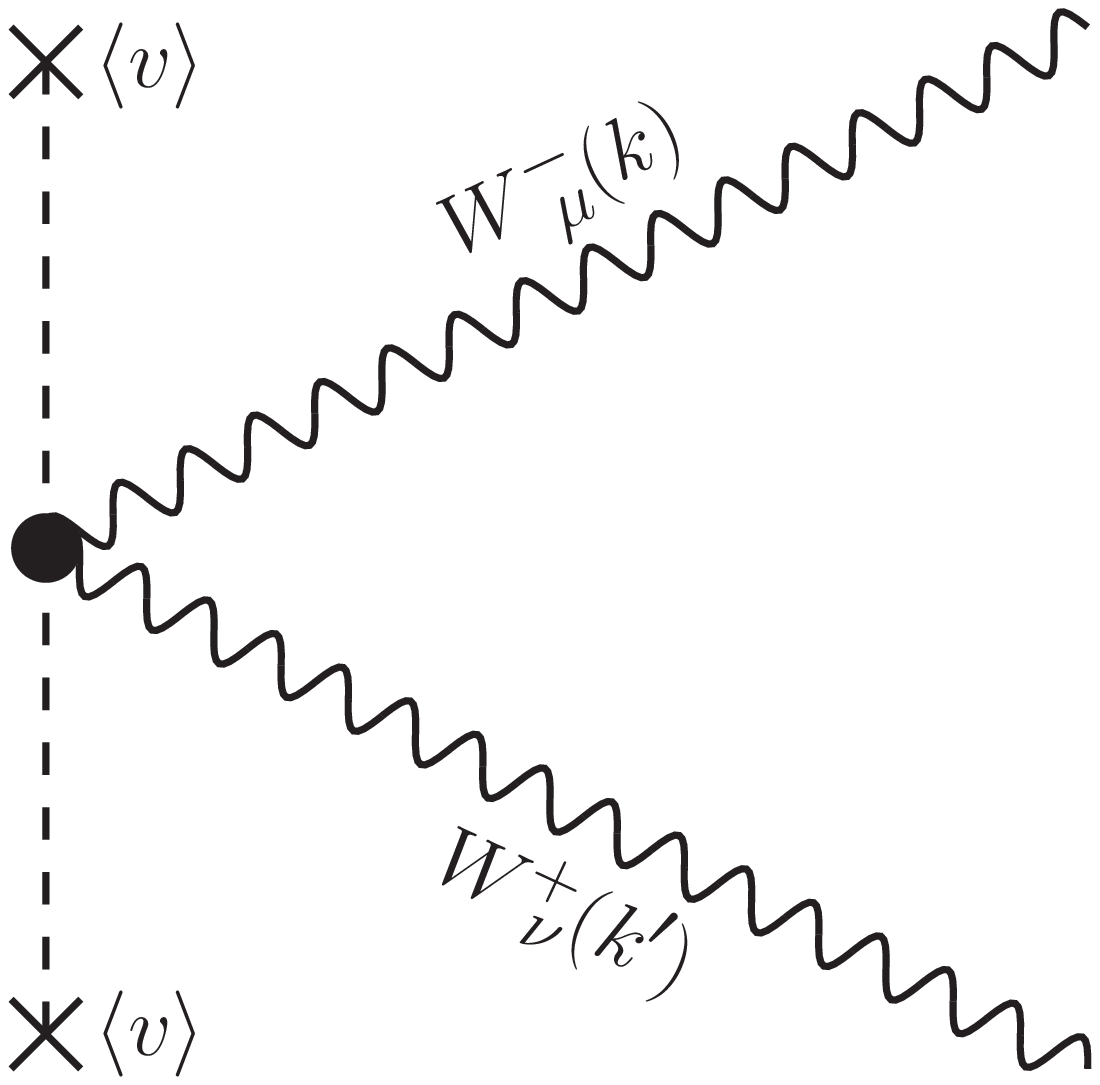}
\caption{Final-state dimension-4 diagrams of the form $W^a_{\ \mu\nu}H^\dagger t^a H$ that couple to tensor WIMP factors. We define $P=p+p'$. The vertex factors $[Zf\bar{f}]$, $[\gamma\bar{f}f]$, $[ZWW]$, and $[\gamma WW]$ are defined in the text.\label{fig:WmunuHHp}}
\end{center}
\end{figure*}

Now we turn to final-state SM operators coupling to tensor DM operators. They will be of the form $B$ or $\widetilde{B}$ multiplied by $H^\dagger  Y_HH$, and $W^a$ or $\widetilde{W}^a$ multiplied by $H^\dagger t^a H$.  These processes will be more complicated to compute because the final states $Zh$, $W^+W^-$ and $f\bar{f}$ may be produced through production of a single $s$-channel photon or $Z$ (see Figs.\ \ref{fig:BmunuHHp} and \ref{fig:WmunuHHp}).  An operator just containing a single $B$ or a $\widetilde{B}$ can also produce a diboson final state by producing a single photon or $Z$, which then produces the two-body final states $Zh$, $W^+W^-$ and $f\bar{f}$.

Because the processes in this subsection are more complicated, we describe the calculation in a little more detail.  We first express the fields involved in the operators in the unitary gauge:
\begin{eqnarray}
B_{\mu\nu} H^\dagger Y_H H & = & \phantom{-} \frac{1}{2} \frac{\left(\vev+h\right)^2}{2} B_{\mu\nu} = \frac{\left(\vev+h\right)^2}{4} C_{YA} \, A_{\mu\nu}, \nonumber \\
W^a_{\ \mu\nu}H^\dagger t^a H & = & -\frac{1}{2} \frac{\left(\vev+h\right)^2}{2} W^3_{\mu\nu} = -\frac{\left(\vev+h\right)^2}{4} \left( C_{3A} \, A_{\mu\nu} - 2ig W^+_{[\mu} W^-_{\nu]} \right),
\end{eqnarray}
where $A$ includes both the photon and the $Z$ boson, $A_{\mu\nu}\equiv 2\partial_{[\mu} A_{\nu]}$, $C_{Y\gamma}=C_{3Z}=\cos\theta_W, -C_{YZ}=C_{3\gamma}=\sin\theta_W$, and $\vev/\sqrt{2}$ is the Higgs vev. We will make use of the facts that $\vev$ and the $W$ and $Z$ masses, $m_W$ and $m_Z$, are related by $m_W=m_Z \cos\theta_W = g\vev/2$, and the $SU(2)$ coupling strength $g$ is related to the EM coupling $e$ via $g=e/\sin\theta_W$.  In this subsection we will not separately consider the operators $\widetilde{B}_{\mu\nu} H^\dagger Y_H H$ and $\widetilde{W}^a_{\ \mu\nu}H^\dagger t^a H$.  It is much more convenient just to use the results for $B_{\mu\nu} H^\dagger Y_H H$ and $W^a_{\ \mu\nu}H^\dagger t^a H$ and let the epsilon tensor operate on the WIMP factors.

First consider production of $\gamma(k) h(k')$ from an operator with the SM factor $B_{\mu\nu} H^\dagger Y_H H$.  There is only vertex production in this case (see Fig.\ \ref{fig:BmunuHHp}).  Summing over the polarization of the photon,
\begin{equation}
\sum_{r, r'}\left|\mathcal{M}\right|^2 = -\vev^2\cos^2\theta_W g_{\nu\nub} k_\mu k_\mub + \{\mu \nu\}\{\bar{\mu} \bar{\nu}\}.
\label{eq:BHHgh}
\end{equation}
Note that we use $+\{\mu \nu\} (\{\bar{\mu} \bar{\nu}\})$ to indicate the addition of all terms symmetric in $\{\mu \nu\}$ ($\{\bar{\mu} \bar{\nu}\}$), which will not contribute when combined with the tensor WIMP factor, which is antisymmetric in those indices.  Production of $\gamma(k) h(k')$ from an operator with the SM factor $W^a_{\ \mu\nu} H^\dagger t^a H$ leads to a SM factor which differs only by the EW mixing term:
\begin{equation}
\sum_{r, r'}\left|\mathcal{M}\right|^2 = -\vev^2\sin^2\theta_W g_{\nu\nub} k_\mu k_\mub + \{\mu \nu\}\{\bar{\mu} \bar{\nu}\}. \label{eq:WHHgh}
\end{equation}

Now turn to processes where there is an $s$-channel contribution.  We start with the final state $Zh$ from the operator $B_{\mu\nu}H^\dagger Y_H H$.  The matrix element has contributions from vertex production and $s$-channel production mediated by a $Z$-boson with momentum $P_\mu=(p+p')_\mu=(k+k')_\mu$ (this definition of $P^\mu$ implies that $P^2=s$). The total is 
\begin{equation}
i\mathcal{M} = \vev\sin\theta_W  k_\mu \ \epsilon^{\ast r}_\nu\! (k) -i\frac{\vev^2}{2}\sin\theta_W \, iP_\mu (-i)\frac{g_{\nu\lambda} - P_\nu P_\lambda/m_Z^2}{s-m_Z^2} \: i\frac{g^2 \vev}{2 \cos^2\theta_W} g^{\lambda\rho}  \ \epsilon^{\ast r}_\rho\! (k) \ +\{\mu \nu\} .
\end{equation}
After some manipulation, the sum over the polarization of the absolute value squared of the matrix element is  
\begin{equation}
\sum_r \left|\mathcal{M}\right|^2 = \frac{\vev^2}{(s-m_Z^2)^2} \sin^2\theta_W \left[ - s^2 g_{\nu\nub} \left(k + \frac{m_Z^2}{s} k' \right)_\mu \left(k + \frac{m_Z^2}{s} k' \right)_\mub + m_Z^2 k_\mu k_\mub k'_\nu k'_\nub  \right] + \{\mu \nu\}\{\bar{\mu} \bar{\nu}\}.
\label{eq:BHHzh}
\end{equation}

The result for the $Zh$ final state from the operator $W^a_{\ \mu\nu}H^\dagger t^a H$ is the same as the previous expression with the replacement $\sin^2\theta_W\rightarrow \cos^2\theta_W$:
\begin{equation}
\sum_r \left|\mathcal{M}\right|^2 = \frac{\vev^2}{(s-m_Z^2)^2} \cos^2\theta_W \left[ - s^2 g_{\nu\nub} \left(k + \frac{m_Z^2}{s} k' \right)_\mu \left(k + \frac{m_Z^2}{s} k' \right)_\mub + m_Z^2 k_\mu k_\mub k'_\nu k'_\nub  \right]+ \{\mu \nu\}\{\bar{\mu} \bar{\nu}\}.
\label{eq:WHHzh}
\end{equation}

Now consider the final state $f\bar{f}$, for which there are only $s$-channel processes (see Figs.\ \ref{fig:BmunuHHp} and \ref{fig:WmunuHHp}).  The amplitude for production of $f(k)\bar{f}(k')$ from $B_{\mu\nu}H^\dagger Y_H H$ is 
\begin{eqnarray}
i\mathcal{M} & = & i\frac{\vev^2}{2}\bar{u}_f^s(k) \left\{ \cos\theta_W iP_\mu(-i) \frac{g_{\nu\lambda}}{s}\,[\gamma f\bar{f}]^\lambda -\sin\theta_W iP_\mu(-i) \frac{g_{\nu\lambda}-P_\nu P_\lambda / m_Z^2}{s-m_Z^2}\,[Z f\bar{f}]^\lambda \right\} v_f^{s'}(k')
\nonumber \\ 
&=& -\frac{1}{2}\frac{\vev m_Z}{s-m_Z^2} \sin\theta_W \, P_\mu \: \bar{u}_f^s(k) \gamma_\nu \left( A_{fB} \pm \gamma^5/2 \right) v_f^{s'}(k') + \{\mu \nu\},
\end{eqnarray}
with $+$ for up-type quarks and neutrinos, and $-$ for down-type quarks and charged leptons.  The vertex factors 
\begin{eqnarray}
[\gamma f\bar{f}]^\lambda & = & i g \sin\theta_W Q_f\, \gamma^\lambda  \nonumber \\
\ [Zf\bar{f}]^\lambda & = &  \frac{ig}{\cos\theta_W}\gamma^\lambda\left[-Q_f\sin^2\theta_W\pm \frac{1}{4}(1-\gamma^5)\right]
\end{eqnarray} 
describe the coupling of photons and $Z$ bosons to $f\bar{f}$ (again, $+$ for up-type quarks and neutrinos, and $-$ for down-type quarks and charged leptons).  The factor $A_{fB}$ is defined as $A_{fB}\equiv2Q_f\left(1-m_W^2/s\right)\mp1/2$, with $-$ ($+$) for neutrinos and up-type quarks (charged leptons and down-type quarks).
This leads to a sum over spin states of 
\begin{equation}
\sum_{s, s'} \left|\mathcal{M}\right|^2 =\frac{\vev^2\, m_Z^2 \sin^2\theta_W}{2(s-m_Z^2)^2}  \left\{ -(4A_{fB}^2+1) k_\mu k_\mub k'_\nu k'_\nub + \left[m_f^2- \left(A_{fB}^2+\frac{1}{4} \right) s\right] g_{\nu\nub} P_\mu P_\mub \right\} +  \{\mu \nu\}\{\bar{\mu} \bar{\nu}\}.
\label{eq:BHHff}
\end{equation}

The production of $f\bar{f}$ from the operator $W^a_{\ \mu\nu}H^\dagger t^a H$ is similar.  The matrix element is
\begin{eqnarray}
i\mathcal{M} & = & i\frac{\vev^2}{2}\bar{u}_f^s(k) \left\{\sin\theta_W iP_\mu(-i) \frac{g_{\nu\lambda}}{s}\,[\gamma f\bar{f}]^\lambda +\cos\theta_W iP_\mu(-i) \frac{g_{\nu\lambda}-P_\nu P_\lambda / m_Z^2}{s-m_Z^2}\,[Z f\bar{f}]^\lambda \right\} v_f^{s'}(k')
\nonumber \\ 
&=& -\frac{1}{2}\frac{\vev m_Z}{s-m_Z^2} \cos\theta_W \, P_\mu \: \bar{u}_f^s(k) \gamma_\nu \left( A_{fW} \pm \gamma^5/2 \right) v_f^{s'}(k') +\{\mu \nu\},
\end{eqnarray}
where $A_{fW}$ is defined as $A_{fW}\equiv2Q_f\sin^2\theta_Wm_Z^2/s\mp1/2$, with $-$ ($+$) for neutrinos and up-type quarks (charged leptons and down-type quarks).  This leads to
\begin{equation}
\sum_{s, s'} \left|\mathcal{M}\right|^2 =\frac{\vev^2\, m_Z^2 \cos^2\theta_W}{2(s-m_Z^2)^2}  \left\{ -(4A_{fW}^2+1) k_\mu k_\mub k'_\nu k'_\nub + \left[m_f^2- \left(A_{fW}^2+\frac{1}{4} \right) s\right] g_{\nu\nub} P_\mu P_\mub \right\} + \{\mu \nu\}\{\bar{\mu} \bar{\nu}\}.
\label{eq:WHHff}
\end{equation}

Finally, let us turn to production of $W^+W^-$.  Production from the operator $B_{\mu\nu}H^\dagger Y_HH$ can only proceed via an $s$-channel process mediated by a photon or a $Z$-boson with momentum $P_\mu=(p+p')_\mu=(k+k')_\mu$.  The relevant coupling of the photon or $Z$ to the $W^+W^-$ final state is given by the factors
\begin{eqnarray}
[\gamma W^+W^-]^{\lambda\rho\sigma} & = & ig\sin\theta_W \left[g^{\rho\sigma}(k'-k)^\lambda + g^{\lambda\rho}(P+k)^\sigma - g^{\lambda\sigma}(P+k')^\rho\right] \nonumber \\
\ [ZW^+W^-]^{\lambda\rho\sigma} & = & ig\cos\theta_W \left[g^{\rho\sigma}(k'-k)^\lambda + g^{\lambda\rho}(P+k)^\sigma - g^{\lambda\sigma}(P+k')^\rho\right].
\end{eqnarray}
The matrix element is
\begin{equation}
i\mathcal{M}= i\frac{\vev^2}{2}iP_\mu(-i)\left\{\cos\theta_W\frac{g_{\nu\lambda}}{s}[\gamma W^+W^-]^{\lambda\rho\sigma} -\sin\theta_W\frac{g_{\nu\lambda}-P_\nu P_\lambda/m_Z^2}{s-m_Z^2}[ZW^+W^-]^{\lambda\rho\sigma}\right\}\epsilon^{\ast r}_\rho\! (k) \ \epsilon^{\ast r'}_\sigma\! (k')  +\{\mu \nu\}.
\end{equation}
Summing over polarizations leads to
\begin{equation}
\sum_{r,r'}\left|\mathcal{M}\right|^2= \dfrac{\vev^2m_Z^2\sin^2\theta_W}{\left(s-m_Z^2\right)^2}\left[\left(1-4\dfrac{m_W^2}{s}+12\dfrac{m_W^4}{s^2}\right)k_\mu k_\mub k'_\nu k'_\nub -2\left(1-4\dfrac{m_W^2}{s}\right)m_W^2g_{\nu\nub}P_\nu P_\nub\right] + \{\mu \nu\}\{\bar{\mu} \bar{\nu}\}.
\label{eq:BHHww}
\end{equation}

The $W^+W^-$ final state from the operator $W^a_{\ \mu\nu}H^\dagger t^a H$ is more complicated still because the final state can be produced directly from the source vertex (see Fig.\ \ref{fig:WmunuHHp}).  This produces an additional term in the amplitude, which results in a total of
\begin{eqnarray}
i\mathcal{M} & = &  \vev m_W\epsilon^{\ast r}_\mu\! (k) \ \epsilon^{\ast r'}_\nu\! (k') -i\frac{\vev^2}{2}iP_\mu(-i)\left\{\sin\theta_W\frac{g_{\nu\lambda}}{P^2}[\gamma W^+W^-]^{\lambda\rho\sigma} \right. \nonumber \\
& & \left. +\cos\theta_W\frac{g_{\nu\lambda}-P_\nu P_\lambda/m_Z^2}{P^2-m_Z^2}[ZW^+W^-]^{\lambda\rho\sigma}\right\}\epsilon^{\ast r}_\rho\! (k) \ \epsilon^{\ast r'}_\sigma\! (k')+\{\mu \nu\} \ .
\end{eqnarray}
After tedious calculation,
\begin{eqnarray}
\sum_{r,r'}\left|\cal{M}\right|^2 & = & \dfrac{\vev^2}{m_W^2} \left\{ m_W^2 g_{\nu\nub} \left[ m_W^2 g_{\mu\mub} - k_\mu k_\mub - k'_\mu k'_\mub + 2U_W(1-U_W)\left(1-\dfrac{4m_W^2}{s}\right) P_\mu P_\mub \right] \right. \nonumber \\
&&\left.+ \left[ 1 - 2U_W \left(1-\dfrac{2m_W^2}{s}\right) + U_W^2 \left(1-4\dfrac{m_W^2}{s}+12\dfrac{m_W^4}{s^2}\right) \right] k_\mu k_\mub k'_\nu k'_\nub \right\} + \{\mu \nu\}\{\bar{\mu} \bar{\nu}\}.
\label{eq:WHHww}
\end{eqnarray}
The factor $U_W$ is defined as $U_W\equiv 1+m_W^2/(s-m_Z^2)$.

The final-state contributions given in Eqns.\ (\ref{eq:BHHgh}), (\ref{eq:WHHgh}), (\ref{eq:BHHzh}),  (\ref{eq:WHHzh}), (\ref{eq:WHHff}), (\ref{eq:BHHff}), (\ref{eq:BHHww}), and (\ref{eq:WHHww}) are given in Table \ref{gaugehiggs4}.  All the terms have the correct structure to couple to $\bar{\chi}\gamma^{\mu\nu} \chi$. 

We note that since we include SM gauge-boson interactions, operators of the form $\bar{\chi}\gamma^{\mu\nu}\chi B_{\mu\nu}$ and $\bar{\chi}\gamma^{\mu\nu}\chi \widetilde{B}_{\mu\nu}$ lead to final states $Zh$, $f\bar{f}$, and $W^+W^-$.  The diagrams for these processes can be constructed by removing the Higgs vev's from the diagrams with a photon or $Z$ propagator in Fig.\ \ref{fig:BmunuHHp}.  Of course none of these processes lead to a photon in the 2 body final state. 

The final state $Z(k)h(k')$ matrix element from the SM term $B_{\mu\nu}$ is
\begin{eqnarray}
\frac{16m_Z^2}{\vev^2 \, (s-m_Z^2)^2} \sin^2\theta_W \left[  -m_Z^2 g_{\nu\nub} P_\mu P_\mub + k_\mu k_\mub k'_\nu k'_\nub  \right] \ ,
\end{eqnarray}
plus terms symmetric in $\{\mu\nu\}$ or $\{\mub\nub\}$. 

The contribution to final states $W^+W^-$ and $f\bar{f}$ from the operator $B_{\mu\nu}$ can be obtained from the operator $B_{\mu\nu}H^\dagger Y_H H$ by removing a factor of $\left(\vev/\sqrt{2}\right)^4/4$ (the factor of $4$ is from removing $Y_H^2$).

The operator $B_{\mu\nu}$ is included in Table \ref{gaugehiggs4}.

\renewcommand*\arraystretch{2.3}
\begin{table}
\caption{\label{gaugehiggs4} SM "tensor" contributions.  The polarizations have been summed over where appropriate.   Since the terms have two indices, they can couple to tensor WIMP operators. Here, we only list the contributions to $\left|\mathcal{M}\right|^2$ for $B_{\mu\nu}$ and $W^a_{\ \mu\nu}$.  For $\widetilde{B}_{\mu\nu}$ and $\widetilde{W}^a_{\ \mu\nu}$, it is easier to contract the epsilon tensors with the SM term.  This has the effect of changing $-M^2$ to $+M^2$ in the last line of Table \ref{WIMPmatrix}. For the results of annihilation into di-boson final states in this table, the addition of all terms symmetric in $ \{\mu \nu\}$ and $\{\bar{\mu} \bar{\nu}\}$ is understood. }
\begin{ruledtabular}
\begin{tabular}{ccll}
Operators & Appears in $\mathcal{M}$ as & \hspace*{96pt}Contribution to $\left|\mathcal{M}\right|^2$ & Notes \\ 
\hline\hline
\multirow{4}{*}{$B_{\mu\nu} Y_H \, H^\dagger H$}
&
$\gamma_{\mu\nu} h$
&
$-\vev^2\cos^2\theta_Wg_{\nu\nub}k_\mu k_\mub$
&
\\
&
$Z_{\mu\nu} h$
&
$\dfrac{\vev^2}{\left(s-m_Z^2\right)^2}\sin^2\theta_W\left[-s^2g_{\nu\nub}\left(k+\dfrac{m_Z^2}{s}k'\right)_\mu\left(k+\dfrac{m_Z^2}{s}k'\right)_\mub + m_Z^2k_\mu k_\mub k'_\nu k'_\nub\right]$
&
\\
&
$\sum_ff\bar{f}$
&
$\dfrac{\vev^2m_Z^2\sin^2\theta_W}{2\left(s-m_Z^2\right)^2} \left\{ -\left(4A^2_{fB}+1\right)k_\mu k_\mub k'_\nu k'_\nub + \left[m_f^2-\left(A^2_{fB}+\dfrac{1}{4}\right)s\right]g_{\nu\nub}P_\mu P_\mub\right\}$
&
$a,b$
\\
&
$W^+W^-$
&
$\dfrac{\vev^2m_Z^2\sin^2\theta_W}{\left(s-m_Z^2\right)^2}\left[\left(1-4\dfrac{m_W^2}{s}+12\dfrac{m_W^4}{s^2}\right)k_\mu k_\mub k'_\nu k'_\nub -2\left(1-4\dfrac{m_W^2}{s}\right)m_W^2g_{\nu\nub}P_\nu P_\nub\right]$
&
$a$
\\
[2ex]
\hline
\multirow{5}{*}{$W^a_{\ \mu\nu} H^\dagger t^a H$}
&
$\gamma_{\mu\nu} h $ 
&
$-\vev^2\sin^2\theta_Wg_{\nu\nub}k_\mu k_\mub$
&
\\
&
$Z_{\mu\nu}h$
&
$\dfrac{\vev^2}{\left(s-m_Z^2\right)^2}\cos^2\theta_W\left[-s^2g_{\nu\nub}\left(k+\dfrac{m_Z^2}{s}k'\right)_\mu\left(k+\dfrac{m_Z^2}{s}k'\right)_\mub + m_Z^2k_\mu k_\mub k'_\nu k'_\nub\right]$
&
\\
&
$\sum_ff\bar{f}$
&
$\dfrac{\vev^2m_Z^2\cos^2\theta_W}{2\left(s-m_Z^2\right)^2} \left\{ -\left(4A^2_{fW}+1\right)k_\mu k_\mub k'_\nu k'_\nub + \left[m_f^2-\left(A^2_{fW}+\dfrac{1}{4}\right)s\right]g_{\nu\nub}P_\mu P_\mub\right\}$
&
$a,c$
\\
[2ex]
&
\multirow{2}{*}{$W^+ W^-$}
&
\parbox[l]{309pt}{$\dfrac{\vev^2}{m_W^2} \left\{ m_W^2 g_{\nu\nub} \left[ m_W^2 g_{\mu\mub} - k_\mu k_\mub - k'_\mu k'_\mub + 2U_W(1-U_W)\left(1-\dfrac{4m_W^2}{s}\right) P_\mu P_\mub \right] \right.$ \\
$ \left.+ \left[ 1 - 2U_W \left(1-\dfrac{2m_W^2}{s}\right) + U_W^2 \left(1-4\dfrac{m_W^2}{s}+12\dfrac{m_W^4}{s^2}\right) \right] k_\mu k_\mub k'_\nu k'_\nub \right\}$}
&
\multirow{2}{*}{$a,d$}
\\
[3ex]
\hline
\multirow{4}{*}{$B_{\mu\nu}$}
&
$Z_{\mu\nu} h$
&
$\dfrac{16}{\vev^2 \left(s-m_Z^2\right)^2}\sin^2\theta_W\left[-s^2g_{\nu\nub}\left(k+\dfrac{m_Z^2}{s}k'\right)_\mu\left(k+\dfrac{m_Z^2}{s}k'\right)_\mub + m_Z^2k_\mu k_\mub k'_\nu k'_\nub\right]$
&
\\
&
$\sum_ff\bar{f}$
&
$\dfrac{16 m_Z^2\sin^2\theta_W}{\vev^2 2\left(s-m_Z^2\right)^2} \left\{ -\left(4A^2_{fB}+1\right)k_\mu k_\mub k'_\nu k'_\nub + \left[m_f^2-\left(A^2_{fB}+\dfrac{1}{4}\right)s\right]g_{\nu\nub}P_\mu P_\mub\right\}$
&
$a,b$
\\
&
$W^+W^-$
&
$\dfrac{16 m_Z^2\sin^2\theta_W}{\vev^2 \left(s-m_Z^2\right)^2}\left[\left(1-4\dfrac{m_W^2}{s}+12\dfrac{m_W^4}{s^2}\right)k_\mu k_\mub k'_\nu k'_\nub -2\left(1-4\dfrac{m_W^2}{s}\right)m_W^2g_{\nu\nub}P_\nu P_\nub\right]$
&
$a$
\\
[2ex]
\end{tabular}
$^a \ P_\mu\equiv(p+p')_\mu=(k+k')_\mu$, $P^2=s$ \\
$^b \ A_{fB}\equiv2Q_f\left(1-m_W^2/s\right)\mp1/2$, with $-$ ($+$) for neutrinos and up-type quarks (charged leptons and down-type quarks). \\
$^c \ A_{fW}\equiv2Q_f\sin^2\theta_Wm_Z^2/s\mp1/2$, with $-$ ($+$) for neutrinos and up-type quarks (charged leptons and down-type quarks).
$^d \ U_W\equiv 1+m_W^2/(s-m_Z^2)$.
\end{ruledtabular}
\end{table}

\section{Results \label{bunchotables}}

The annihilation cross section, $\sigma(s)$, for WIMPs of mass $M$ into a two-body final state with masses $m$ and $m'$ is given in terms of a dimensionless factor $\Sigma$, which depends on $s=(p+p')^2=(k+k')^2$, the WIMP mass $M$, and the final state masses $m$ and $m'$ as
\begin{equation}
\sigma(s) = \dfrac{1}{32\pi M^2} \sqrt{\dfrac{4M^2}{s}}
\sqrt{\dfrac{M^2}{s-4M^2}}\sqrt{1-\dfrac{(m+m')^2}{s}}\sqrt{1-\dfrac{(m-m')^2}{s}} \ \Sigma(s;M,m,m').
\end{equation}  
Physically, $\Sigma(s;M,m,m')$ is given in terms of the matrix element $\mathcal{M}$ by
\begin{eqnarray}
\Sigma(s;M,m,m') & = & \int \frac{d\Omega}{4\pi} \ 
                       \frac{1}{4}\sum_{s,s'}\sum_{r[,r']}
                       \left|\mathcal{M}\right|^2 \qquad \mathrm{fermonic\ WIMPs} \nonumber \\
                 & = & \int \frac{d\Omega}{4\pi} \ \sum_{r[,r']}
                   \left|\mathcal{M}\right|^2\qquad \phantom{\frac{1}{4}\sum_{s,s'}}
                       \mathrm{bosonic\ WIMPs} ,
\end{eqnarray}
where initial spin states $s$ and $s'$ are averaged over for fermionic WIMPs, and  polarizations $r$ and $r'$ of final-state bosons are summed over (if one of the final states is a spinless boson, the sum over $r'$ is omitted).  The integral $\int d\Omega$ is the solid angle integration in the CoM frame with the extra factor of $1/2$ understood if the two final state particles are identical.  The $\Sigma(s;M;m,m')$ factors for the various terms and processes are given in tables below.

The nonrelativistic cross section, $\left[\sigma v\right]_\mathrm{NR}$, is obtained by the substitution $s\rightarrow 4 M^2$, unless $s$ appears in the combination $s-4M^2$, in which case $s-4M^2\rightarrow M^2v^2$.  This leads to the substitution $1-4M^2/s\rightarrow v^2/4$.  The nonrelativistic cross section is usually expressed as a an expansion in terms of $v^2$: $\left[\sigma v\right]_\mathrm{NR}=a+bv^2+\cdots$.  Therefore,
\begin{equation}
\left[\sigma v\right]_\mathrm{NR}=\dfrac{1}{32\pi M^2}\sqrt{1-\dfrac{(m+m')^2}{4M^2}}\sqrt{1-\dfrac{(m-m')^2}{4M^2}} \ \Sigma(s=4M^2 \textrm{ or } s-4M^2=M^2v^2;M,m,m')=a+bv^2.
\end{equation}

For present annihilation in the galactic center to produce the purported 130 GeV line, we need $\left[\sigma v\right]_\mathrm{NR} \sim 10^{-27} \cm^3\s^{-1}=8.59\times10^{-11}\GeV^{-2}$.   Annihilation in the galactic center has $v\sim10^{-3}$,  so we will assume annihilation at rest, and also assume that if $\Sigma\propto 1-4M^2/s=v^2/4$, that present-day annihilation will be negligible.  For $\bar{\chi}\chi$ annihilation to particles of mass $m$ and $m'$, the energy of the particle of mass $m$ is $E=M+m^2/4M-m'^2/4M$ and the energy of the particle of mass $m'$ is $E=M+m'^2/4M-m^2/4M$. Possible final states we consider are $\gamma\gamma$, $\gamma Z$, $\gamma h$, $ZZ$, $W^+W^-$, $hh$, and $Zh$.  The 130 GeV line could result from annihilation of a particle of mass 130 GeV for photons from the $\gamma\gamma$ final state, mass 144 GeV for photons from the $ \gamma Z$ final state, or mass 155 GeV for photons from the $\gamma h$ final state. For some operators, the branching fraction into $\gamma Z$ or $\gamma h$ is larger than the fraction into $\gamma\gamma$.  The energies of the other annihilation products are needed for the calculation of the continuum, which are given in Table \ref{energies} for WIMP masses of 130 GeV, 144 GeV, and 155 GeV.  The values of $\Sigma(s;M,m,m')$ as a function of $\Lambda$ for arbitrary WIMP mass are indicated in the Tables for the various operators. (Of course it is understood that the WIMP mass must be above threshold for a given process.)

\renewcommand*\arraystretch{1.5}
\begin{table}
\caption{\label{energies} Energies of products of WIMP annihilation.}
\begin{ruledtabular}
\begin{tabular}{cccccccccc}
\multirow{2}{*}{$M$} & 
$\gamma \gamma$ &
\multicolumn{2}{c}{$\gamma Z$} &
\multicolumn{2}{c}{$\gamma h$} &
$ZZ$; $W^+W^-$ &
\multicolumn{2}{c}{$Z h$} \\ 
&
$E_\gamma$ & 
$E_\gamma$ &
$E_Z$ &
$E_\gamma$ &
$E_h$ &
$E_Z$; $E_W$ &
$E_Z$ &
$E_h$ \\
\hline
130 GeV & 130 GeV & 114 GeV & 146 GeV & 100 GeV & 160 GeV & 130 GeV & 116 Gev & 144 GeV \\
144 GeV & 144 GeV & 130 GeV & 158 GeV & 117 GeV & 171 GeV & 144 GeV & 131 GeV & 157 GeV \\
155 GeV & 155 GeV & 142 GeV & 168 GeV & 130 GeV & 180 GeV & 155 GeV & 143 GeV & 167 GeV
\end{tabular}
\end{ruledtabular}
\end{table}

Also shown in Tables \ref{resultone}--\ref{resultten} are the values of $\Omega h^2$ as a function of $\Lambda$ for the various operators. The relic density depends on the total annihilation cross section.  In terms of $a$  and $b$, the early-universe freeze-out temperature, $T_F$, may be expressed in terms of $x_F = M/T_F$ by \cite{kolb1994early}
\begin{equation}
x_F = \ln\left[c(c+2)\sqrt{\frac{45}{8}}\frac{M M_{Pl}(a+6b/x_F)}{2\pi^3\sqrt{g_*(x_F)}}\right] = \ln\left[1.19\times10^{19}\GeV^2(a+6b/x_F)\right],
\end{equation}
where for relic density calculation $a$ and $b$ are the values for the total cross section.  We use $c$, the free parameter to fit the numerical results, of $c=1$, and assume the effective number of degrees of freedom at freeze out is $g_*(x_F)=106$. Here we have assumed $M=130$ GeV, but $x_F$ only has a logarithmic dependence on $M$.  The present value of $\Omega h^2$ is given in terms of $a$, $b$, and $x_F$ by \cite{kolb1994early}
\begin{equation}
\Omega h^2 = \frac{1.09\times 10^9 x_F \GeV^{-1}}{M_{Pl}\sqrt{g_*(x_F)}(a+3b/x_F)} = 0.11 \frac{7.88\times 10^{-11}x_F \GeV^{-2}}{a+3b/x_F}.
\end{equation}
If $b=0$, pure $s$-wave annihilation, then $\left[\sigma v\right]_\mathrm{NR}=a$, and 
\begin{equation}
\left.\Omega h^2\right|_{b=0} = 0.11\, \frac{1.88\times10^{-9}\, \GeV^{-2}}{a} + \ \mathrm{corrections\ logarithmic\ in} \ M \ \mathrm{and}\ a.
\end{equation}
If $a=0$, pure $p$-wave annihilation, then $\left[\sigma v\right]_\mathrm{NR}=bv^2=\frac{3}{2}bx^{-1}$, and
\begin{equation}
\left.\Omega h^2\right|_{a=0} = 0.11\, \frac{1.58\times10^{-8}\, \GeV^{-2}}{b}  + \ \mathrm{corrections\ logarithmic\ in} \ M \ \mathrm{and}\ b.
\end{equation}
If the 130 GeV photon line with a signal strength of $10^{-27}$cm$^3$s$^{-1}$ is confirmed, our tables can be used to derive further information about dark-matter annihilation. For a subset of the operators tabulated here, the scales $\Lambda$  in the last two columns of the tables are similar. This means both the thermal freeze out and the indirect signal can be described mainly by one of these operators.  For example, this is the case for operator $\bar \chi i \gamma^5 \chi W^a_{\mu \nu} \widetilde{W}^{a \mu \nu}$ (see Tables \ref{linestableST} and \ref{linestableV} for more examples). 

There are related signals and constraints. We have calculated the production rates for all possible di-boson states. In addition to the channel that gives the leading photon line signal, in many cases there is a subdominant channel which gives a second photon line. Observing such a line with predicted energy and signal strength  is a definitive confirmation of the discovery and the underlying dynamics.

\begin{figure*}
\begin{center}
\includegraphics[width=.4\linewidth,keepaspectratio]{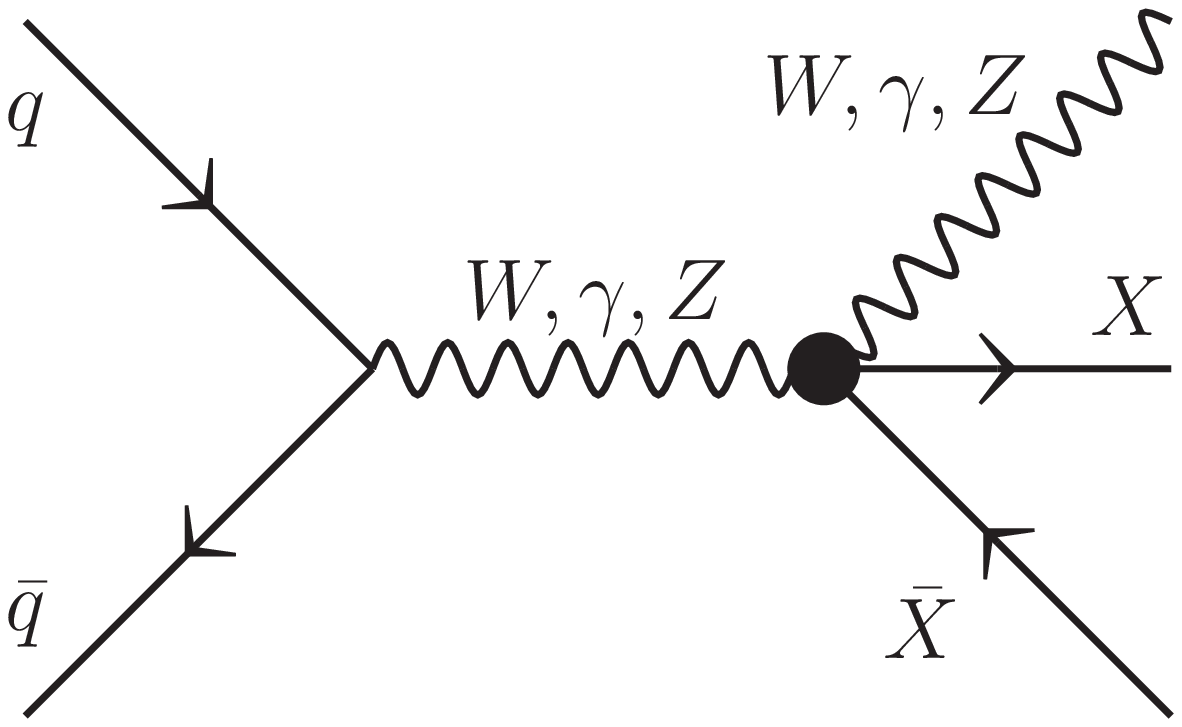} \hspace*{36pt}
\includegraphics[width=.4\linewidth,keepaspectratio]{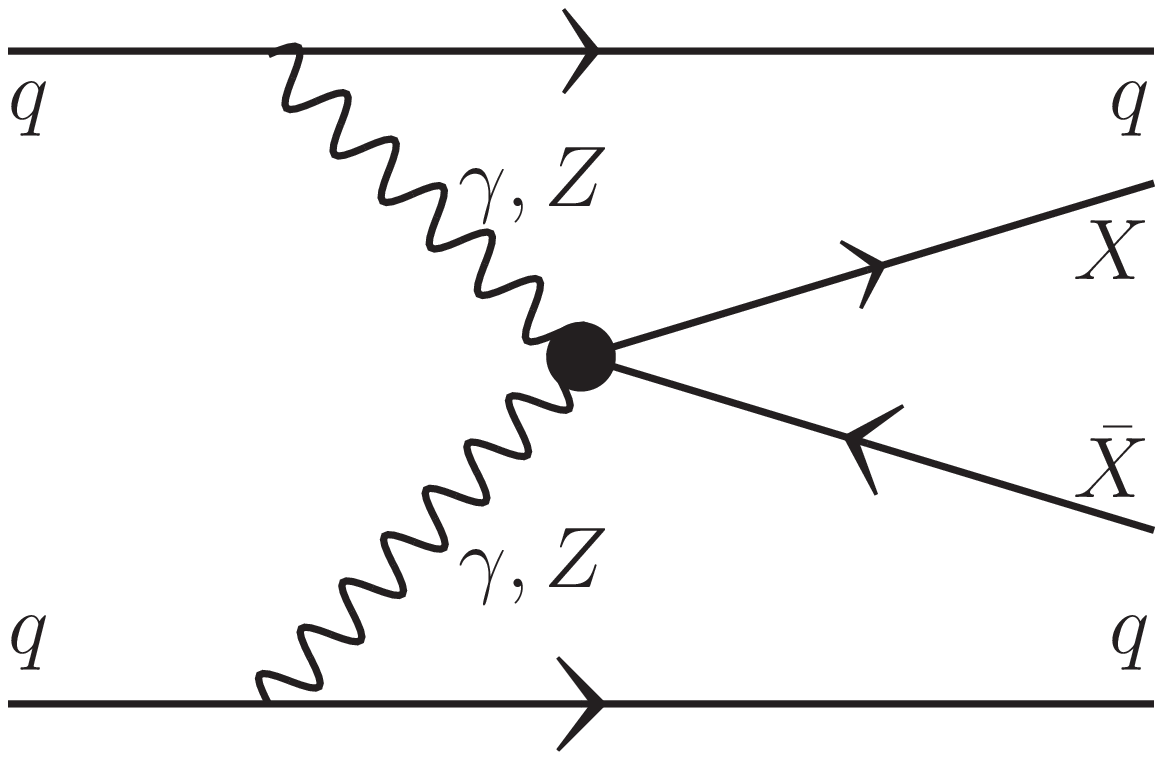}
\caption{\label{fig:LHC} Feynman diagrams for production of a WIMP pair and a monophoton/mono-$W$/mono-$Z$ (left diagram) or WIMP pair and a jet (right diagram).  The WIMP pair would not be detected; its existence would be inferred by the missing transverse energy.  In the right diagram (vector-boson fusion) at least one of the final-state quarks must produce a hard jet.}
\end{center}
\end{figure*}

There are also channels which do not contain a photon, such as $ZZ$, $Zh$, $W^+ W^-$, $hh$, and $f \bar f$.  The strengths of these channels relative to the photonic ones are also fixed by the operator, and they provide important additional signals. One of the most important signals is the continuum photon radiation off charge particles from the dark-matter annihilation. In our case, this is most relevant for the decay products of $W$, $Z$, $h$. If dark matter can directly annihilate into SM fermions, \textit{e.g.,} for the operator $\bar{\chi}\gamma^{\mu\nu}\chi B_{\mu\nu}Y_HH^\dagger H$, the continuum radiation from this operator is also relevant. If a single operator dominates the dark-matter annihilation, the relation between the signal strength of the photon line(s) and that of the continuum radiation is fixed.  If a photon line is observed and there is no obvious excess of the continuum radiation, such relations can be immediately used to put limits on potential operators responsible for dark-matter annihilation. For example, assuming the 130 GeV photon line is indeed a dark-matter signal, the constraints from the continuum radiation on the annihilation cross section of various possible SM final states have been studied in Ref.\ \cite{Cohen:2012me}. For dark-matter annihilation channels to $ZZ$ and $W^+W^-$, the most conservative estimate constrain their rate to be less then about $100 \times \sigma_{\gamma}$. At the same time, an estimate taking into account the shape of the distribution gives a much stronger limit of about $\sigma_{WW, ZZ} < 10 \times \sigma_\gamma $. For example, the $\phi^\dagger \phi WW$ operators, with $\sigma_{WW} \sim 20 \sigma_{\gamma} $, is already interesting in this regard.  Ref.~\cite{Cohen:2012me} also gives limits on annihilation channels directly into a pair of SM fermions. The strongest limit is on annihilation to $b \bar b$,  $\sigma_{b \bar b} < 10 \sigma_\gamma$, with a similar constraint on $\tau^+ \tau^-$. The operator $\bar \chi \gamma^{\mu \nu} \chi W^a_{\mu \nu} H^{\dagger} t^a H $ could already have interesting limits in this case.  

Since most of our operators contain at least two SM bosons, its scattering with nuclei can only proceed through one-loop processes. Therefore, the reach from direct detection is limited. The exceptions are the dimension-7 operators in Table XVII and Table XVIII. In fact, substituting  both Higgs fields with their vev's, these are directly related to the dark-matter dipole operators studied in Refs.\ \cite{Banks:2010eh,Fortin:2011hv}. In this case, direct detection experiments already set strong limit on the size of such operators. The limit on operator containing $\widetilde{B}_{\mu \nu}$ is much stronger. However, even for the operator containing $B_{\mu \nu}$, direct detection limits imply that $\Lambda$ is larger than what is required for producing the appropriate relic abundance of dark matter. For it to be a viable DM candidate species, additional annihilation channels would be necessary. 

Weak-scale dark matter can be produced directly at the LHC. There have been experimental searches \cite{ATLAS:2012ky,ATLAS-CONF-2012-147,Chatrchyan:2012me,CMS-PAS-EXO-12-048} following the ``Maverick'' effective field theory approach \cite{Beltran:2010ww,Goodman:2010yf,Goodman:2010ku,Fox:2011pm,Fox:2012ee,Rajaraman:2011wf,Fox:2011fx,Lin:2013sca}. These references assume an effective field theory where the dark matter couples to quarks.  In this paper we have assumed that the DM sector couples to the SM sector via electroweak gauge bosons and Higgs.  The calculation in this case involves different processes than in previous analyses where it was assumed that the DM couples to the SM sector through quarks.  For the couplings explored in this paper are two basic channels (see Fig.\ \ref{fig:LHC}). The first is a $2 \to 3$ process mediated by an off-shell SM boson, with another boson in the final state. For example, this could produce mono-photon $+$ missing transverse energy (MET) and mono-$Z$ $+$ MET events.  The ratio of the cross sections for these processes are set for a given operator.  The second channel, vector boson fusion,  starts with 2 SM gauge bosons radiating from the incoming quarks, resulting $j+$MET or $jj+$MET events. Studies have been carried out in both the mono-$Z$/mono-$\gamma$ \cite{Carpenter:2012rg} and the vector-boson fusion \cite{Cotta:2012nj} channels. Both channels yield interesting limits. At the same time, they have only focused on a limited set of operators and final states.

\renewcommand*\arraystretch{2.25}

\begin{table}
\caption{\label{resultone} Results for dimension-6 scalar operators for scalar-WIMP annihilation. Indicated values of $\Lambda$ for $\Omega h^2$ and $\left[\sigma v\right]_\mathrm{NR}$ assume $M=130\ \GeV$ for the 130 GeV line from $\gamma\gamma$ final state, and $M=144\ \GeV$ if the 130 GeV line arises from the $\gamma Z$ final state. The $*$-superscript denotes the process for the 130 GeV line. }
\begin{ruledtabular}
\begin{tabular}{cclcc}
operator 
& 
final state 
&
$\phantom{\cos^2\theta_W}\phantom{\cos^2\theta_W}\Sigma(s,M,m_1,m_2)$ 
&
$\Omega_{\chi\chi}h^2/0.11$ 
&
$\dfrac{\left[\sigma v\right]_\mathrm{NR}}{10^{-27}\cm^3\s^{-1}}$
\\ 
[1ex]
\hline
%
\multirow{3}{*}{$\phi^\dagger \phi \: B_{\mu\nu} 
                                  B^{\mu\nu}$} 
& 
$\gamma\gamma^*$ 
&
$4\dfrac{s^2}{\Lambda^4}\cos^4\theta_W$ 
&
\multirow{3}{*}{$\left(\dfrac{1487\ \GeV}{\Lambda}\right)^4$}
&
$\left(\dfrac{2936\ \GeV}{\Lambda}\right)^4$
\\
&
$\gamma Z$ 
&
$8\dfrac{s^2}{\Lambda^4}\cos^2\theta_W\sin^2\theta_W
\left(1-\dfrac{m_Z^2}{s}\right)^2$ 
&
&
$0.403\times\left[\gamma\gamma\right]$
\\
& 
$ZZ$ 
&
$4\dfrac{s^2}{\Lambda^4}\sin^4\theta_W
\phantom{\cos^2\theta_W}
\left(1-\dfrac{4m_Z^2}{s}+6\dfrac{m_Z^4}{s^2}\right)$ 
&
&
$0.038\times\left[\gamma\gamma\right]$
\\
[2ex]
\hline
%
\multirow{3}{*}{$\phi^\dagger \phi \: B_{\mu\nu} 
                       \widetilde{B}^{\mu\nu}$}
&
$\gamma\gamma^*$ 
&
$4\dfrac{s^2}{\Lambda^4}\cos^4\theta_W$ 
&
\multirow{3}{*}{$\left(\dfrac{1486\ \GeV}{\Lambda}\right)^4$}
&
$\left(\dfrac{2936\ \GeV}{\Lambda}\right)^4$ 
\\
& 
$\gamma Z$ 
&
$8\dfrac{s^2}{\Lambda^4}\cos^2\theta_W\sin^2\theta_W
\left(1-\dfrac{m_Z^2}{s}\right)^2$ 
&
&
$0.403\times\left[\gamma\gamma\right]$
\\
&
$Z Z$ 
&
$4\dfrac{s^2}{\Lambda^4}\sin^4\theta_W
\phantom{\cos^2\theta_W}
\left(1-\dfrac{4m_Z^2}{s}\right)$ 
&
&
$0.032\times\left[\gamma\gamma\right]$
\\
[2ex]
\hline
%
\multirow{4}{*}{$\phi^\dagger \phi \: W^a_{\ \mu\nu} 
                                  W^{a\,\mu\nu}$} 
&
$\gamma\gamma^*$ 
&
$4\dfrac{s^2}{\Lambda^4}\sin^4\theta_W$ 
&
\multirow{4}{*}{$\left(\dfrac{1740\ \GeV}{\Lambda}\right)^4$} 
&
$\left(\dfrac{1604\ \GeV}{\Lambda}\right)^4$ 
\\
&
$\gamma Z$ &
$8\dfrac{s^2}{\Lambda^4}\cos^2\theta_W\sin^2\theta_W
\left(1-\dfrac{m_Z^2}{s}\right)^2$ &
&
$4.516\times\left[\gamma\gamma\right]$ 
\\
&  
$ZZ$&
$4\dfrac{s^2}{\Lambda^4}\cos^4\theta_W
\phantom{\cos^2\theta_W}
\left(1-\dfrac{4m_Z^2}{s}+6\dfrac{m_Z^4}{s^2}\right)$ &
&
$4.782\times\left[\gamma\gamma\right]$
\\
&
$W^+W^-$ &
$8\dfrac{s^2}{\Lambda^4}
\phantom{\cos^2\theta_W}\phantom{\cos^2\theta_W}
\left(1-\dfrac{4m_W^2}{s}+6\dfrac{m_W^4}{s^2}\right)$ &
&
$19.98\times\left[\gamma\gamma\right]$
\\
[2ex]
\hline
%
\multirow{4}{*}{$\phi^\dagger \phi \: W^a_{\ \mu\nu} 
                                      \widetilde{W}^{a\,\mu\nu}$}
&
$\gamma\gamma^*$ 
&
$4\dfrac{s^2}{\Lambda^4}\sin^4\theta_W$ 
&
\multirow{4}{*}{$\left(\dfrac{1705\ \GeV}{\Lambda}\right)^4$}
&
$\left(\dfrac{1604\ \GeV}{\Lambda}\right)^4$ 
\\
&
$\gamma Z$ 
&
$8\dfrac{s^2}{\Lambda^4}\cos^2\theta_W\sin^2\theta_W
\left(1-\dfrac{m_Z^2}{s}\right)^2$ 
&
&
$4.516\times\left[\gamma\gamma\right]$ 
\\
&   
$ZZ$
&
$4\dfrac{s^2}{\Lambda^4}\cos^4\theta_W
\phantom{\cos^2\theta_W}
\left(1-\dfrac{4m_Z^2}{s}\right)$ 
&
&
$4.056\times\left[\gamma\gamma\right]$ 
\\
&
$W^+W^-$
&
$8\dfrac{s^2}{\Lambda^4} 
\phantom{\cos^2\theta_W}\phantom{\cos^2\theta_W}
\left(1-\dfrac{4m_W^2}{s}\right)$
&
&
$18.35\times\left[\gamma\gamma\right]$ 
\\ 
[2ex]
\hline
%
\multirow{4}{*}{$\phi^\dagger \phi \: W^a_{\ \mu\nu} 
                                                 W^{a\,\mu\nu}$}
&
$\gamma\gamma$ 
&
$4\dfrac{s^2}{\Lambda^4}\sin^4\theta_W$ 
&
\multirow{4}{*}{$\left(\dfrac{1890\ \GeV}{\Lambda}\right)^4$}
&
$0.205\times\left[\gamma Z\right]$  
\\
&
$\gamma Z^*$ 
&
$8\dfrac{s^2}{\Lambda^4}\cos^2\theta_W\sin^2\theta_W
\left(1-\dfrac{m_Z^2}{s}\right)^2$ 
&
&
$\left(\dfrac{2509\ \GeV}{\Lambda}\right)^4$
\\
&  
$ZZ$
&
$4\dfrac{s^2}{\Lambda^4}\cos^4\theta_W
\phantom{\cos^2\theta_W}
\left(1-\dfrac{4m_Z^2}{s}+6\dfrac{m_Z^4}{s^2}\right)$ 
&
&
$1.172\times\left[\gamma Z\right]$ 
\\
&
$W^+W^-$
&
$8\dfrac{s^2}{\Lambda^4}
\phantom{\cos^2\theta_W}\phantom{\cos^2\theta_W}
\left(1-\dfrac{4m_W^2}{s}+6\dfrac{m_W^4}{s^2}\right)$
&
&
$4.661\times\left[\gamma Z\right]$ 
\\
[2ex]
\hline
\multirow{4}{*}{$\phi^\dagger \phi \: W^a_{\ \mu\nu} 
                                       \widetilde{W}^{a\,\mu\nu}$}
&
$\gamma\gamma$ 
&
$4\dfrac{s^2}{\Lambda^4}\sin^4\theta_W$ 
&
\multirow{4}{*}{$\left(\dfrac{1866\ \GeV}{\Lambda}\right)^4$}
&
$0.205\times\left[\gamma Z\right]$ 
\\
&
$\gamma Z^*$ 
&
$8\dfrac{s^2}{\Lambda^4}\cos^2\theta_W\sin^2\theta_W
\left(1-\dfrac{m_Z^2}{s}\right)^2$ &
&
$\left(\dfrac{2509\ \GeV}{\Lambda}\right)^4$ 
\\
&  
$ZZ$
&
$4\dfrac{s^2}{\Lambda^4}\cos^4\theta_W
\phantom{\cos^2\theta_W}
\left(1-\dfrac{4m_Z^2}{s}\right)$ 
&
&
$1.065\times\left[\gamma Z\right]$
\\
& 
$W^+W^-$
&
$8\dfrac{s^2}{\Lambda^4} 
\phantom{\cos^2\theta_W}\phantom{\cos^2\theta_W}
\left(1-\dfrac{4m_W^2}{s}\right)$
&
& 
$4.427\times\left[\gamma Z\right]$ 
\\
[2ex] 
\end{tabular}
\end{ruledtabular}
\end{table}

\begin{table}
\caption{\label{resulttwo} Results for dimension-7 scalar operators for fermionic-WIMP annihilation.  Indicated values of $\Lambda$ for $\Omega h^2$ assume $M=130\ \GeV$.  Since $\Sigma$ always has a factor of $1-4M^2/s\rightarrow M^2v^2$ in the non-relativistic limit, it will not have a measurable signals from present-day annihilation.  }
\begin{ruledtabular}
\begin{tabular}{cclcc}
operator 
& 
final state 
&
$\phantom{\cos^2\theta_W}\phantom{\cos^2\theta_W}\Sigma(s,M,m_1,m_2)$ 
&
$\Omega_{\chi\chi}h^2/0.11$ 
&
$\dfrac{\left[\sigma v\right]_\mathrm{NR}}{10^{-27}\cm^3\s^{-1}}$
\\ 
[1ex]
\hline
\multirow{3}{*}{$\bar{\chi}\chi \: B_{\mu\nu} 
                               B^{\mu\nu}$}
& 
$\gamma\gamma$ 
&
$2\dfrac{s^3}{\Lambda^6}\cos^4\theta_W
\phantom{\cos^2\theta_W}
\left(1-\dfrac{4M^2}{s}\right)$ 
&
\multirow{3}{*}{$\left(\dfrac{412\ \GeV}{\Lambda}\right)^6$}
&
$\left[\sigma v\right]_\mathrm{NR}\propto v^2$
\\
&
$\gamma Z$ 
&
$4\dfrac{s^3}{\Lambda^6}\cos^2\theta_W\sin^2\theta_W
\left(1-\dfrac{4M^2}{s}\right)
\left(1-\dfrac{m_Z^2}{s}\right)^2$ 
&
&
$\left[\sigma v\right]_\mathrm{NR}\propto v^2$
\\
& 
$Z Z$ 
&
$2\dfrac{s^3}{\Lambda^6}\sin^4\theta_W
\phantom{\cos^2\theta_W}
\left(1-\dfrac{4M^2}{s}\right)
\left(1-\dfrac{4m_Z^2}{s}+6\dfrac{m_Z^4}{s^2}\right)$ 
&
&
$\left[\sigma v\right]_\mathrm{NR}\propto v^2$
\\
[2ex]
\hline
\multirow{3}{*}{$\bar{\chi}\chi \: B_{\mu\nu} 
                    \widetilde{B}^{\mu\nu}$ }
&
$\gamma\gamma$ 
&
$2\dfrac{s^3}{\Lambda^6}\cos^4\theta_W
\phantom{\cos^2\theta_W}
\left(1-\dfrac{4M^2}{s}\right)$ 
&
\multirow{3}{*}{$\left(\dfrac{412\ \GeV}{\Lambda}\right)^6$}
&
$\left[\sigma v\right]_\mathrm{NR}\propto v^2$
\\
& 
$\gamma Z$ 
&
$4\dfrac{s^3}{\Lambda^6}\cos^2\theta_W\sin^2\theta_W
\left(1-\dfrac{4M^2}{s}\right)
\left(1-\dfrac{m_Z^2}{s}\right)^2$ 
&
&
$\left[\sigma v\right]_\mathrm{NR}\propto v^2$
\\
&
$Z Z$ 
&
$2\dfrac{s^3}{\Lambda^6}\sin^4\theta_W
\phantom{\cos^2\theta_W}
\left(1-\dfrac{4M^2}{s}\right)\left(1-\dfrac{4m_Z^2}{s}\right)$ 
&
&
$\left[\sigma v\right]_\mathrm{NR}\propto v^2$
\\
[2ex]
\hline
%
\multirow{4}{*}{$\bar{\chi}\chi \: W^a_{\ \mu\nu} 
                                              W^{a\,\mu\nu}$}
&
$\gamma\gamma$ 
&
$2\dfrac{s^3}{\Lambda^6}\sin^4\theta_W
\phantom{\cos^2\theta_W}
\left(1-\dfrac{4M^2}{s}\right)$ 
&
\multirow{4}{*}{$\left(\dfrac{458\ \GeV}{\Lambda}\right)^6$}
&
$\left[\sigma v\right]_\mathrm{NR}\propto v^2$
\\
&
$\gamma Z$ 
&
$4\dfrac{s^3}{\Lambda^6}\cos^2\theta_W\sin^2\theta_W
\left(1-\dfrac{4M^2}{s}\right)
\left(1-\dfrac{m_Z^2}{s}\right)^2$ 
&
&
$\left[\sigma v\right]_\mathrm{NR}\propto v^2$
\\
&  
$ZZ$
&
$2\dfrac{s^3}{\Lambda^6}\cos^4\theta_W
\phantom{\cos^2\theta_W}
\left(1-\dfrac{4M^2}{s}\right)
\left(1-\dfrac{4m_Z^2}{s}+6\dfrac{m_Z^4}{s^2}\right)$ 
&
&
$\left[\sigma v\right]_\mathrm{NR}\propto v^2$
\\
&
$W^+W^-$
&
$4\dfrac{s^3}{\Lambda^6}
\phantom{\cos^2\theta_W}\phantom{\cos^2\theta_W}
\left(1-\dfrac{4M^2}{s}\right)
\left(1-\dfrac{4m_W^2}{s}+6\dfrac{m_W^4}{s^2}\right)$
&
&
$\left[\sigma v\right]_\mathrm{NR}\propto v^2$
\\
[2ex]
\hline
%
\multirow{4}{*}{$\bar{\chi}\chi \: W^a_{\ \mu\nu} 
                                   \widetilde{W}^{a\,\mu\nu}$}
&
$\gamma\gamma$ 
&
$2\dfrac{s^3}{\Lambda^6}\sin^4\theta_W
\phantom{\cos^2\theta_W}
\left(1-\dfrac{4M^2}{s}\right)$ 
&
\multirow{4}{*}{$\left(\dfrac{452\ \GeV}{\Lambda}\right)^6$}
&
$\left[\sigma v\right]_\mathrm{NR}\propto v^2$
\\
&
$\gamma Z$ 
&
$4\dfrac{s^3}{\Lambda^6}\cos^2\theta_W\sin^2\theta_W
\left(1-\dfrac{4M^2}{s}\right)
\left(1-\dfrac{m_Z^2}{s}\right)^2$ 
&
&
$\left[\sigma v\right]_\mathrm{NR}\propto v^2$
\\
&  
$Z Z$
&
$2\dfrac{s^3}{\Lambda^6}\cos^4\theta_W
\phantom{\cos^2\theta_W}
\left(1-\dfrac{4M^2}{s}\right)
\left(1-\dfrac{4m_Z^2}{s}\right)$ 
&
&
$\left[\sigma v\right]_\mathrm{NR}\propto v^2$
\\
&
$W^+W^-$
&
$4\dfrac{s^3}{\Lambda^6} 
\phantom{\cos^2\theta_W}\phantom{\cos^2\theta_W}
\left(1-\dfrac{4M^2}{s}\right)
\left(1-\dfrac{4m_W^2}{s}\right)$ 
&
&
$\left[\sigma v\right]_\mathrm{NR}\propto v^2$
\\
[2ex]
\end{tabular}
\end{ruledtabular}
\end{table}

\begin{table}
\caption{\label{resultthree} Results for dimension-7 scalar operators for fermionic-WIMP annihilation. Indicated values of $\Lambda$ for $\Omega h^2$ and $\left[\sigma v\right]_\mathrm{NR}$ assume $M=130\ \GeV$ for the 130 GeV line from $\gamma\gamma$ final state, and $M=144\ \GeV$ if the 130 GeV line arises from the $\gamma Z$ final state. The $*$-superscript denotes the process for the 130 GeV line.  }
\begin{ruledtabular}
\begin{tabular}{cclcc}
operator 
& 
final state 
&
$\phantom{\cos^2\theta_W}\phantom{\cos^2\theta_W}\Sigma(s,M,m_1,m_2)$ 
&
$\Omega_{\chi\chi}h^2/0.11$ 
&
$\dfrac{\left[\sigma v\right]_\mathrm{NR}}{10^{-27}\cm^3\s^{-1}}$
\\ 
[1ex]
\hline
%
\multirow{3}{*}{$\bar{\chi}i\gamma^5\chi \: B_{\mu\nu} B^{\mu\nu}$}
& 
$\gamma\gamma^*$ 
&
$2\dfrac{s^3}{\Lambda^6}\cos^4\theta_W$ 
&
\multirow{3}{*}{$\left(\dfrac{741\ \GeV}{\Lambda}\right)^6$}
&
$\left(\dfrac{1166\ \GeV}{\Lambda}\right)^6$
\\
&
$\gamma Z$ 
&
$4\dfrac{s^3}{\Lambda^6}\cos^2\theta_W\sin^2\theta_W
\left(1-\dfrac{m_Z^2}{s}\right)^2$ &
&
$0.403\left[\gamma\gamma\right]$
\\
& 
$Z Z$ 
&
$2\dfrac{s^3}{\Lambda^6}\sin^4\theta_W
\phantom{\cos^2\theta_W}
\left(1-\dfrac{4m_Z^2}{s}+6\dfrac{m_Z^4}{s^2}\right)$ &
&
$0.038\left[\gamma\gamma\right]$
\\
[2ex]
\hline
\multirow{3}{*}{$\bar{\chi}i\gamma^5\chi \: B_{\mu\nu} 
                                            \widetilde{B}^{\mu\nu}$}
&
$\gamma\gamma^*$ 
&
$2\dfrac{s^3}{\Lambda^6}\cos^4\theta_W$ 
&
\multirow{3}{*}{$\left(\dfrac{740\ \GeV}{\Lambda}\right)^6$}
&
$\left(\dfrac{1166\ \GeV}{\Lambda}\right)^6$ 
\\
& 
$\gamma Z$ 
&
$4\dfrac{s^3}{\Lambda^6}\cos^2\theta_W\sin^2\theta_W
\left(1-\dfrac{m_Z^2}{s}\right)^2$ 
&
&
$0.403\left[\gamma\gamma\right]$
\\
&
$Z Z$ 
&
$2\dfrac{s^3}{\Lambda^6}\sin^4\theta_W
\phantom{\cos^2\theta_W}
\left(1-\dfrac{4m_Z^2}{s}\right)$ 
&
&
$0.032\left[\gamma\gamma\right]$
\\
[2ex]
\hline
\multirow{4}{*}{$\bar{\chi}i\gamma^5\chi \: W^a_{\ \mu\nu} 
                                                      W^{a\,\mu\nu}$} 
&
$\gamma\gamma^*$ 
&
$2\dfrac{s^3}{\Lambda^6}\sin^4\theta_W$ 
&
\multirow{4}{*}{$\left(\dfrac{823\ \GeV}{\Lambda}\right)^6$}
&
$\left(\dfrac{779\ \GeV}{\Lambda}\right)^6$ 
\\
&
$\gamma Z$ 
&
$4\dfrac{s^3}{\Lambda^6}\cos^2\theta_W\sin^2\theta_W
\left(1-\dfrac{m_Z^2}{s}\right)^2$ 
&
&
$4.516\left[\gamma\gamma\right]$ 
\\
&  
$ZZ$
&
$2\dfrac{s^3}{\Lambda^6}\cos^4\theta_W
\phantom{\cos^2\theta_W}
\left(1-\dfrac{4m_Z^2}{s}+6\dfrac{m_Z^4}{s^2}\right)$
&
&
$4.782\left[\gamma\gamma\right]$
\\
&
$W^+W^-$ &
$4\dfrac{s^3}{\Lambda^6}
\phantom{\cos^2\theta_W}
\phantom{\cos^2\theta_W}
\left(1-\dfrac{4m_W^2}{s}+6\dfrac{m_W^4}{s^2}\right)$ &
&
$19.98\left[\gamma\gamma\right]$
\\
[2ex]
\hline
%
\multirow{4}{*}{$\bar{\chi}i\gamma^5\chi \: W^a_{\ \mu\nu} 
                \widetilde{W}^{a\,\mu\nu}$}
&
$\gamma\gamma^*$ 
&
$2\dfrac{s^3}{\Lambda^6}\sin^4\theta_W$ 
&
\multirow{4}{*}{$\left(\dfrac{812\ \GeV}{\Lambda}\right)^6$}
&
$\left(\dfrac{779\ \GeV}{\Lambda}\right)^6$ 
\\
&
$\gamma Z$ 
&
$4\dfrac{s^3}{\Lambda^6}\cos^2\theta_W\sin^2\theta_W
\left(1-\dfrac{m_Z^2}{s}\right)^2$ 
&
&
$4.516\left[\gamma\gamma\right]$ 
\\
&  
$ZZ$
&
$2\dfrac{s^3}{\Lambda^6}\cos^4\theta_W
\phantom{\cos^2\theta_W}
\left(1-\dfrac{4m_Z^2}{s}\right)$
&
&
$4.056\left[\gamma\gamma\right]$
\\
&
$W^+W^-$
&
$4\dfrac{s^3}{\Lambda^6} 
\phantom{\cos^2\theta_W}
\phantom{\cos^2\theta_W}
\left(1-\dfrac{4m_W^2}{s}\right)$
&
&
$18.35\left[\gamma\gamma\right]$
\\ 
[2ex]
\hline
\multirow{4}{*}{$\bar{\chi}i\gamma^5\chi \: W^a_{\ \mu\nu} 
                                                      W^{a\,\mu\nu}$} 
&
$\gamma\gamma$ 
&
$2\dfrac{s^3}{\Lambda^6}\sin^4\theta_W$ 
&
\multirow{4}{*}{$\left(\dfrac{899\ \GeV}{\Lambda}\right)^6$}
&
$0.205\left[\gamma Z\right]$
\\
&
$\gamma Z^*$ 
&
$4\dfrac{s^3}{\Lambda^6}\cos^2\theta_W\sin^2\theta_W
\left(1-\dfrac{m_Z^2}{s}\right)^2$ 
&
&
$\left(\dfrac{1086\ \GeV}{\Lambda}\right)^6$  
\\
&  
$ZZ$
&
$2\dfrac{s^3}{\Lambda^6}\cos^4\theta_W
\phantom{\cos^2\theta_W}
\left(1-\dfrac{4m_Z^2}{s}+6\dfrac{m_Z^4}{s^2}\right)$
&
&
$1.172\left[\gamma Z\right]$
\\
&
$W^+W^-$ &
$4\dfrac{s^3}{\Lambda^6}
\phantom{\cos^2\theta_W}
\phantom{\cos^2\theta_W}
\left(1-\dfrac{4m_W^2}{s}+6\dfrac{m_W^4}{s^2}\right)$ &
&
$4.661\left[\gamma Z\right]$
\\
[2ex]
\hline
%
\multirow{4}{*}{$\bar{\chi}i\gamma^5\chi \: W^a_{\ \mu\nu} 
                \widetilde{W}^{a\,\mu\nu}$}
&
$\gamma \gamma$ 
&
$2\dfrac{s^3}{\Lambda^6}\sin^4\theta_W$ 
&
\multirow{4}{*}{$\left(\dfrac{892\ \GeV}{\Lambda}\right)^6$}
&
$0.205\left[\gamma Z\right]$ 
\\
&
$\gamma Z^*$ 
&
$4\dfrac{s^3}{\Lambda^6}\cos^2\theta_W\sin^2\theta_W
\left(1-\dfrac{m_Z^2}{s}\right)^2$ 
&
&
$\left(\dfrac{1086\ \GeV}{\Lambda}\right)^6$
\\
&  
$ZZ$
&
$2\dfrac{s^3}{\Lambda^6}\cos^4\theta_W
\phantom{\cos^2\theta_W}
\left(1-\dfrac{4m_Z^2}{s}\right)$
&
&
$1.065\left[\gamma Z\right]$
\\
&
$W^+W^-$
&
$4\dfrac{s^3}{\Lambda^6} 
\phantom{\cos^2\theta_W}
\phantom{\cos^2\theta_W}
\left(1-\dfrac{4m_W^2}{s}\right)$
&
&
$4.427\left[\gamma Z\right]$
\\
[2ex]
\end{tabular}
\end{ruledtabular}
\end{table}

\begin{table}
\caption{\label{resultthreefive} Results for dimension-5 scalar operators for fermionic-WIMP annihilation. Indicated values of $\Lambda$ for $\Omega h^2$ and $\left[\sigma v\right]_\mathrm{NR}$ assume $M=130\ \GeV$. }
\begin{ruledtabular}
\begin{tabular}{cclcc}
operator 
& 
final state 
&
$\Sigma(s,M,m_1,m_2)$ 
&
$\Omega_{\chi\chi}h^2/0.11$ 
&
$\dfrac{\left[\sigma v\right]_\mathrm{NR}}{10^{-27}\cm^3\s^{-1}}$
\\ 
[1ex]
\hline
$\bar{\chi}\chi H^\dagger H$
&
$hh$
&
$\dfrac{1}{8}\dfrac{s}{\Lambda^2}\left(1-\dfrac{4M^2}{s}\right)$
&
$\left(\dfrac{280\ \GeV}{\Lambda}\right)^2$
&
$\left[\sigma v\right]_\mathrm{NR}\propto v^2$
\\
[1ex]
$\bar{\chi}i\gamma^5\chi H^\dagger H$
&
$hh$
&
$\dfrac{1}{8}\dfrac{s}{\Lambda^2}$
&
$\left(\dfrac{1628\ \GeV}{\Lambda}\right)^2$
&
$\left(\dfrac{7609\ \GeV}{\Lambda}\right)^2$
\\
[2ex]
\end{tabular}
\end{ruledtabular}
\end{table}


\renewcommand*\arraystretch{2.4}

\begin{table}
\caption{\label{resultfourpp} Results for dimension-8 vector operators for scalar-WIMP annihilation. These terms can not produce a photon line. A value of $M=155$ GeV is assumed.}
\begin{ruledtabular}
\begin{tabular}{cclcc}
operator 
& 
final state 
&
\hspace*{36pt}$\Sigma(s,M,m_1,m_2)$ 
&
$\Omega_{\chi\chi}h^2/0.11$ 
&
$\dfrac{\left[\sigma v\right]_\mathrm{NR}}{10^{-27}\cm^3\s^{-1}}$ 
\\ 
[1ex]
\hline
%
\parbox[l]{118pt}{$\left(\phi^\dagger \partial^\mu \phi + h.c. \right)
\times$ \\
$\left(B_{\lambda\mu} Y_H \, H^\dagger D^\lambda H + h.c. \right)$}
&
$Zh$ 
&
$\dfrac{1}{16}\dfrac{\vev^2s^3}{\Lambda^8}\sin^2\theta_W\dfrac{m_Z^2}{s}\beta^2_{Zh}$
&
$\left(\dfrac{219\ \GeV}{\Lambda}\right)^8$
&
$\left(\dfrac{322\ \GeV}{\Lambda}\right)^8$
\\
[2ex]
\hline
\parbox[l]{118pt}{$\left(\phi^\dagger \partial^\mu \phi + h.c. \right) \times $ \\
$\left(W^a_{\ \lambda\mu}  \, H^\dagger t^aD^\lambda H + h.c. \right)$}
& 
$Zh$
&
$\dfrac{1}{16}\dfrac{\vev^2s^3}{\Lambda^8}\cos^2\theta_W\dfrac{m_Z^2}{s}\beta^2_{Zh}$
&
$\left(\dfrac{254\ \GeV}{\Lambda}\right)^8$
&
$\left(\dfrac{375\ \GeV}{\Lambda}\right)^8$
\\
[2ex]
\end{tabular}
$\beta_{Zh}^2=\left[1-(m_h+m_Z)^2/s\right]\left[1-(m_h-m_Z)^2/s\right]$
\end{ruledtabular}
\end{table}

\begin{table}
\caption{\label{resultfourpm} Results for dimension-8 vector operators for scalar-WIMP annihilation. Indicated values of $\Lambda$ for $\Omega h^2$ and $\left[\sigma v\right]_\mathrm{NR}$ assume $M=144\ \GeV$ necessary to produce a 130 GeV line in the $\gamma Z$ final state.}
\begin{ruledtabular}
\begin{tabular}{cclcc}
operator 
& 
final state 
&
\hspace*{36pt}$\Sigma(s,M,m_1,m_2)$ 
&
$\Omega_{\chi\chi}h^2/0.11$ 
&
$\dfrac{\left[\sigma v\right]_\mathrm{NR}}{10^{-27}\cm^3\s^{-1}}$ 
\\ 
[1ex]
\hline
\multirow{2}{*}{\parbox[l]{118pt}{$\left(\phi^\dagger \partial^\mu \phi + h.c. \right) 
\times$ \\
$i\left(B_{\lambda\mu} Y_H \, H^\dagger D^\lambda H - h.c. \right)$}}
& 
$\gamma Z$
&
$\dfrac{1}{8}\dfrac{\vev^2s^3}{\Lambda^8}\cos^2\theta_W\dfrac{m_Z^2}{s}\left(1-\dfrac{m_Z^2}{s}\right)^2$
&
\multirow{2}{*}{$\left(\dfrac{323\ \GeV}{\Lambda}\right)^8$}
&
$\left(\dfrac{438\ \GeV}{\Lambda}\right)^8$
\\
&
$ZZ$
&
$\dfrac{3}{8}\dfrac{\vev^2s^3}{\Lambda^8}\sin^2\theta_W\dfrac{m_Z^2}{s}$
&
&
$0.952\left[\gamma Z\right]$
\\
[2ex]
\hline
\multirow{2}{*}{\parbox[l]{118pt}{$\left(\phi^\dagger \partial^\mu \phi + h.c. \right) 
\times$ \\
$i\left(\widetilde{B}_{\lambda\mu} Y_H \, H^\dagger D^\lambda H - h.c. \right)$}}
& 
$\gamma Z$
&
$\dfrac{1}{8}\dfrac{\vev^2s^3}{\Lambda^8}\cos^2\theta_W\dfrac{m_Z^2}{s}\left(1-\dfrac{m_Z^2}{s}\right)^2$
&
\multirow{2}{*}{$\left(\dfrac{310\ \GeV}{\Lambda}\right)^8$}
&
$\left(\dfrac{438\ \GeV}{\Lambda}\right)^8$
\\
&
$ZZ$
&
$\dfrac{1}{4}\dfrac{\vev^2s^3}{\Lambda^8}\sin^2\theta_W\dfrac{m_Z^2}{s}\left(1-\dfrac{4m_Z^2}{s}\right)$
&
&
$0.380\left[\gamma Z\right]$
\\
[2ex]
\hline
\multirow{3}{*}{\parbox[l]{118pt}{$\left(\phi^\dagger \partial^\mu \phi + h.c. \right)
\times$ \\
$i\left(W^a_{\ \lambda\mu}  \, H^\dagger t^aD^\lambda H - h.c. \right)$}}
& 
$\gamma Z$
&
$\dfrac{1}{8}\dfrac{\vev^2s^3}{\Lambda^8}\sin^2\theta_W\dfrac{m_Z^2}{s}
\left(1-\dfrac{m_Z^2}{s}\right)^2$
&
\multirow{3}{*}{$\left(\dfrac{387\ \GeV}{\Lambda}\right)^8$}
&
$\left(\dfrac{377\ \GeV}{\Lambda}\right)^8$
\\
&
$ZZ$
&
$\dfrac{3}{8}\dfrac{\vev^2s^3}{\Lambda^8}\cos^2\theta_W\dfrac{m_Z^2}{s}$
&
&
$10.67\left[\gamma Z\right]$
\\
&
$W^+W^-$
&
$\dfrac{3}{4}\dfrac{\vev^2s^3}{\Lambda^8}\phantom{\sin^2\theta_W}\dfrac{m_W^2}{s}$
&
&
$15.90\left[\gamma Z\right]$
\\
[2ex]
\hline
\multirow{3}{*}{\parbox[l]{118pt}{$\left(\phi^\dagger \partial^\mu \phi + h.c. \right) 
\times$ \\
$i\left(\widetilde{W}^a_{\ \lambda\mu}  \, H^\dagger t^aD^\lambda H - h.c. \right)$}}
& 
$\gamma Z$
&
$\dfrac{1}{8}\dfrac{\vev^2s^3}{\Lambda^8}\sin^2\theta_W\dfrac{m_Z^2}{s}
\left(1-\dfrac{m_Z^2}{s}\right)^2$
&
\multirow{3}{*}{$\left(\dfrac{361\ \GeV}{\Lambda}\right)^8$}
&
$\left(\dfrac{377\ \GeV}{\Lambda}\right)^8$
\\
&
$ZZ$
&
$\dfrac{1}{4}\dfrac{\vev^2s^3}{\Lambda^8}\cos^2\theta_W\dfrac{m_Z^2}{s}
\left(1-\dfrac{4m_Z^2}{s}\right)$
&
&
$4.261\left[\gamma Z\right]$
\\
&
$W^+W^-$
&
$\dfrac{1}{2}\dfrac{\vev^2s^3}{\Lambda^8}\phantom{\sin^2\theta_W}\dfrac{m_W^2}{s}
\left(1-\dfrac{4m_W^2}{s}\right)$
&
&
$10.60\left[\gamma Z\right]$
\\
[2ex]
\end{tabular}
\end{ruledtabular}
\end{table}

\begin{table}
\caption{\label{resultfivemp}  Results for dimension-8 vector operators for scalar-WIMP annihilation. Indicated values of $\Lambda$ for $\Omega h^2$ and $\left[\sigma v\right]_\mathrm{NR}$ assume $M=155\ \GeV$ necessary to produce a 130 GeV line in the $\gamma h$ final state. }
\begin{ruledtabular}
\begin{tabular}{cclcc}
operator 
& 
final state 
&
\hspace*{48pt}$\Sigma(s,M,m_1,m_2)$ 
&
$\Omega_{\chi\chi}h^2/0.11$ 
&
$\dfrac{\left[\sigma v\right]_\mathrm{NR}}{10^{-27}\cm^3\s^{-1}}$ 
\\ 
[1ex]
\hline
%
\multirow{3}{*}{\parbox[l]{114pt}{$i\left(\phi^\dagger \partial^\mu \phi - h.c. \right) 
\times$ \\
$\left(B_{\lambda\mu} Y_H \, H^\dagger D^\lambda H + h.c. \right)$}}
& 
$\gamma h$
&
$\dfrac{1}{24}\dfrac{\vev^2s^3}{\Lambda^8}\cos^2\theta_W\left(1-\dfrac{4M^2}{s}\right)\left(1-\dfrac{m_h^2}{s}\right)^2$
&
\multirow{3}{*}{$\left(\dfrac{230\ \GeV}{\Lambda}\right)^8$} 
&
$\left[\sigma v\right]_\mathrm{NR}\propto v^2$
\\
&
\multirow{2}{*}{$Zh$}
&
\multirow{2}{*}{\parbox{137pt}{\hspace*{-15pt}$\dfrac{1}{24}\dfrac{\vev^2s^3}{\Lambda^8}\sin^2\theta_W\left(1-\dfrac{4M^2}{s}\right)$ \\
$\quad \times \left[\beta_{Zh}^2\left(1+\dfrac{m_Z^2}{2s}\right)+\dfrac{6m_Z^2m_h^2}{s^2}\right]$}}
&
&
\multirow{2}{*}{$\left[\sigma v\right]_\mathrm{NR}\propto v^2$}
\\
&
&
&
\\
[2ex]
\hline
\multirow{2}{*}{\parbox[l]{114pt}{$i\left(\phi^\dagger \partial^\mu \phi - h.c. \right)
\times $ \\
$\left(\widetilde{B}_{\lambda\mu} Y_H \, H^\dagger D^\lambda H + h.c. \right)$}}
& 
$\gamma h$
&
$\dfrac{1}{24}\dfrac{\vev^2s^3}{\Lambda^8}\cos^2\theta_W\left(1-\dfrac{4M^2}{s}\right)\left(1-\dfrac{m_h^2}{s}\right)^2$
&
\multirow{2}{*}{$\left(\dfrac{230\ \GeV}{\Lambda}\right)^8$} 
&
$\left[\sigma v\right]_\mathrm{NR}\propto v^2$
\\
&
$Zh$
&
$\dfrac{1}{24}\dfrac{\vev^2s^3}{\Lambda^8}\sin^2\theta_W
\left(1-\dfrac{4M^2}{s}\right)\beta_{Zh}^2$
&
&
$\left[\sigma v\right]_\mathrm{NR}\propto v^2$
\\
[2ex]
\hline
\multirow{5}{*}{\parbox[l]{114pt}{$i\left(\phi^\dagger \partial^\mu \phi - h.c. \right) \times $ \\
$\left(W^a_{\ \lambda\mu}  \, H^\dagger t^aD^\lambda H + h.c. \right)$}}
& 
$\gamma h$
&
$\dfrac{1}{24}\dfrac{\vev^2s^3}{\Lambda^8}\sin^2\theta_W
\left(1-\dfrac{4M^2}{s}\right)\left(1-\dfrac{m_h^2}{s}\right)^2$
&
\multirow{5}{*}{$\left(\dfrac{255\ \GeV}{\Lambda}\right)^8$} 
&
$\left[\sigma v\right]_\mathrm{NR}\propto v^2$
\\
&
$Zh$
&
\multirow{2}{*}{\parbox{136pt}{\hspace*{-12pt}$\dfrac{1}{24}\dfrac{\vev^2s^3}{\Lambda^8}\cos^2\theta_W\left(1-\dfrac{4M^2}{s}\right)$\\
$\quad\times\left[\beta_{Zh}^2\left(1+\dfrac{m_Z^2}{2s}\right)+\dfrac{6m_Z^2m_h^2}{s^2}\right]$}}
&
&
\multirow{2}{*}{$\left[\sigma v\right]_\mathrm{NR}\propto v^2$}
\\
& & & & \\
&
$W^+W^-$
&
\multirow{2}{*}{\parbox{175pt}{$\dfrac{1}{24}\dfrac{\vev^2s^3}{\Lambda^8}\phantom{\sin^2\theta_W}
\left(1-\dfrac{4M^2}{s}\right)\left(1-\dfrac{4m_W^2}{s}\right)$ \\
$\times\left(1+\dfrac{3m_W^2}{s}\right)$}}
&
&
\multirow{2}{*}{$\left[\sigma v\right]_\mathrm{NR}\propto v^2$}
\\
& & & & 
\\
[2ex]
\hline
\multirow{3}{*}{\parbox[l]{114pt}{$i\left(\phi^\dagger \partial^\mu \phi - h.c. \right) 
\times$ \\
$\left(\widetilde{W}^a_{\ \lambda\mu}  \, H^\dagger t^aD^\lambda H + h.c. \right)$}}
& 
$\gamma h$
&
$\dfrac{1}{24}\dfrac{\vev^2s^3}{\Lambda^8}\sin^2\theta_W
\left(1-\dfrac{4M^2}{s}\right)\left(1-\dfrac{m_h^2}{s}\right)$
&
\multirow{3}{*}{$\left(\dfrac{277\ \GeV}{\Lambda}\right)^8$}
&
$\left[\sigma v\right]_\mathrm{NR}\propto v^2$
\\
&
$Zh$
&
$\dfrac{1}{24}\dfrac{\vev^2s^3}{\Lambda^8}\cos^2\theta_W
\left(1-\dfrac{4M^2}{s}\right)\beta_{Zh}^2$
&
&
$\left[\sigma v\right]_\mathrm{NR}\propto v^2$
\\
&
$W^+W^-$
&
$\dfrac{1}{12}\dfrac{\vev^2s^3}{\Lambda^8}\phantom{\sin^2\theta_W}
\left(1-\dfrac{4M^2}{s}\right)\left(1+\dfrac{2m_W^2}{s}\right)$
&
&
$\left[\sigma v\right]_\mathrm{NR}\propto v^2$
\\
[2ex]
\end{tabular}
$\beta_{Zh}^2=\left[1-(m_h+m_Z)^2/s\right]\left[1-(m_h-m_Z)^2/s\right]$
\end{ruledtabular}
\end{table}

\begin{table}
\caption{\label{resultfivemm}  Results for dimension-8 vector operators for scalar-WIMP annihilation. Indicated values of $\Lambda$ for $\Omega h^2$ and $\left[\sigma v\right]_\mathrm{NR}$ assume $M=144\ \GeV$ necessary to produce a 130 GeV line in the $\gamma Z$ final state. }
\begin{ruledtabular}
\begin{tabular}{cclcc}
operator 
& 
final state 
&
\hspace*{48pt}$\Sigma(s,M,m_1,m_2)$ 
&
$\Omega_{\chi\chi}h^2/0.11$ 
&
$\dfrac{\left[\sigma v\right]_\mathrm{NR}}{10^{-27}\cm^3\s^{-1}}$ 
\\ 
[1ex]
\hline
\multirow{3}{*}{\parbox[l]{114pt}{$i\left(\phi^\dagger \partial^\mu \phi - h.c. \right) 
\times$ \\
$\left(B_{\lambda\mu} Y_H \, H^\dagger D^\lambda H - h.c. \right)$}}
& 
\multirow{2}{*}{$\gamma Z$}
&
\multirow{2}{*}{\parbox{171pt}{$\dfrac{1}{24}\dfrac{\vev^2s^3}{\Lambda^8}\cos^2\theta_W\left(1-\dfrac{4M^2}{s}\right)\left(1+\dfrac{m_Z^2}{s}\right)$ \\  $\times\left(1-\dfrac{m_Z^2}{s}\right)^2$}}
&
\multirow{3}{*}{$\left(\dfrac{230\ \GeV}{\Lambda}\right)^8$}
&
\multirow{2}{*}{$\left[\sigma v\right]_\mathrm{NR}\propto v^2$}
\\
& & & & \\
&
$ZZ$
&
$\dfrac{1}{24}\dfrac{\vev^2s^3}{\Lambda^8}\sin^2\theta_W\left(1-\dfrac{4M^2}{s}\right)\left(1-\dfrac{4m_Z^2}{s}\right)$
&
&
$\left[\sigma v\right]_\mathrm{NR}\propto v^2$
\\
[2ex]
\hline
\multirow{3}{*}{\parbox[l]{114pt}{$i\left(\phi^\dagger \partial^\mu \phi - h.c. \right) 
\times$ \\
$\left(\widetilde{B}_{\lambda\mu} Y_H \, H^\dagger D^\lambda H - h.c. \right)$}}
& 
\multirow{2}{*}{$\gamma Z$}
&
\multirow{2}{*}{\parbox{171pt}{$\dfrac{1}{24}\dfrac{\vev^2s^3}{\Lambda^8}\cos^2\theta_W\left(1-\dfrac{4M^2}{s}\right)\left(1+\dfrac{m_Z^2}{s}\right)$ \\  $\times\left(1-\dfrac{m_Z^2}{s}\right)^2$}}
&
\multirow{3}{*}{$\left(\dfrac{228\ \GeV}{\Lambda}\right)^8$}
&
\multirow{2}{*}{$\left[\sigma v\right]_\mathrm{NR}\propto v^2$}
\\
& & & & 
\\
&
$ZZ$
&
$\dfrac{1}{24}\dfrac{\vev^2s^3}{\Lambda^8}\sin^2\theta_W\left(1-\dfrac{4M^2}{s}\right)\left(1-\dfrac{4m_Z^2}{s}\right)^2$
&
&
$\left[\sigma v\right]_\mathrm{NR}\propto v^2$
\\
[2ex]
\hline
\multirow{4}{*}{\parbox[l]{114pt}{$i\left(\phi^\dagger \partial^\mu \phi - h.c. \right) 
\times$ \\
$\left(W^a_{\ \lambda\mu}  \, H^\dagger t^aD^\lambda H - h.c. \right)$}}
& 
\multirow{2}{*}{$\gamma Z$}
&
\multirow{2}{*}{\parbox{171pt}{$\dfrac{1}{24}\dfrac{\vev^2s^3}{\Lambda^8}\sin^2\theta_W\left(1-\dfrac{4M^2}{s}\right)\left(1+\dfrac{m_Z^2}{s}\right)$ \\  $\times\left(1-\dfrac{m_Z^2}{s}\right)^2$}}
&
\multirow{4}{*}{$\left(\dfrac{255\ \GeV}{\Lambda}\right)^8$}
&
\multirow{2}{*}{$\left[\sigma v\right]_\mathrm{NR}\propto v^2$}
\\
 & & & & \\
&
$ZZ$
&
$\dfrac{1}{24}\dfrac{\vev^2s^3}{\Lambda^8}\cos^2\theta_W\left(1-\dfrac{4M^2}{s}\right)\left(1-\dfrac{4m_Z^2}{s}\right)$ 
&
&
$\left[\sigma v\right]_\mathrm{NR}\propto v^2$
\\
&
$W^+W^-$
&
$\dfrac{1}{12}\dfrac{\vev^2s^3}{\Lambda^8}\phantom{\sin^2\theta_W}\left(1-\dfrac{4M^2}{s}\right)\left(1-\dfrac{4m_W^2}{s}\right)$ 
&
&
$\left[\sigma v\right]_\mathrm{NR}\propto v^2$
\\
[2ex]
\hline
\multirow{4}{*}{\parbox[l]{114pt}{$i\left(\phi^\dagger \partial^\mu \phi - h.c. \right)  \times$ \\
$\left(\widetilde{W}^a_{\ \lambda\mu}  \, H^\dagger t^aD^\lambda H - h.c. \right)$}}
& 
\multirow{2}{*}{$\gamma Z$}
&
\multirow{2}{*}{\parbox{171pt}{$\dfrac{1}{24}\dfrac{\vev^2s^3}{\Lambda^8}\sin^2\theta_W\left(1-\dfrac{4M^2}{s}\right)\left(1+\dfrac{m_Z^2}{s}\right)$ \\  $\times\left(1-\dfrac{m_Z^2}{s}\right)^2$}}
&
\multirow{4}{*}{$\left(\dfrac{244\ \GeV}{\Lambda}\right)^8$}
&
\multirow{2}{*}{$\left[\sigma v\right]_\mathrm{NR}\propto v^2$}
\\
 & & & & \\
&
$ZZ$
&
$\dfrac{1}{24}\dfrac{\vev^2s^3}{\Lambda^8}\cos^2\theta_W\left(1-\dfrac{4M^2}{s}\right)\left(1-\dfrac{4m_Z^2}{s}\right)^2$ 
&
&
$\left[\sigma v\right]_\mathrm{NR}\propto v^2$
\\
&
$W^+W^-$
&
$\dfrac{1}{12}\dfrac{\vev^2s^3}{\Lambda^8}\phantom{\sin^2\theta_W}\left(1-\dfrac{4M^2}{s}\right)\left(1-\dfrac{4m_W^2}{s}\right)^2$ 
&
&
$\left[\sigma v\right]_\mathrm{NR}\propto v^2$
\\
[2ex]
\end{tabular}
\end{ruledtabular}
\end{table}

\begin{table}
\caption{\label{resultsixp} Results for dimension-8 vector operators for Fermionic-WIMP annihilation. Indicated values of $\Lambda$ for $\Omega h^2$ and $\left[\sigma v\right]_\mathrm{NR}$ assume $M=155\ \GeV$ necessary to produce a 130 GeV line in the $\gamma h$ final state.}
\begin{ruledtabular}
\begin{tabular}{cclcc}
operator 
& 
final state 
&
\hspace*{48pt}$\Sigma(s,M,m_1,m_2)$ 
&
$\Omega h^2/0.11$ 
&
$\dfrac{\left[\sigma v\right]_\mathrm{NR}}{10^{-27}\cm^3\s^{-1}}$ 
\\ 
[1ex]
\hline
%
\multirow{3}{*}{\parbox[l]{114pt}{$\bar{\chi}\gamma^\mu\chi\ \ \times $ \\
$\left(B_{\lambda\mu} Y_H \, H^\dagger D^\lambda H + h.c. \right)$}}
& 
$\gamma h$
&
$\dfrac{1}{24}\dfrac{\vev^2s^3}{\Lambda^8}\cos^2\theta_W\left(1+\dfrac{2M^2}{s}\right)\left(1-\dfrac{m_h^2}{s}\right)^2$
&
\multirow{3}{*}{$\left(\dfrac{377\ \GeV}{\Lambda}\right)^8$}
&
$\left(\dfrac{541\ \GeV}{\Lambda}\right)^8$
\\
[2ex]
&
\multirow{2}{*}{$Zh$} 
&
\multirow{2}{*}{\parbox[l]{192pt}{\hspace*{-70pt}$\dfrac{1}{24}\dfrac{\vev^2s^3}{\Lambda^8}\sin^2\theta_W \left(1+\dfrac{2M^2}{s}\right) $ 
\\
$\times  \left[\beta_{Zh}^2\left(1+\dfrac{m_Z^2}{2s}\right)+\dfrac{6m_Z^2m_h^2}{s^2}\right]$}}
&
&
\multirow{2}{*}{$0.222[\gamma h]$}
\\
&
&
&
&
\\
[2ex]
\hline
\multirow{2}{*}{\parbox[l]{114pt}{$\bar{\chi}\gamma^\mu\chi\ \ \times$ \\
$\left(\widetilde{B}_{\lambda\mu} Y_H \, H^\dagger D^\lambda H + h.c. \right)$}}
& 
$\gamma h$
&
$\dfrac{1}{24}\dfrac{\vev^2s^3}{\Lambda^8}\cos^2\theta_W\left(1+\dfrac{2M^2}{s}\right)\left(1-\dfrac{m_h^2}{s}\right)^2$
&
\multirow{2}{*}{$\left(\dfrac{375\ \GeV}{\Lambda}\right)^8$}
&
$\left(\dfrac{541\ \GeV}{\Lambda}\right)^8$
\\
&
$Zh$
&
$\dfrac{1}{24}\dfrac{\vev^2s^3}{\Lambda^8}\sin^2\theta_W \left(1+\dfrac{2M^2}{s}\right)\beta_{Zh}^2 $ 
&
&
$0.184[\gamma h]$
\\
[2ex]
\hline
\multirow{5}{*}{\parbox[l]{114pt}{$\bar{\chi}\gamma^{\mu}\chi\ \ \times $ \\
$\left(W^a_{\ \lambda\mu}  \, H^\dagger t^aD^\lambda H + h.c. \right)$}}
& 
$\gamma h$
&
$\dfrac{1}{24}\dfrac{\vev^2s^3}{\Lambda^8}\sin^2\theta_W\left(1+\dfrac{2M^2}{s}\right)\left(1-\dfrac{m_h^2}{s}\right)^2$
&
\multirow{5}{*}{$\left(\dfrac{442\ \GeV}{\Lambda}\right)^8$}
&
$\left(\dfrac{465\ \GeV}{\Lambda}\right)^8$
\\
&
\multirow{2}{*}{$Zh$ }
&
\multirow{2}{*}{\parbox[l]{192pt}{\hspace*{-70pt}$\dfrac{1}{24}\dfrac{\vev^2s^3}{\Lambda^8}\cos^2\theta_W \left(1+\dfrac{2M^2}{s}\right) $ 
\\
$\times  \left[\beta_{Zh}^2\left(1+\dfrac{m_Z^2}{2s}\right)+\dfrac{6m_Z^2m_h^2}{s^2}\right]$}}
&
&
\multirow{2}{*}{$2.493[\gamma h]$}
\\
&
&
&
&
\\
&
\multirow{2}{*}{$W^+W^-$}
&
\multirow{2}{*}{\parbox[l]{176pt}{$\dfrac{1}{12}\dfrac{\vev^2s^3}{\Lambda^8}\phantom{\sin^2\theta_W} \left(1+\dfrac{2M^2}{s}\right)\left(1-\dfrac{4m_W^2}{s}\right) $ 
\\
$\quad\times \left(1+\dfrac{3m_W^2}{s}\right)$}}
&
&
\multirow{2}{*}{$11.12[\gamma h]$}
\\
&
&
&
&
\\
[2ex]
\hline
\multirow{3}{*}{\parbox[l]{114pt}{$\bar{\chi}\gamma^{\mu}\chi\ \ \times $ \\
$ \left(\widetilde{W}^a_{\ \lambda\mu}  \, H^\dagger t^aD^\lambda H + h.c. \right)$}}
& 
$\gamma h$
&
$\dfrac{1}{24}\dfrac{\vev^2s^3}{\Lambda^8}\sin^2\theta_W\left(1+\dfrac{2M^2}{s}\right)\left(1-\dfrac{m_h^2}{s}\right)^2$
&
\multirow{3}{*}{$\left(\dfrac{451\ \GeV}{\Lambda}\right)^8$}
&
$\left(\dfrac{465\ \GeV}{\Lambda}\right)^8$
\\
&
$Zh$
&
$\dfrac{1}{24}\dfrac{\vev^2s^3}{\Lambda^8}\cos^2\theta_W\left(1+\dfrac{2M^2}{s}\right)\beta_{Zh}^2$
&
&
$2.061[\gamma h]$
\\
&
$W^+W^-$ 
&
$\dfrac{1}{12}\dfrac{\vev^2s^3}{\Lambda^8}\phantom{\sin^2\theta_W}\left(1+\dfrac{2M^2}{s}\right)\left(1+\dfrac{2m_W^2}{s}\right)$
&
&
$14.36[\gamma h]$
\\
[2ex]
\end{tabular}
\end{ruledtabular}
$\beta_{Zh}^2=\left[1-(m_h+m_Z)^2/s\right]\left[1-(m_h-m_Z)^2/s\right]$
\end{table}

\begin{table}
\caption{\label{resultsixm} Results for dimension-8 vector operators for Fermionic-WIMP annihilation. Indicated values of $\Lambda$ for $\Omega h^2$ and $\left[\sigma v\right]_\mathrm{NR}$ assume $M=144\ \GeV$ necessary to produce a 130 GeV line in the $\gamma Z$ final state.}
\begin{ruledtabular}
\begin{tabular}{cclcc}
operator 
& 
final state 
&
\hspace*{48pt}$\Sigma(s,M,m_1,m_2)$ 
&
$\Omega h^2/0.11$ 
&
$\dfrac{\left[\sigma v\right]_\mathrm{NR}}{10^{-27}\cm^3\s^{-1}}$ 
\\ 
[1ex]
\hline

\multirow{4}{*}{\parbox[l]{118pt}{$\bar{\chi}\gamma^\mu\chi\ \ \times $ \\
$i\left(B_{\lambda\mu} Y_H \, H^\dagger D^\lambda H - h.c. \right)$}}
& 
\multirow{2}{*}{$\gamma Z$}
&
\multirow{2}{*}{\parbox[l]{171pt}{$\dfrac{1}{24}\dfrac{\vev^2s^3}{\Lambda^8}\cos^2\theta_W\left(1+\dfrac{2M^2}{s}\right)\left(1+\dfrac{m_Z^2}{s}\right)$ \\ $\times \left(1-\dfrac{m_Z^2}{s}\right)^2$}}
&
\multirow{3}{*}{$\left(\dfrac{375\ \GeV}{\Lambda}\right)^8$}
&
\multirow{2}{*}{$\left(\dfrac{542\ \GeV}{\Lambda}\right)^8$}
\\
& & & & \\
&
$ZZ$
&
$\dfrac{1}{24}\dfrac{\vev^2s^3}{\Lambda^8}\sin^2\theta_W\left(1+\dfrac{2M^2}{s}\right)\left(1-\dfrac{4m_Z^2}{s}\right)$
&
&
$0.173[\gamma Z]$
\\
[2ex]
\hline
\multirow{3}{*}{\parbox[l]{118pt}{$\bar{\chi}\gamma^\mu\chi\ \ \times $ \\
$i\left(\widetilde{B}_{\lambda\mu} Y_H \, H^\dagger D^\lambda H - h.c. \right)$}}
& 
\multirow{2}{*}{$\gamma Z$}
&
\multirow{2}{*}{\parbox[l]{171pt}{$\dfrac{1}{24}\dfrac{\vev^2s^3}{\Lambda^8}\cos^2\theta_W\left(1+\dfrac{2M^2}{s}\right)\left(1+\dfrac{m_Z^2}{s}\right)$ \\ $\times \left(1-\dfrac{m_Z^2}{s}\right)^2$}}
&
\multirow{3}{*}{$\left(\dfrac{373\ \GeV}{\Lambda}\right)^8$}
&
\multirow{2}{*}{$\left(\dfrac{542\ \GeV}{\Lambda}\right)^8$}
\\
& & & & \\
&
$ZZ$
&
$\dfrac{1}{24}\dfrac{\vev^2s^3}{\Lambda^8}\sin^2\theta_W\left(1+\dfrac{2M^2}{s}\right)\left(1-\dfrac{4m_Z^2}{s}\right)^2$
&
&
$0.103[\gamma Z]$
\\
[2ex]
\hline
\multirow{4}{*}{\parbox[l]{118pt}{$\bar{\chi}\gamma^\mu\chi\ \ \times $ \\
$i\left(W^a_{\ \lambda\mu}  \, H^\dagger t^aD^\lambda H - h.c. \right)$}}
& 
\multirow{2}{*}{$\gamma Z$}
&
\multirow{2}{*}{\parbox[l]{171pt}{$\dfrac{1}{24}\dfrac{\vev^2s^3}{\Lambda^8}\sin^2\theta_W\left(1+\dfrac{2M^2}{s}\right)\left(1+\dfrac{m_Z^2}{s}\right)$ \\ $\times \left(1-\dfrac{m_Z^2}{s}\right)^2$}}
&
\multirow{4}{*}{$\left(\dfrac{417\ \GeV}{\Lambda}\right)^8$}
&
\multirow{2}{*}{$\left(\dfrac{466\ \GeV}{\Lambda}\right)^8$}
\\
& & & & \\
&
$ZZ$
&
$\dfrac{1}{24}\dfrac{\vev^2s^3}{\Lambda^8}\cos^2\theta_W\left(1+\dfrac{2M^2}{s}\right)\left(1-\dfrac{4m_Z^2}{s}\right)$
&
&
$1.936[\gamma Z]$
\\
&
$W^+W^-$ 
&
$\dfrac{1}{12}\dfrac{\vev^2s^3}{\Lambda^8}\phantom{\sin^2\theta_W}
\left(1+\dfrac{2M^2}{s}\right)\left(1-\dfrac{4m_W^2}{s}\right)$
&
&
$6.196[\gamma Z]$
\\
[2ex]
\hline
\multirow{4}{*}{\parbox[l]{118pt}{$\bar{\chi}\gamma^\mu\chi\ \ \times $  \\
$i\left(\widetilde{W}^a_{\ \lambda\mu}  \, H^\dagger t^aD^\lambda H - h.c. \right)$}}
& 
\multirow{2}{*}{$\gamma Z$}
&
\multirow{2}{*}{\parbox[l]{171pt}{$\dfrac{1}{24}\dfrac{\vev^2s^3}{\Lambda^8}\sin^2\theta_W\left(1+\dfrac{2M^2}{s}\right)\left(1+\dfrac{m_Z^2}{s}\right)$ \\ $\times \left(1-\dfrac{m_Z^2}{s}\right)^2$}}
&
\multirow{4}{*}{$\left(\dfrac{399\ \GeV}{\Lambda}\right)^8$}
&
\multirow{2}{*}{$\left(\dfrac{466\ \GeV}{\Lambda}\right)^8$}
\\
& & & & \\
&
$ZZ$
&
$\dfrac{1}{24}\dfrac{\vev^2s^3}{\Lambda^8}\cos^2\theta_W\left(1+\dfrac{2M^2}{s}\right)\left(1-\dfrac{4m_Z^2}{s}\right)^2$
&
&
$1.160[\gamma Z]$
\\
&
$W^+W^-$ 
&
$\dfrac{1}{12}\dfrac{\vev^2s^3}{\Lambda^8}\phantom{\sin^2\theta_W}\left(1+\dfrac{2M^2}{s}\right)\left(1-\dfrac{4m_W^2}{s}\right)^2$
&
&
$4.264[\gamma Z]$
\\
[2ex]
\end{tabular}
\end{ruledtabular}
\end{table}

\begin{table}
\caption{\label{resultsevenp} Results for dimension-8 axial-vector operators for Fermionic-WIMP annihilation. Indicated values of $\Lambda$ for $\Omega h^2$ and 
$\left[\sigma v\right]_\mathrm{NR}$ assume $M=155\ \GeV$ necessary to produce a 130 GeV line in the $\gamma h$ final state.}
\begin{ruledtabular}
\begin{tabular}{cclcc}
operator 
& 
final state 
&
\hspace*{56pt}$\Sigma(s,M,m_1,m_2)$ 
&
$\Omega h^2/0.11$ 
&
$\dfrac{\left[\sigma v\right]_\mathrm{NR}}{10^{-27}\cm^3\s^{-1}}$ 
\\ 
[1ex]
\hline
%
\multirow{3}{*}{\parbox[l]{114pt}{$\bar{\chi}\gamma^{\mu5}\chi\ \ \times $ \\
$\left(B_{\lambda\mu} Y_H \, H^\dagger D^\lambda H + h.c. \right)$}}
& 
$\gamma h$
&
$\dfrac{1}{24}\dfrac{\vev^2s^3}{\Lambda^8}\cos^2\theta_W\left(1-\dfrac{4M^2}{s}\right)\left(1-\dfrac{m_h^2}{s}\right)^2$
&
\multirow{3}{*}{$\left(\dfrac{238\ \GeV}{\Lambda}\right)^8$}
&
$\left[\sigma v\right]_\mathrm{NR}\propto v^2$
\\
[1ex]
&
\multirow{2}{*}{$Zh$} 
&
\multirow{2}{*}{\parbox{177pt}{$\dfrac{1}{24}\dfrac{\vev^2s^3}{\Lambda^8}\sin^2\theta_W\left\{
\left(1-\dfrac{4M^2}{s}\right)\left[\dfrac{6m_Z^2m_h^2}{s^2}\right.\right. $ 
\\ $\left.\left. + \beta_{Zh}^2\left(1+\dfrac{m_Z^2}{2s}\right)\right] +\beta_{Zh}^2\dfrac{3m_Z^2M^2}{s^2}\right\} $ }}
&
&
\multirow{2}{*}{$\left(\dfrac{296\ \GeV}{\Lambda}\right)^8$}
\\
&
&
&
&
\\
[2ex]
\hline
\multirow{2}{*}{\parbox[l]{114pt}{$\bar{\chi}\gamma^{\mu5}\chi\ \ \times$ \\
$\left(\widetilde{B}_{\lambda\mu} Y_H \, H^\dagger D^\lambda H + h.c. \right)$}}
& 
$\gamma h$
&
$\dfrac{1}{24}\dfrac{\vev^2s^3}{\Lambda^8}\cos^2\theta_W\left(1-\dfrac{4M^2}{s}\right)\left(1-\dfrac{m_h^2}{s}\right)^2$
&
\multirow{2}{*}{$\left(\dfrac{230\ \GeV}{\Lambda}\right)^8$}
&
$\left[\sigma v\right]_\mathrm{NR}\propto v^2$
\\
&
$Zh$
&
$\dfrac{1}{24}\dfrac{\vev^2s^3}{\Lambda^8}\sin^2\theta_W\left(1-\dfrac{4M^2}{s}\right)\beta_{Zh}^2$
&
&
$\left[\sigma v\right]_\mathrm{NR}\propto v^2$
\\
[2ex]
\hline
\multirow{5}{*}{\parbox[l]{114pt}{$\bar{\chi}\gamma^{\mu5}\chi\ \ \times $ \\
$\left(W^a_{\ \lambda\mu}  \, H^\dagger t^aD^\lambda H + h.c. \right)$}}
& 
$\gamma h$
&
$\dfrac{1}{24}\dfrac{\vev^2s^3}{\Lambda^8}\sin^2\theta_W\left(1-\dfrac{4M^2}{s}\right)\left(1-\dfrac{m_h^2}{s}\right)^2$
&
\multirow{5}{*}{$\left(\dfrac{277\ \GeV}{\Lambda}\right)^8$}
&
$\left[\sigma v\right]_\mathrm{NR}\propto v^2$
\\
&
\multirow{2}{*}{$Zh$}
&
\multirow{2}{*}{\parbox{177pt}{$\dfrac{1}{24}\dfrac{\vev^2s^3}{\Lambda^8}\cos^2\theta_W\left\{
\left(1-\dfrac{4M^2}{s}\right)\left[\dfrac{6m_Z^2m_h^2}{s^2}\right.\right. $ 
\\ $\left.\left. + \beta_{Zh}^2\left(1+\dfrac{m_Z^2}{2s}\right)\right] +\beta_{Zh}^2\dfrac{3m_Z^2M^2}{s^2}\right\} $ }}
&
&
\multirow{2}{*}{$\left(\dfrac{326\ \GeV}{\Lambda}\right)^8$}
\\
& & & & \\
&
\multirow{2}{*}{$W^+W^-$}
&
\multirow{2}{*}{\parbox{177pt}{$\dfrac{1}{12}\dfrac{\vev^2s^3}{\Lambda^8}\phantom{\cos^2\theta_W}
\left(1-\dfrac{4M^2}{s}\right)\left(1-\dfrac{4m_W^2}{s}\right) $ 
\\ $\times\left(1+\dfrac{3m_W^2}{s}\right)$ }}
&
&
\multirow{2}{*}{$\left[\sigma v\right]_\mathrm{NR}\propto v^2$}
\\
& & & &
\\
[2ex]
\hline
\multirow{3}{*}{\parbox[l]{114pt}{$\bar{\chi}\gamma^{\mu5}\chi\ \ \times $ \\
$ \left(\widetilde{W}^a_{\ \lambda\mu}  \, H^\dagger t^aD^\lambda H + h.c. \right)$}}
& 
$\gamma h$
&
$\dfrac{1}{24}\dfrac{\vev^2s^3}{\Lambda^8}\sin^2\theta_W\left(1-\dfrac{4M^2}{s}\right)\left(1-\dfrac{m_h^2}{s}\right)^2$
&
\multirow{3}{*}{$\left(\dfrac{276\ \GeV}{\Lambda}\right)^8$}
&
$\left[\sigma v\right]_\mathrm{NR}\propto v^2$
\\
&
$Zh$
&
$\dfrac{1}{24}\dfrac{\vev^2s^3}{\Lambda^8}\cos^2\theta_W\left(1-\dfrac{4M^2}{s}\right)\beta_{Zh}^2$
&
&
$\left[\sigma v\right]_\mathrm{NR}\propto v^2$
\\
&
$W^+W^-$ 
&
$\dfrac{1}{12}\dfrac{\vev^2s^3}{\Lambda^8}\phantom{\cos^2\theta_W}
\left(1-\dfrac{4M^2}{s}\right)\left(1+\dfrac{2m_W^2}{s}\right)$
&
&
$\left[\sigma v\right]_\mathrm{NR}\propto v^2$
\\
[2ex]
%
\end{tabular}
\end{ruledtabular}
$\beta_{Zh}^2=\left[1-(m_h+m_Z)^2/s\right]\left[1-(m_h-m_Z)^2/s\right]$
\end{table}

\begin{table}
\caption{\label{resultsevenm} Results for dimension-8 axial-vector operators for Fermionic-WIMP annihilation. Indicated values of $\Lambda$ for $\Omega h^2$ and 
$\left[\sigma v\right]_\mathrm{NR}$ assume $M=144\ \GeV$ necessary to produce a 130 GeV line in the $\gamma Z$ final state.}
\begin{ruledtabular}
\begin{tabular}{cclcc}
operator 
& 
final state 
&
\hspace*{72pt}$\Sigma(s,M,m_1,m_2)$ 
&
$\Omega h^2/0.11$ 
&
$\dfrac{\left[\sigma v\right]_\mathrm{NR}}{10^{-27}\cm^3\s^{-1}}$ 
\\ 
[1ex]
\hline
\multirow{4}{*}{\parbox[l]{118pt}{$\bar{\chi}\gamma^{\mu5}\chi\ \ \times $ \\
$i\left(B_{\lambda\mu} Y_H \, H^\dagger D^\lambda H - h.c. \right)$}}
& 
\multirow{2}{*}{$\gamma Z$}
&
\multirow{2}{*}{\parbox[l]{184pt}{$\dfrac{1}{24}\dfrac{\vev^2s^3}{\Lambda^8}\cos^2\theta_W\left[\left(1-\dfrac{4M^2}{s}\right)\left(1+\dfrac{m_Z^2}{s}\right)\right.$ \\
$\left. \times \left(1-\dfrac{m_Z^2}{s}\right)^2 + \dfrac{6m_Z^2M^2}{s^2}\left(1-\dfrac{m_Z^2}{s}\right)^2\right]$}}
&
\multirow{4}{*}{$\left(\dfrac{301\ \GeV}{\Lambda}\right)^8$}
&
\multirow{2}{*}{$\left(\dfrac{402\ \GeV}{\Lambda}\right)^8$}
\\
[1ex]
&
&
&
&
\\
[1ex]
&
\multirow{2}{*}{$ZZ$}
&
\multirow{2}{*}{\parbox[l]{184pt}{$\dfrac{1}{24}\dfrac{\vev^2s^3}{\Lambda^8}\sin^2\theta_W\left[\left(1-\dfrac{4M^2}{s}\right)\left(1-\dfrac{4m_Z^2}{s}\right)\right.$ \\
$\left. + \dfrac{18m_Z^2M^2}{s^2}\right]$}}
&
&
\multirow{2}{*}{$0.952\left[\gamma Z \right]$}
\\
&
&
&
&
\\
[2ex]
\hline
\multirow{4}{*}{\parbox[l]{118pt}{$\bar{\chi}\gamma^{\mu5}\chi\ \ \times $ \\
$i\left(\widetilde{B}_{\lambda\mu} Y_H \, H^\dagger D^\lambda H - h.c. \right)$}}
& 
\multirow{2}{*}{$\gamma Z$}
&
\multirow{2}{*}{\parbox[l]{184pt}{$\dfrac{1}{24}\dfrac{\vev^2s^3}{\Lambda^8}\cos^2\theta_W\left[\left(1-\dfrac{4M^2}{s}\right)\left(1+\dfrac{m_Z^2}{s}\right)\right.$ \\
$\left. \times \left(1-\dfrac{m_Z^2}{s}\right)^2 + \dfrac{6m_Z^2M^2}{s^2}\left(1-\dfrac{m_Z^2}{s}\right)^2\right]$}}
&
\multirow{4}{*}{$\left(\dfrac{286\ \GeV}{\Lambda}\right)^8$}
&
\multirow{2}{*}{$\left(\dfrac{402\ \GeV}{\Lambda}\right)^8$}
\\
[1ex]
&
&
&
&
\\
[1ex]
&
\multirow{2}{*}{$ZZ$}
&
\multirow{2}{*}{\parbox[l]{186pt}{$\dfrac{1}{24}\dfrac{\vev^2s^3}{\Lambda^8}\sin^2\theta_W\left[\left(1-\dfrac{4M^2}{s}\right)\left(1-\dfrac{4m_Z^2}{s}\right)^2\right.$ \\
$\left. + \left(1-\dfrac{4m_Z^2}{s}\right)\dfrac{12m_Z^2M^2}{s^2}\right]$}}
&
&
\multirow{2}{*}{$0.228\left[\gamma Z \right]$}
\\
&
&
&
&
\\
[2ex]
\hline
\multirow{6}{*}{\parbox[l]{118pt}{$\bar{\chi}\gamma^{\mu5}\chi\ \ \times $ \\
$i\left(W^a_{\ \lambda\mu}  \, H^\dagger t^aD^\lambda H - h.c. \right)$}}
& 
\multirow{2}{*}{$\gamma Z$}
&
\multirow{2}{*}{\parbox[l]{184pt}{$\dfrac{1}{24}\dfrac{\vev^2s^3}{\Lambda^8}\sin^2\theta_W\left[\left(1-\dfrac{4M^2}{s}\right)\left(1+\dfrac{m_Z^2}{s}\right)\right.$ \\
$\left. \times \left(1-\dfrac{m_Z^2}{s}\right)^2 + \dfrac{6m_Z^2M^2}{s^2}\left(1-\dfrac{m_Z^2}{s}\right)^2\right]$}}
&
\multirow{6}{*}{$\left(\dfrac{368\ \GeV}{\Lambda}\right)^8$}
&
\multirow{2}{*}{$\left(\dfrac{345\ \GeV}{\Lambda}\right)^8$}
\\
& & & & \\
&
\multirow{2}{*}{$ZZ$}
&
\multirow{2}{*}{\parbox[l]{184pt}{$\dfrac{1}{24}\dfrac{\vev^2s^3}{\Lambda^8}\cos^2\theta_W\left[\left(1-\dfrac{4M^2}{s}\right)\left(1-\dfrac{4m_Z^2}{s}\right)\right.$ \\
$\left. + \dfrac{18m_Z^2M^2}{s^2}\right]$}}
&
&
\multirow{2}{*}{$10.67\left[\gamma Z \right]$}
\\
& & & & \\
&
\multirow{2}{*}{$W^+W^-$}
&
\multirow{2}{*}{\parbox[l]{184pt}{$\dfrac{1}{12}\dfrac{\vev^2s^3}{\Lambda^8}\phantom{\sin^2\theta_W}\left[\left(1-\dfrac{4M^2}{s}\right)\left(1-\dfrac{4m_W^2}{s}\right)\right.$ \\
$\left. + \dfrac{18m_W^2M^2}{s^2}\right]$}}
&
&
\multirow{2}{*}{$23.09\left[\gamma Z \right]$}
\\
& & & &
\\
[2ex]
\hline
\multirow{6}{*}{\parbox[l]{118pt}{$\bar{\chi}\gamma^{\mu5}\chi\ \ \times $  \\
$i\left(\widetilde{W}^a_{\ \lambda\mu}  \, H^\dagger t^aD^\lambda H - h.c. \right)$}}
& 
\multirow{2}{*}{$\gamma Z$}
&
\multirow{2}{*}{\parbox[l]{184pt}{$\dfrac{1}{24}\dfrac{\vev^2s^3}{\Lambda^8}\sin^2\theta_W\left[\left(1-\dfrac{4M^2}{s}\right)\left(1+\dfrac{m_Z^2}{s}\right)\right.$ \\
$\left. \times \left(1-\dfrac{m_Z^2}{s}\right)^2 + \dfrac{6m_Z^2M^2}{s^2}\left(1-\dfrac{m_Z^2}{s}\right)^2\right]$}}
&
\multirow{6}{*}{$\left(\dfrac{335\ \GeV}{\Lambda}\right)^8$}
&
\multirow{2}{*}{$\left(\dfrac{345\ \GeV}{\Lambda}\right)^8$}
\\
& & & & \\
&
\multirow{2}{*}{$ZZ$}
&
\multirow{2}{*}{\parbox[l]{187pt}{$\dfrac{1}{24}\dfrac{\vev^2s^3}{\Lambda^8}\cos^2\theta_W\left[\left(1-\dfrac{4M^2}{s}\right)\left(1-\dfrac{4m_Z^2}{s}\right)^2\right.$ \\
$\left. + \left(1-\dfrac{4m_Z^2}{s}\right)\dfrac{12m_Z^2M^2}{s^2}\right]$}}
&
&
\multirow{2}{*}{$2.552\left[\gamma Z \right]$}
\\
& & & & \\
&
$W^+W^-$ 
&
\multirow{2}{*}{\parbox[l]{189pt}{$\dfrac{1}{12}\dfrac{\vev^2s^3}{\Lambda^8}\phantom{\cos^2\theta_W}\left[\left(1-\dfrac{4M^2}{s}\right)\left(1-\dfrac{4m_W^2}{s}\right)^2\right.$ \\
$\left. + \left(1-\dfrac{4m_W^2}{s}\right)\dfrac{12m_W^2M^2}{s^2}\right]$}}
&
&
\multirow{2}{*}{$10.60\left[\gamma Z \right]$}
\\
& & & & 
\\
[2ex]
%
\end{tabular}
\end{ruledtabular}
\end{table}

\begin{table}
\caption{\label{resulteight} Results for dimension-7 tensor operators for fermionic-WIMP annihilation.  Indicated values of $\Lambda$ for $\Omega h^2$ and $\left[\sigma v\right]_\mathrm{NR}$ assume $M=155\ \GeV$ necessary to produce a 130 GeV line in the $\gamma h$ final state.}
\begin{ruledtabular}
\begin{tabular}{cclcc}
operator 
& 
final state 
&
\hspace*{48pt}$\Sigma(s,M,m_1,m_2)$ 
&
$\Omega h^2/0.11$ 
&
$\dfrac{\left[\sigma v\right]_\mathrm{NR}}{10^{-27}\cm^3\s^{-1}}$
\\ 
[1ex]
\hline
%
\multirow{8}{*}{$\bar{\chi}\gamma^{\mu\nu}\chi \: B_{\mu\nu} Y_HH^\dagger H$}
&
$\gamma h$
&
\parbox{197pt}{\hspace*{-24pt}$\dfrac{1}{6}\dfrac{\vev^2s^2}{\Lambda^6}\cos^2\theta_W 
\left(1-\dfrac{m_h^2}{s}\right)^2\left(1+\dfrac{2M^2}{s}\right)$}
&
\multirow{8}{*}{$\left(\dfrac{579\ \GeV}{\Lambda}\right)^6$}
&
{$\left(\dfrac{820\ \GeV}{\Lambda}\right)^6$}
\\
[2ex]
&
\multirow{3}{*}{$Zh$}
&
\multirow{3}{*}{\parbox{197pt}{\hspace*{-52pt}$\dfrac{1}{6}\dfrac{\vev^2s^2}{\Lambda^6}\sin^2\theta_W \left[ \beta_{Zh}^2\left(1+\dfrac{2M^2}{s}\right)  \right.$ \\  
$+\dfrac{m_Z^2}{4s}\left(1+\dfrac{8M^2}{s}\right)\left(1-\dfrac{m_Z^2}{s}\right)^{-2}
\left(25-\dfrac{14m_Z^2}{s}   \right.$ \\ $\left. \left.
-\dfrac{14m_h^2}{s}+\dfrac{10m_Z^2m_h^2}{s^2}+\dfrac{m_h^4}{s^2}+\dfrac{m_Z^4}{s^2}
\right)\right]$} }
&
&
\multirow{3}{*}{$0.591\left[\gamma h\right]$}
\\
& & & \\
& & & \\
[1.5ex]
&
\multirow{2}{*}{$W^+W^-$}
&
\multirow{2}{*}{\parbox{197pt}{$\dfrac{1}{24}\dfrac{\vev^2m_Z^2s}{\Lambda^6}\sin^2\theta_W\left(1+\dfrac{8M^2}{s}\right)\left(1-\dfrac{m_Z^2}{s}\right)^{-2}$ \\
$\times\left(1-\dfrac{4m_W^2}{s}\right)\left(1+\dfrac{23m_W^2}{s}+\dfrac{12m_W^4}{s^2}\right)$ }}
&
&
\multirow{2}{*}{$0.043\left[\gamma h\right]$}
\\
& & & \\
[2ex]
&
\multirow{2}{*}{$\sum_f f\bar{f}$}
&
\multirow{2}{*}{\parbox{211pt}{\hspace*{-74pt}$\dfrac{1}{24}\dfrac{\vev^2m_Z^2s}{\Lambda^6}\sin^2\theta_W\left(1+\dfrac{8M^2}{s}\right)$ \\
$\times\left(1-\dfrac{m_Z^2}{s}\right)^{-2}\sum_f \left[4A_{fB}^2+1+\dfrac{4m_f^2}{s}\left(2A_{fB}^2-1\right)\right]$ }}
&
&
\multirow{2}{*}{$1.055\left[\gamma h\right]$}
\\
& & & \\
[2ex]
\hline
%
\multirow{8}{*}{$\bar{\chi}\gamma^{\mu\nu}\chi \: \widetilde{B}_{\mu\nu} 
                                                                Y_HH^\dagger H$}
&
$\gamma h$
&
\parbox{197pt}{\hspace*{-24pt}$\dfrac{1}{6}\dfrac{\vev^2s^2}{\Lambda^6}\cos^2\theta_W 
\left(1-\dfrac{m_h^2}{s}\right)^2\left(1+\dfrac{2M^2}{s}\right)$}
&
\multirow{8}{*}{$\left(\dfrac{506\ \GeV}{\Lambda}\right)^6$}
&
{$\left(\dfrac{820\ \GeV}{\Lambda}\right)^6$}
\\
[2ex]
&
\multirow{3}{*}{$Zh$}
&
\multirow{3}{*}{\parbox{197pt}{\hspace*{-52pt}$\dfrac{1}{6}\dfrac{\vev^2s^2}{\Lambda^6}\sin^2\theta_W \left[ \beta_{Zh}^2\left(1+\dfrac{2M^2}{s}\right)  \right.$ \\  
$+\dfrac{m_Z^2}{4s}\left(1-\dfrac{4M^2}{s}\right)\left(1-\dfrac{m_Z^2}{s}\right)^{-2}
\left(25-\dfrac{14m_Z^2}{s}   \right.$ \\ $\left. \left.
-\dfrac{14m_h^2}{s}+\dfrac{10m_Z^2m_h^2}{s^2}+\dfrac{m_h^4}{s^2}+\dfrac{m_Z^4}{s^2}
\right)\right]$} }
&
&
\multirow{3}{*}{$0.184\left[\gamma h\right]$}
\\
& & & \\
& & & \\
[1.5ex]
&
\multirow{2}{*}{$W^+W^-$}
&
\multirow{2}{*}{\parbox{197pt}{$\dfrac{1}{24}\dfrac{\vev^2m_Z^2s}{\Lambda^6}\sin^2\theta_W\left(1-\dfrac{4M^2}{s}\right)\left(1-\dfrac{m_Z^2}{s}\right)^{-2}$ \\
$\times\left(1-\dfrac{4m_W^2}{s}\right)\left(1+\dfrac{23m_W^2}{s}+\dfrac{12m_W^4}{s^2}\right)$ }}
&
&
\multirow{2}{*}{$\left[\sigma v\right]_\mathrm{NR}\propto v^2$}
\\
& & & \\
[2ex]
&
\multirow{2}{*}{$\sum_f f\bar{f}$}
&
\multirow{2}{*}{\parbox{211pt}{$\hspace*{-74pt}\dfrac{1}{24}\dfrac{\vev^2m_Z^2s}{\Lambda^6}\sin^2\theta_W\left(1-\dfrac{4M^2}{s}\right)$ \\
$\times\left(1-\dfrac{m_Z^2}{s}\right)^{-2}\sum_f\left[4A_{fB}^2+1+\dfrac{4m_f^2}{s}\left(2A_{fB}^2-1\right)\right]$ }}
&
&
\multirow{2}{*}{$\left[\sigma v\right]_\mathrm{NR}\propto v^2$}
\\
& & & \\
[2ex]
\end{tabular}
\end{ruledtabular}
$\beta_{Zh}^2=\left[1-(m_h+m_Z)^2/s\right]\left[1-(m_h-m_Z)^2/s\right]$\\
$A_{fB}=2Q_f\left(1-m_W^2/s\right)\mp1/2$, with $-$ ($+$) for neutrinos and up-type quarks (electrons and down-type quarks).
\end{table}

\begin{table}
\caption{\label{resultnine} Results for dimension-7 tensor operators for fermionic-WIMP annihilation (continued).  Indicated values of $\Lambda$ for $\Omega h^2$ and $\left[\sigma v\right]_\mathrm{NR}$ assume $M=155\ \GeV$ necessary to produce a 130 GeV line in the $\gamma h$ final state.}
\begin{ruledtabular}
\begin{tabular}{cclcc}
operator 
& 
final state 
&
\hspace*{48pt}$\Sigma(s,M,m_1,m_2)$ 
&
$\Omega h^2/0.11$ 
&
$\dfrac{\left[\sigma v\right]_\mathrm{NR}}{10^{-27}\cm^3\s^{-1}}$
\\ 
[1ex]
\hline
%
\multirow{10}{*}{$\bar{\chi}\gamma^{\mu\nu}\chi \: W^a_{\ \mu\nu}  H^\dagger t^a H$}
&
$\gamma h$
&
\parbox{197pt}{\hspace*{-24pt}$\dfrac{1}{6}\dfrac{\vev^2s^2}{\Lambda^6}\sin^2\theta_W 
\left(1-\dfrac{m_h^2}{s}\right)^2\left(1+\dfrac{2M^2}{s}\right)$}
&
\multirow{10}{*}{$\left(\dfrac{766\ \GeV}{\Lambda}\right)^6$}
&
{$\left(\dfrac{671\ \GeV}{\Lambda}\right)^6$}
\\
[2ex]
&
\multirow{3}{*}{$Zh$}
&
\multirow{3}{*}{\parbox{197pt}{\hspace*{-52pt}$\dfrac{1}{6}\dfrac{\vev^2s^2}{\Lambda^6}\cos^2\theta_W \left[ \beta_{Zh}^2\left(1+\dfrac{2M^2}{s}\right)  \right.$ \\  
$+\dfrac{m_Z^2}{4s}\left(1+\dfrac{8M^2}{s}\right)\left(1-\dfrac{m_Z^2}{s}\right)^{-2}
\left(25-\dfrac{14m_Z^2}{s}   \right.$ \\ $\left. \left.
-\dfrac{14m_h^2}{s}+\dfrac{10m_Z^2m_h^2}{s^2}+\dfrac{m_h^4}{s^2}+\dfrac{m_Z^4}{s^2}
\right)\right]$} }
&
&
\multirow{3}{*}{$6.629\left[\gamma h\right]$}
\\
& & & \\
& & & \\
[1.5ex]
&
\multirow{4}{*}{$W^+W^-$}
&
\multirow{4}{*}{\parbox{208pt}{\hspace*{-12pt}$\dfrac{\vev^2s^2}{\Lambda^6}\left\{\dfrac{1}{3}\left(1+\dfrac{2M^2}{s}\right)\left(1-\dfrac{m_W^2}{s}\right)-\dfrac{6m_W^2M^2}{s^2}\right.$\\
$+\dfrac{1}{8}\left(1-\dfrac{4m_W^2}{s}\right)\left(1+\dfrac{8M^2}{s}\right)\left[1+\dfrac{m_W^2}{3s}\left(12\phantom{\dfrac{m_W^2}{s}} \right.\right.$ \\
$ +2\left(1-\dfrac{m_Z^2}{s}\right)^{-1}\left(13+12\dfrac{m_W^2}{s}\right)+\left(1-\dfrac{m_Z^2}{s}\right)^{-2}$\\
$\left.\left.\left.\times\left(1+\dfrac{23m_W^2}{s}+\dfrac{12m_W^4}{s^2}\right)\right) \right]\right\}$
}}
&
&
\multirow{4}{*}{$23.25\left[\gamma h\right]$}
\\
& & & \\
& & & \\
& & & \\
[2ex]
&
\multirow{2}{*}{$\sum_f f\bar{f}$}
&
\multirow{2}{*}{\parbox{215pt}{$\hspace*{-103pt}\dfrac{1}{24}\dfrac{\vev^2m_W^2s}{\Lambda^6}\left(1+\dfrac{8M^2}{s}\right)$ \\
$\times\left(1-\dfrac{m_Z^2}{s}\right)^{-2}\sum_f\left[4A_{fW}^2+1+\dfrac{4m_f^2}{s}\left(2A_{fW}^2-1\right)\right]$ }}
&
&
\multirow{2}{*}{$17.31\left[\gamma h\right]$}
\\
& & & \\
[2ex]
\hline
%
\multirow{10}{*}{$\bar{\chi}\gamma^{\mu\nu}\chi \: \widetilde{W}^a_{\ \mu\nu}  H^\dagger t^a H$}
&
$\gamma h$
&
\parbox{197pt}{\hspace*{-24pt}$\dfrac{1}{6}\dfrac{\vev^2s^2}{\Lambda^6}\sin^2\theta_W 
\left(1-\dfrac{m_h^2}{s}\right)^2\left(1+\dfrac{2M^2}{s}\right)$}
&
\multirow{10}{*}{$\left(\dfrac{638\ \GeV}{\Lambda}\right)^6$}
&
{$\left(\dfrac{671\ \GeV}{\Lambda}\right)^6$}
\\
[2ex]
&
\multirow{3}{*}{$Zh$}
&
\multirow{3}{*}{\parbox{197pt}{\hspace*{-52pt}$\dfrac{1}{6}\dfrac{\vev^2s^2}{\Lambda^6}\cos^2\theta_W \left[ \beta_{Zh}^2\left(1+\dfrac{2M^2}{s}\right)  \right.$ \\  
$+\dfrac{m_Z^2}{4s}\left(1-\dfrac{4M^2}{s}\right)\left(1-\dfrac{m_Z^2}{s}\right)^{-2}
\left(25-\dfrac{14m_Z^2}{s}   \right.$ \\ $\left. \left.
-\dfrac{14m_h^2}{s}+\dfrac{10m_Z^2m_h^2}{s^2}+\dfrac{m_h^4}{s^2}+\dfrac{m_Z^4}{s^2}
\right)\right]$} }
&
&
\multirow{3}{*}{$2.061\left[\gamma h\right]$}
\\
& & & \\
& & & \\
[1.5ex]
&
\multirow{4}{*}{$W^+W^-$}
&
\multirow{4}{*}{\parbox{208pt}{\hspace*{-12pt}$\dfrac{\vev^2s^2}{\Lambda^6}\left\{\dfrac{1}{3}\left(1+\dfrac{2M^2}{s}\right)\left(1-\dfrac{m_W^2}{s}\right)+\dfrac{6m_W^2M^2}{s^2}\right.$\\
$+\dfrac{1}{8}\left(1-\dfrac{4m_W^2}{s}\right)\left(1-\dfrac{4M^2}{s}\right)\left[1+\dfrac{m_W^2}{3s}\left(12\phantom{\dfrac{m_W^2}{s}} \right.\right.$ \\
$ +2\left(1-\dfrac{m_Z^2}{s}\right)^{-1}\left(13+12\dfrac{m_W^2}{s}\right)+\left(1-\dfrac{m_Z^2}{s}\right)^{-2}$\\
$\left.\left.\left. \times\left(1+\dfrac{23m_W^2}{s}+\dfrac{12m_W^4}{s^2}\right)\right) \right]\right\}$
}}
&
&
\multirow{4}{*}{$12.66\left[\gamma h\right]$}
\\
& & & \\
& & & \\
& & & \\
[2ex]
&
\multirow{2}{*}{$\sum_f f\bar{f}$}
&
\multirow{2}{*}{\parbox{215pt}{$\hspace*{-103pt}\dfrac{1}{24}\dfrac{\vev^2m_W^2s}{\Lambda^6}\left(1-\dfrac{4M^2}{s}\right)$ \\
$\times\left(1-\dfrac{m_Z^2}{s}\right)^{-2}\sum_f\left[4A_{fW}^2+1+\dfrac{4m_f^2}{s}\left(2A_{fW}^2-1\right)\right]$ }}
&
&
\multirow{2}{*}{$\left[\sigma v\right]_\mathrm{NR}\propto v^2$}
\\
& & & \\
[2ex]
\end{tabular}
\end{ruledtabular}
$\beta_{Zh}^2=\left[1-(m_h+m_Z)^2/s\right]\left[1-(m_h-m_Z)^2/s\right]$ \\
$A_{fW}=2Q_f\sin^2\theta_Wm_Z^2/s\mp1/2$, with $-$ ($+$) for neutrinos and up-type quarks (electrons and down-type quarks).
\end{table}

\renewcommand*\arraystretch{2.5}
\begin{table}
\caption{\label{resultten} Results for dimension-5 tensor operators for fermionic-WIMP annihilation.  Indicated values of $\Lambda$ for $\Omega h^2$ and $\left[\sigma v\right]_\mathrm{NR}$ assume $M=155\ \GeV$.}
\begin{ruledtabular}
\begin{tabular}{cclcc}
operator 
& 
final state 
&
\hspace*{48pt}$\Sigma(s,M,m_1,m_2)$ 
&
$\Omega h^2/0.11$ 
&
$\dfrac{\left[\sigma v\right]_\mathrm{NR}}{10^{-27}\cm^3\s^{-1}}$
\\ 
[1ex]
\hline
%
\multirow{5}{*}{$\bar{\chi}\gamma^{\mu\nu}\chi \: B_{\mu\nu}$}
&
$Zh$
&
$\dfrac{2}{3}\dfrac{m_Z^2s}{\vev^2\Lambda^2}\sin^2\theta_W \left(1+\dfrac{8M^2}{s}\right)\left( \beta_{Zh}^2+\dfrac{12M^2}{s}\right)\left(1-\dfrac{m_Z^2}{s}\right)^{-2}$ 
&
\multirow{5}{*}{$\left(\dfrac{8418\ \GeV}{\Lambda}\right)^2$}
&
$\left(\dfrac{9635\ \GeV}{\Lambda}\right)^2$
\\
[1.5ex]
&
\multirow{2}{*}{$W^+W^-$}
&
\multirow{2}{*}{\parbox{197pt}{\hspace*{-18pt}$\dfrac{2}{3}\dfrac{m_Z^2s}{\vev^2\Lambda^2}\sin^2\theta_W\left(1+\dfrac{8M^2}{s}\right)\left(1-\dfrac{m_Z^2}{s}\right)^{-2}$ \\
$\times\left(1-\dfrac{4m_W^2}{s}\right)\left(1+\dfrac{23m_W^2}{s}+\dfrac{12m_W^4}{s^2}\right)$ }}
&
&
\multirow{2}{*}{$0.651\left[Zh\right]$}
\\
& & & \\
[2ex]
&
\multirow{2}{*}{$\sum_f f\bar{f}$}
&
\multirow{2}{*}{\parbox{211pt}{\hspace*{-92pt}$\dfrac{2}{3}\dfrac{m_Z^2s}{\vev^2\Lambda^2}\sin^2\theta_W\left(1+\dfrac{8M^2}{s}\right)$ \\
$\times\left(1-\dfrac{m_Z^2}{s}\right)^{-2}\sum_f \left[4A_{fB}^2+1+\dfrac{4m_f^2}{s}\left(2A_{fB}^2-1\right)\right]$ }}
&
&
\multirow{2}{*}{$16.01\left[Zh\right]$}
\\
& & & \\
[2ex]
\hline
%
%
\multirow{5}{*}{$\bar{\chi}\gamma^{\mu\nu}\chi \: \widetilde{B}_{\mu\nu}$}
&
$Zh$
&
$\dfrac{2}{3}\dfrac{m_Z^2s}{\vev^2\Lambda^2}\sin^2\theta_W \left(1-\dfrac{4M^2}{s}\right)\left( \beta_{Zh}^2+\dfrac{12M^2}{s}\right)\left(1-\dfrac{m_Z^2}{s}\right)^{-2}$ 
&
\multirow{5}{*}{$\left(\dfrac{837\ \GeV}{\Lambda}\right)^2$}
&
$\left[\sigma v\right]_\mathrm{NR}\propto v^2$
\\
[1.5ex]
&
\multirow{2}{*}{$W^+W^-$}
&
\multirow{2}{*}{\parbox{197pt}{\hspace*{-18pt}$\dfrac{2}{3}\dfrac{m_Z^2s}{\vev^2\Lambda^2}\sin^2\theta_W\left(1-\dfrac{4M^2}{s}\right)\left(1-\dfrac{m_Z^2}{s}\right)^{-2}$ \\
$\times\left(1-\dfrac{4m_W^2}{s}\right)\left(1+\dfrac{23m_W^2}{s}+\dfrac{12m_W^4}{s^2}\right)$ }}
&
&
\multirow{2}{*}{$\left[\sigma v\right]_\mathrm{NR}\propto v^2$}
\\
& & & \\
[2ex]
&
\multirow{2}{*}{$\sum_f f\bar{f}$}
&
\multirow{2}{*}{\parbox{211pt}{$\hspace*{-92pt}\dfrac{2}{3}\dfrac{m_Z^2s}{\vev^2\Lambda^2}\sin^2\theta_W\left(1-\dfrac{4M^2}{s}\right)$ \\
$\times\left(1-\dfrac{m_Z^2}{s}\right)^{-2}\sum_f\left[4A_{fB}^2+1+\dfrac{4m_f^2}{s}\left(2A_{fB}^2-1\right)\right]$ }}
&
&
\multirow{2}{*}{$\left[\sigma v\right]_\mathrm{NR}\propto v^2$}
\\
& & & \\
[2ex]
\end{tabular}
\end{ruledtabular}
$\beta_{Zh}^2=\left[1-(m_h+m_Z)^2/s\right]\left[1-(m_h-m_Z)^2/s\right]$\\
$A_{fB}=2Q_f\left(1-m_W^2/s\right)\mp1/2$, with $-$ ($+$) for neutrinos and up-type quarks (electrons and down-type quarks).
\end{table}

\section{Lines \label{lines}}

In this section we discuss the indirect detection signal without the assumption of  $M=130$ GeV.  We first note that there are two simple limits, $M< M_Z/2$, and $M\gg m_h$.  In the first case the total annihilation cross section is only into the $\gamma \gamma$ final state.  In the second case, the total annihilation cross section is the sum of cross sections into all possible final states, $\gamma\gamma + \cdots$. At the same time, in the limit $M\gg m_h$, we can ignore the masses of final-state particles. Therefore, the relative signal strengths of different annihilation channels are independent of $M$ in this limit. 

In both limiting cases and in the non-relativistic limit $v \to 0$, 
\begin{eqnarray}
\left[\sigma_\mathrm{TOTAL}v\right]_\mathrm{NR} & = & \frac{1}{32\pi M^2} \Sigma_\mathrm{TOTAL}(s=4M^2;M;0;0) \nonumber \\
\left[\sigma_{\gamma i}v\right]_\mathrm{NR}     & = & \frac{1}{32\pi M^2} \Sigma_{\gamma i}(s=4M^2;M;0;0)
\nonumber \\
& = &  \left[\sigma v\right]_{\rm TOTAL} \frac{\Sigma_{\gamma i}(s=4M^2;M;0;0)}{\Sigma_\mathrm{TOTAL}(s=4M^2;M;0;0)}.
\end{eqnarray}
We have assumed that the annihilation is pure $s$ wave.  If it is pure $p$ wave then the lines from clusters will be very weak because of the small value of $v^2$.  To have $\Omega h^2=0.11$ requires $\left[\sigma v\right]_{\rm TOTAL} =1.88\times10^{-9}\, \GeV^{-2}$, and using $10^{-27}\cm^3\s^{-1}= 8.57\times10^{-11}\GeV^{-2}$, then
\begin{equation}
\label{rats}
\frac{\left[\sigma_{\gamma i} v\right]_\mathrm{NR}} {10^{-27}\,\cm^3\s^{-1}} = 22 \
\frac{\Sigma_{\gamma i}(s=4M^2;M;0;0)}    {\Sigma_{\rm TOTAL}(s=4M^2;M;0;0)} ,
\end{equation}
where $i=\gamma,\ Z,\ \mathrm{or}\ h$.

If $M< M_Z/2$, then the only available di-boson annihilation channel is into $\gamma\gamma$.  Another possible annihilation channel in this case is into $f \bar{f}$, where $f$ includes all SM fermions except the top quark. Note that at tree level and to the leading order in $1/\Lambda$, the operators under our consideration can mediate annihilation either into $\gamma \gamma$ or $f \bar f$, but not both. Therefore, if the $\gamma \gamma$ final state is available, the the ratio of the $\Sigma$'s in Eq.\ (\ref{rats}) is unity and $\left[\sigma_{\gamma \gamma}v\right]_\mathrm{NR}/10^{-27}\,\cm^3\s^{-1} = 22$ is a robust prediction. 

For intermediate masses, the relative signal strengths of photon lines would depend on the dark-matter mass $M$. In Tables \ref{linestableST} and \ref{linestableV}, we collect results for the two limiting cases,  as well as an example of an intermediate dark-matter mass (we choose $M=130$ GeV). We see that the number of photon lines and their relative strengths provide sensitive diagnostic handles on the dynamics of dark-matter annihilation since different operators correspond to different numbers of photon lines and different values of $\left[\sigma_{\gamma\gamma} v\right]_\mathrm{NR}/10^{-27}\cm^3\s^{-1}$.

Since there is an uncertainty of the WIMP density at the galactic center (and the rate is proportional to the square of the density!), one might argue that any operator with $s$-wave annihilation might give a detectable signal.  

We stress that in our case where the DM couples to electroweak gauge and Higgs bosons, the gamma-ray lines and the gamma-ray continuum play a similar role to the role that direct detection plays in the case that DM couples to quarks.

\renewcommand*\arraystretch{1.5}
\begin{table}
\caption{\label{linestableST} The cross section for $\gamma$-ray producing processes assuming $\Lambda$ is the value necessary for scalar/pseudoscalar and tensor WIMP operators to result in $\Omega h^2=0.11$ in the dark-matter species.  For $M<m_Z/2$, for all operators that have nonvanishing branching to two photons, $\left[\sigma _{\gamma \gamma}v\right]_\mathrm{NR}/10^{-27}\,\cm^3\s^{-1} = 22$, and that is the only $\gamma$-ray producing process.}
\begin{ruledtabular}
\begin{tabular}{c|ccc|cc|cc}
\multirow{2}{*}{operator} 
& 
\multicolumn{3}{c|}{$\dfrac{\left[\sigma_{\gamma\gamma} v\right]_\mathrm{NR}}{10^{-27}\cm^3\s^{-1}}$}
&
\multicolumn{2}{c|}{$\dfrac{\left[\sigma_{\gamma Z} v\right]_\mathrm{NR}}{10^{-27}\cm^3\s^{-1}}$}
&
\multicolumn{2}{c}{$\dfrac{\left[\sigma_{\gamma h} v\right]_\mathrm{NR}}{10^{-27}\cm^3\s^{-1}}$}
\\ 
& 
$M \leq m_Z/2$
&
$M=130\ \GeV$
&
$M\gg m_h$
&
$M=130\ \GeV$
&
$M\gg m_h$
&
$M=130\ \GeV$
&
$M\gg m_h$
\\
\hline
$\Lambda^{-2}\phi^\dagger \phi \: B_{\mu\nu} B^{\mu\nu}$
&
\multirow{4}{*}{22}
&
\multirow{4}{*}{15}
&
\multirow{4}{*}{13}
&
\multirow{4}{*}{6}
&
\multirow{4}{*}{8}
&
\multirow{4}{*}{0}
&
\multirow{4}{*}{0}
\\
$\Lambda^{-2}\phi^\dagger \phi \: B_{\mu\nu} \widetilde{B}^{\mu\nu}$
& & & & & & 
\\
$\Lambda^{-3}\bar{\chi}i\gamma^5\chi \: B_{\mu\nu} B^{\mu\nu}$
& & & & & &
\\
$\Lambda^{-3}\bar{\chi}i\gamma^5\chi \: B_{\mu\nu} 
                                            \widetilde{B}^{\mu\nu}$
& & & & & &
\\
\hline
$\Lambda^{-2}\phi^\dagger \phi \: W^a_{\ \mu\nu} 
                                  W^{a\,\mu\nu}$ 
& 
\multirow{4}{*}{22}
& 
\multirow{4}{*}{0.7-0.8}
& 
\multirow{4}{*}{0.4}
& 
\multirow{4}{*}{3-4}
& 
\multirow{4}{*}{3}
& 
\multirow{4}{*}{0}
&
\multirow{4}{*}{0}
\\
$\Lambda^{-2}\phi^\dagger \phi \: W^a_{\ \mu\nu} 
                       \widetilde{W}^{a\,\mu\nu}$
& & & & & & 
\\
$\Lambda^{-3}\bar{\chi}i\gamma^5\chi \: W^a_{\ \mu\nu} 
                                                      W^{a\,\mu\nu}$
& & & & & & 
\\
$\Lambda^{-3}\bar{\chi}i\gamma^5\chi \: W^a_{\ \mu\nu} 
                \widetilde{W}^{a\,\mu\nu}$
& & & & & & 
\\
\hline
$\Lambda^{-3}\bar{\chi}\gamma^{\mu\nu}\chi \: B_{\mu\nu} Y_HH^\dagger H$
&
\multirow{2}{*}{0}
&
\multirow{2}{*}{0}
&
\multirow{2}{*}{0}
&
\multirow{2}{*}{0}
&
\multirow{2}{*}{0}
&
8
&
\multirow{2}{*}{17}
\\
$\Lambda^{-3}\bar{\chi}\gamma^{\mu\nu}\chi \: \widetilde{B}_{\mu\nu} Y_HH^\dagger H$
&
&
&
&
&
&
18
&
\\
\hline
$\Lambda^{-3}\bar{\chi}\gamma^{\mu\nu}\chi \: W^a_{\ \mu\nu}  H^\dagger t^a H$
&
\multirow{2}{*}{0}
&
\multirow{2}{*}{0}
&
\multirow{2}{*}{0}
&
\multirow{2}{*}{0}
&
\multirow{2}{*}{0}
&
\multirow{2}{*}{0.5-1}
&
\multirow{2}{*}{1-2}
\\
$\Lambda^{-3}\bar{\chi}\gamma^{\mu\nu}\chi \: \widetilde{W}^a_{\ \mu\nu}  H^\dagger t^a H$
&
&
&
&
&
&
&
\end{tabular}
\end{ruledtabular}
\end{table}
\begin{table}
\caption{\label{linestableV} The cross section for $\gamma$-ray producing processes assuming $\Lambda$ is the value necessary for vector WIMP operators to result in $\Omega h^2=0.11$ in the dark-matter species.  For $M<m_Z/2$, for all operators that have nonvanishing branching to two photons, $\left[\sigma _{\gamma \gamma}v\right]_\mathrm{NR}/10^{-27}\,\cm^3\s^{-1} = 22$, and that is the only $\gamma$-ray producing process.}
\begin{ruledtabular}
\begin{tabular}{c|ccc|cc|cc}
\multirow{2}{*}{operator} 
& 
\multicolumn{3}{c|}{$\dfrac{\left[\sigma_{\gamma\gamma} v\right]_\mathrm{NR}}{10^{-27}\cm^3\s^{-1}}$}
&
\multicolumn{2}{c|}{$\dfrac{\left[\sigma_{\gamma Z} v\right]_\mathrm{NR}}{10^{-27}\cm^3\s^{-1}}$}
&
\multicolumn{2}{c}{$\dfrac{\left[\sigma_{\gamma h} v\right]_\mathrm{NR}}{10^{-27}\cm^3\s^{-1}}$}
\\ 
& 
$M \leq m_Z/2$
&
$M=130\ \GeV$
&
$M\gg m_h$
&
$M=130\ \GeV$
&
$M\gg m_h$
&
$M=130\ \GeV$
&
$M\gg m_h$
\\
\hline
\multirow{2}{*}{\parbox[l]{118pt}{$\Lambda^{-4}\left(\phi^\dagger \partial^\mu \phi + h.c. \right) \times$ \\
$i\left(B_{\lambda\mu} Y_H \, H^\dagger D^\lambda H - h.c. \right)$}}
& 
\multirow{4}{*}{0}
&
\multirow{4}{*}{0}
&
\multirow{4}{*}{0}
&
\multirow{4}{*}{11-16}
&
\multirow{4}{*}{$\propto m_Z^2/4M^2$}
&
\multirow{4}{*}{0}
&
\multirow{4}{*}{0}
\\
& & & & & & & 
\\
\multirow{2}{*}{\parbox[l]{118pt}{$\Lambda^{-4}\left(\phi^\dagger \partial^\mu \phi + h.c. \right) \times$ \\
$i\left(\widetilde{B}_{\lambda\mu} Y_H \, H^\dagger D^\lambda H - h.c. \right)$}}
& & & & & & &
\\
& & & & & & &
\\
\hline
\multirow{2}{*}{\parbox[l]{118pt}{$\Lambda^{-4}\left(\phi^\dagger \partial^\mu \phi + h.c. \right)\times$ \\
$i\left(W^a_{\ \lambda\mu}\, H^\dagger t^aD^\lambda H - h.c. \right)$}}
& 
\multirow{4}{*}{0}
&
\multirow{4}{*}{0}
&
\multirow{4}{*}{0}
&
\multirow{4}{*}{0.8-1.4}
&
\multirow{4}{*}{$\propto m_Z^2/4M^2$}
&
\multirow{4}{*}{0}
&
\multirow{4}{*}{0}
\\
& & & & & & & 
\\
\multirow{2}{*}{\parbox[l]{118pt}{$\Lambda^{-4}\left(\phi^\dagger \partial^\mu \phi + h.c. \right)\times$ \\
$i\left(\widetilde{W}^a_{\ \lambda\mu}\, H^\dagger t^aD^\lambda H - h.c. \right)$}}
& & & & & & &
\\
& & & & & & &
\\
\hline
\multirow{2}{*}{\parbox[l]{118pt}{$\Lambda^{-4}\bar{\chi}\gamma^\mu\chi \times$ \\
$i\left(B_{\lambda\mu} Y_H \, H^\dagger D^\lambda H + h.c. \right)$}}
& 
\multirow{4}{*}{0}
&
\multirow{4}{*}{0}
&
\multirow{4}{*}{0}
&
\multirow{4}{*}{0}
&
\multirow{4}{*}{0}
&
\multirow{4}{*}{18-19}
&
\multirow{4}{*}{17}
\\
& & & & & & & 
\\
\multirow{2}{*}{\parbox[l]{118pt}{$\Lambda^{-4}\bar{\chi}\gamma^\mu\chi \times$ \\
$i\left(\widetilde{B}_{\lambda\mu} Y_H \, H^\dagger D^\lambda H + h.c. \right)$}}
& & & & & & &
\\
& & & & & & &
\\
\hline
\multirow{2}{*}{\parbox[l]{118pt}{$\Lambda^{-4}\bar{\chi}\gamma^\mu\chi\times$ \\
$i\left(W^a_{\ \lambda\mu}\, H^\dagger t^aD^\lambda H + h.c. \right)$}}
& 
\multirow{4}{*}{0}
&
\multirow{4}{*}{0}
&
\multirow{4}{*}{0}
&
\multirow{4}{*}{0}
&
\multirow{4}{*}{0}
&
\multirow{4}{*}{1.3-1.5}
&
\multirow{4}{*}{1.7}
\\
& & & & & & & 
\\
\multirow{2}{*}{\parbox[l]{118pt}{$\Lambda^{-4}\bar{\chi}\gamma^\mu\chi\times$ \\
$i\left(\widetilde{W}^a_{\ \lambda\mu}\, H^\dagger t^aD^\lambda H + h.c. \right)$}}
& & & & & & &
\\
& & & & & & &
\\
\hline
\multirow{2}{*}{\parbox[l]{118pt}{$\Lambda^{-4}\bar{\chi}\gamma^\mu\chi \times$ \\
$i\left(B_{\lambda\mu} Y_H \, H^\dagger D^\lambda H - h.c. \right)$}}
& 
\multirow{4}{*}{0}
&
\multirow{4}{*}{0}
&
\multirow{4}{*}{0}
&
\multirow{4}{*}{0}
&
\multirow{4}{*}{0}
&
\multirow{4}{*}{19-20}
&
\multirow{4}{*}{17}
\\
& & & & & & & 
\\
\multirow{2}{*}{\parbox[l]{118pt}{$\Lambda^{-4}\bar{\chi}\gamma^\mu\chi \times$ \\
$i\left(\widetilde{B}_{\lambda\mu} Y_H \, H^\dagger D^\lambda H - h.c. \right)$}}
& & & & & & &
\\
& & & & & & &
\\
\hline
\multirow{2}{*}{\parbox[l]{118pt}{$\Lambda^{-4}\bar{\chi}\gamma^\mu\chi\times$ \\
$i\left(W^a_{\ \lambda\mu}\, H^\dagger t^aD^\lambda H - h.c. \right)$}}
& 
\multirow{4}{*}{0}
&
\multirow{4}{*}{0}
&
\multirow{4}{*}{0}
&
\multirow{4}{*}{0}
&
\multirow{4}{*}{0}
&
\multirow{4}{*}{2-4}
&
\multirow{4}{*}{1.7}
\\
& & & & & & & 
\\
\multirow{2}{*}{\parbox[l]{118pt}{$\Lambda^{-4}\bar{\chi}\gamma^\mu\chi\times$ \\
$i\left(\widetilde{W}^a_{\ \lambda\mu}\, H^\dagger t^aD^\lambda H - h.c. \right)$}}
& & & & & & &
\\
& & & & & & &
\\
\hline
\multirow{2}{*}{\parbox[l]{118pt}{$\Lambda^{-4}\bar{\chi}\gamma^{\mu5}\chi \times$ \\
$i\left(B_{\lambda\mu} Y_H \, H^\dagger D^\lambda H - h.c. \right)$}}
& 
\multirow{4}{*}{0}
&
\multirow{4}{*}{0}
&
\multirow{4}{*}{0}
&
\multirow{4}{*}{0}
&
\multirow{4}{*}{0}
&
\multirow{4}{*}{10-15}
&
\multirow{4}{*}{$\propto m_Z^2/4M^2$}
\\
& & & & & & & 
\\
\multirow{2}{*}{\parbox[l]{118pt}{$\Lambda^{-4}\bar{\chi}\gamma^{\mu5}\chi \times$ \\
$i\left(\widetilde{B}_{\lambda\mu} Y_H \, H^\dagger D^\lambda H - h.c. \right)$}}
& & & & & & &
\\
& & & & & & &
\\
\hline
\multirow{2}{*}{\parbox[l]{118pt}{$\Lambda^{-4}\bar{\chi}\gamma^\mu\chi\times$ \\
$i\left(W^a_{\ \lambda\mu}\, H^\dagger t^aD^\lambda H - h.c. \right)$}}
& 
\multirow{4}{*}{0}
&
\multirow{4}{*}{0}
&
\multirow{4}{*}{0}
&
\multirow{4}{*}{0}
&
\multirow{4}{*}{0}
&
\multirow{4}{*}{0.6-1.3}
&
\multirow{4}{*}{$\propto m_Z^2/4M^2$}
\\
& & & & & & & 
\\
\multirow{2}{*}{\parbox[l]{118pt}{$\Lambda^{-4}\bar{\chi}\gamma^\mu\chi\times$ \\
$i\left(\widetilde{W}^a_{\ \lambda\mu}\, H^\dagger t^aD^\lambda H - h.c. \right)$}}
& & & & & & &
\\
& & & & & & &
\\
\end{tabular}
\end{ruledtabular}
\end{table}

\section{Conclusions \label{conclusions}}

In this paper we considered the scenario in which dark matter dominantly annihilates into Standard Model di-boson final states, including at least one SM gauge boson. There are many possible ways of realizing this scenario. We choose to take the effective field theory approach in which unknown new physics has been integrated out. In the resulting EFT, we have operators of the form $\Lambda^{4-d} J_{\rm SM} \cdot J_{\rm DM}$, where $d$ is the dimension of operator $J_{\rm SM} \cdot J_{\rm DM}$. The dependence of the unknown new physics is through a single suppression scale $\Lambda$.

We considered both scalar and fermionic dark-matter candidates, which we chose to be a Standard Model singlet.  We presented a complete list of effective operators up to dimension 8, consistent with all of the Standard Model gauge symmetries. Since a light Higgs boson has already been discovered, we have included it in the low-energy effective field theory as well, and used the formulation in which the Standard Model electroweak gauge symmetries are linearly realized. Since each operator we considered is gauge invariant and Hermitian, it does not have to be related to other operators. In particular, any single operator we considered could give the dominant contribution to dark-matter annihilation. This is the simplifying assumption we made in our phenomenological analysis. 

For each of the operators in this list we presented detailed calculations of the annihilation cross section into different possible final states, including $\gamma \gamma$, $\gamma Z$, $\gamma h$, $ZZ$, $W^+ W^-$, and $Zh$. We then computed the scale $\Lambda$ necessary to produce the correct dark-matter relic abundance. Motivated by the possible evidence of an 130 GeV photon line in the Fermi data,  we also calculated the scale necessary to produce such a signal. We found several examples in which both the photon line signal and the thermal relic abundance can be simultaneously produced by a single effective operator. This could offer a compelling simple explanation if the Fermi photon line is confirmed.  In many cases, such operators also predict the existence of additional photon lines and other correlated indirect-detection signals, possibly providing further confirmation of the dark-matter signal and information about the relevant operator.

In general, even without the assumption about the 130 GeV photon line signal, the study of this set of operators reveals many interesting patterns of indirect-detection signals.  In this paper, as a first step, we presented the relative photon line signal strengths as a function of the dark-matter mass. We found that the relative strengths can be very different for different operators. Such differences can yield valuable information about the underlying dark-matter annihilation process if a signal is observed. Connections can also be made with direct detection and LHC searches. We will leave this to a future study.


\acknowledgments

This work was supported in part by the Kavli Institute for Cosmological Physics at the University of Chicago through grant NSF PHY-1125897 and an endowment from the Kavli Foundation and its founder Fred Kavli.  L.T.W. is supported by  the DOE Early Career Award under grant de-sc0003930. E.W.K. would like to thank the Institute of Theoretical Physics of the University of Heidelberg for their hospitality.

\bibliography{diboson}

\end{document}